\documentclass[a4paper,fleqn,usenatbib]{mnras}

\usepackage{newtxtext,newtxmath}

\usepackage[T1]{fontenc}
\usepackage{ae,aecompl}

\usepackage{graphicx}	
\usepackage{amsmath}	
\usepackage{amssymb}	

\newcommand{\sampsiz}{138}
\newcommand{\uplim}{0.015}  
\newcommand{\medval}{21.59} 
\newcommand{\avfval}{0.005} 
\newcommand{\medzval}{2.78} 

\newcommand{\pcm}{\,cm$^{-2}$}	
\newcommand{\kms}{\,km\,s$^{-1}$}	
\newcommand{\pcc}{\,cm$^{-3}$}	
\newcommand{\HI}{{\ion{H}{I}}}	
\newcommand{\HII}{{\ion{H}{II}}}	
\newcommand{\NH}{\ensuremath{{N_{\HI}}}}
\newcommand{\fesc}{\ensuremath{f_{\rm esc}}}
\newcommand{\avfesc}{\ensuremath{\langle{f_{\rm esc}}\rangle}}
\newcommand{\lognhc}{\ensuremath{{\rm log}(\NH/{\rm cm}^{-2})}}	
\newcommand{\lognh}{\ensuremath{{\rm log}(\NH)}}	

\newcommand{\msol}{{M$_\odot$}}
\newcommand{\lya}{{Ly$\alpha$}}
\newcommand{\lyb}{{Ly$\beta$}}
\newcommand{\iraf}{{\sc Iraf}}
\newcommand*{\Num}{N\textsuperscript{{o}}}

\newcommand{\JensenA}{(1)}
\newcommand{\FynboA}{(2)}
\newcommand{\VreeswijkA}{(3)}
\newcommand{\HjorthA}{(4)}
\newcommand{\FynboB}{(5)}
\newcommand{\MollerA}{(6)}
\newcommand{\ShinA}{(7)}
\newcommand{\VreeswijkB}{(8)}
\newcommand{\JakobssonA}{(9)}
\newcommand{\FynboC}{(10)}
\newcommand{\BergerA}{(11)}
\newcommand{\TotaniA}{(12)}
\newcommand{\ChenA}{(13)}
\newcommand{\CharyA}{(14)}
\newcommand{\FerreroA}{(15)}
\newcommand{\This}{(16)}

\newcommand{\WisemanA}{(17)}
\newcommand{\PatelA}{(18)}
\newcommand{\deUPA}{(19)}
\newcommand{\KuinA}{(20)}
\newcommand{\DAvanzoA}{(21)}
\newcommand{\SavaglioA}{(22)}
\newcommand{\LevesqueA}{(23)}
\newcommand{\SelsingA}{(24)}
\newcommand{\CucchiaraA}{(25)}

\newcommand{\Zafar}{(26)}
\newcommand{\CucchiaraB}{(27)}
\newcommand{\JeongA}{(28)}
\newcommand{\ChornockA}{(29)}
\newcommand{\MelandriA}{(30)}
\newcommand{\PuglieseA}{(31)}

\newcommand{\ChenB}{(32)}
\newcommand{\PerleyA}{(33)}
\newcommand{\SchulzeA}{(34)}
\newcommand{\GreinerA}{(35)}
\newcommand{\McGuireA}{(36)}
\newcommand{\TanvirA}{(37)}
\newcommand{\McGuireB}{(38)}
\newcommand{\ThoeneA}{(39)}
\newcommand{\FriisA}{(40)}
\newcommand{\PerleyC}{(41)}
\newcommand{\LaskarA}{(42)}
\newcommand{\PerleyB}{(43)}
\newcommand{\MyersA}{(44)}

\title[The escape fraction of ionizing radiation]{The  fraction of ionizing radiation from massive stars that escapes to the intergalactic medium}

\author[N. R. Tanvir et al.]{N. R. Tanvir,$^{1}$\thanks{E-mail: nrt3@le.ac.uk (NRT)}
J. P. U. Fynbo,$^{2}$
A. de Ugarte Postigo,$^{3}$
J. Japelj,$^{4}$
K. Wiersema,$^{1}$
\newauthor
D. Malesani,$^{2}$
D. A. Perley,$^{5}$
A. J. Levan,$^{6}$
J. Selsing,$^2$
S. B. Cenko,$^{8,9}$
\newauthor
D. A. Kann,$^3$
B. Milvang-Jensen,$^{2}$
E. Berger,$^{7}$
Z. Cano,$^{3}$
R. Chornock,$^{10}$
\newauthor
S. Covino,$^{11}$
A. Cucchiara,$^{12}$
V. D'Elia,$^{13,14}$
P. Goldoni,$^{15}$
A. Gomboc,$^{16}$
\newauthor
K. E. Heintz,$^{17,2}$
J. Hjorth,$^2$
L. Izzo,$^3$ 
P. Jakobsson,$^{17}$
L. Kaper,$^{4}$
T. Kr\"uhler,$^{18}$
\newauthor
T. Laskar,$^{19,20}$
M. Myers,$^{20}$
S. Piranomonte,$^{13}$
G. Pugliese,$^4$ 
R. S\'anchez-Ram\'irez,$^3$
\newauthor
S. Schulze,$^{21}$
M. Sparre,$^{22,2}$
E. R. Stanway,$^{6}$
G. Tagliaferri,$^{11}$
C. C. Th\"one,$^3$
\newauthor
S. Vergani,$^{23}$
P. M. Vreeswijk,$^{21}$
R. A. M. J. Wijers,$^4$
D. Watson,$^{2}$
 and D. Xu,$^{24}$\\
\\
Affiliations are listed at the end of the paper}

\date{Accepted XXX. Received YYY; in original form ZZZ}

\pubyear{2018}

\begin{document}
\label{firstpage}
\pagerange{\pageref{firstpage}--\pageref{lastpage}}
\maketitle

\begin{abstract}
The part played by stars in the ionization of the intergalactic medium (IGM) remains an open question. 
A key issue is the proportion of the stellar ionizing radiation that escapes the galaxies in which it is produced. 
Spectroscopy of gamma-ray burst (GRB) afterglows can be used to determine the neutral hydrogen column-density, $\NH$, 
in their host galaxies and hence the opacity to extreme ultra-violet (EUV) radiation along the lines-of-sight to the bursts.
Thus, making the reasonable assumption that long-duration GRB locations are representative of the sites of massive stars that
dominate EUV production, one can calculate an average escape fraction of ionizing radiation in a way that is independent of
galaxy size, luminosity or underlying spectrum.
Here we present a sample of $\NH$ measures for \sampsiz\ GRBs 
in the range $1.6<z<6.7$
and use it to establish an average escape fraction at the Lyman limit of ${\avfesc}\approx\avfval$,
with a 98\% confidence upper limit of ${\avfesc}\approx\uplim$.
This analysis suggests that stars provide a small contribution to the ionizing radiation budget of the IGM
at $z<5$, where the bulk of the bursts lie. 
At higher redshifts, $z>5$, firm conclusions are limited by the small size of the GRB sample (7/\sampsiz),
but any decline in average \HI\ column-density seems to be modest. 
We also find no indication of a significant correlation of \NH\ with galaxy UV luminosity 
or host stellar mass, for the subset of events for which these are available.
We discuss in some detail a number of selection effects and potential biases.
Drawing on a range of evidence we argue that such effects, while not negligible,
are unlikely to produce systematic errors (in either direction) of more than a factor $\sim2$ in 
\fesc, and so would not affect the primary conclusions.
Given that many GRB hosts are low metallicity, high specific star-formation rate, dwarf galaxies, these results present a particular
problem for the hypothesis that such galaxies dominated the reionization of the universe.
\end{abstract}

\begin{keywords}
dark ages, reionization, first stars -- gamma-ray burst: general -- galaxies: ISM -- intergalactic medium
\end{keywords}



\section{Introduction}

A key question for our understanding of the reionization of hydrogen in the intergalactic
medium (IGM) is the extent to which  ionizing extreme ultraviolet (EUV) radiation from 
massive stars escapes from the galaxies in which it is produced.
This can be parameterised 
by the escape fraction, \fesc, the proportion of photons produced by stars at
the Lyman limit wavelength ($\lambda=912$\,\AA) that leave the virial radius of their host galaxy.
Only if the average escape fraction, \avfesc, is sufficiently high in the era of reionization 
\citep[EoR; $7\lesssim z\lesssim9$;][]{Planck2016}, 
i.e. \avfesc\ at least 0.1--0.2, is it likely that this phase change was predominantly driven
by EUV star-light \citep[e.g.][]{Ouchi2009,Bouwens2012,Finkelstein2012,Robertson2015,Faisst2016}. 
Otherwise some other significant source of ionizing radiation is required, such as
a large population of faint quasars \citep[][but see \citet{Hassan2018} for counter arguments]{Madau2015,Khaire2016}, X-ray binaries \citep{Mirabel2011,Fragos2013,Knevitt2014,Madau2017} 
or decaying/annihilating particles \citep{Sciama1982,Hansen2004}.

Direct searches for Lyman continuum emission below 912\,\AA\ in the rest frame are compromised
by absorption due to neutral gas in the intergalactic medium (the \lya\ forest), and essentially
impossible above $z\sim4$ as the IGM absorption becomes near total -- the so-called Gunn-Peterson trough \citep{Gunn1965}.
Observations at lower redshifts are still difficult, and there have been extensive efforts searching
for such continuum emission from star-forming galaxies at $z=2$--4 in recent years
\citep[e.g][]{Steidel2001,Shapley2006,Vanzella2010,Vanzella2012,Nestor2013,Mostardi2013,Vanzella2015,Japelj2017,Marchi2017}.
Results have been conflicting, particularly due to rare cases of low redshift galaxies aligning by
chance with higher redshift targets \citep[e.g.][]{Vanzella2012,Siana2015}, but given the scarcity of high escape fraction systems
it appears that  {\fesc} is not high on average \citep{Grazian2017,Rutkowski2017}. 
This is consistent with quasars being the primary source of EUV radiation maintaining a reionized IGM at $z<4$.
However, at least some individual cases at $z \gtrsim 3$ appear to have very high escape fractions 
${{\fesc}}\gtrsim0.5$ \citep{DeBarros2016,Vanzella2016,Shapley2016}, and so might be analogues of galaxies in the EoR.

To account for the discrepancy between the expected and observed level of the escape fraction, it has been often suggested that 
\avfesc\ may actually increase with decreasing galaxy luminosity and/or with increasing redshift \citep[e.g.][]
{Razoumov2010,Ciardi2012,Kuhlen2012,Fontanot2014,HXu2016,Anderson2017}. 
Observationally, it is hard to reach sufficiently stringent constraints on the 
escape fraction for faint galaxy populations to investigate its dependence on luminosity 
 \citep[e.g.][]{Japelj2017}, while due to IGM absorption the claim of
 a changing escape fraction with redshift can only be investigated 
 through  secondary means and simulations \citep[e.g.][]{Zackrisson2013,Sharma2016}.
The challenge facing simulators is to model in sufficient detail the complex baryonic physics and radiative transfer 
given limited resolution, while also sampling a range of galaxies and environments.
Typically, models of an instantaneous burst of star formation incorporating only single-star stellar evolution 
produce the large bulk of their EUV within a few Myr,
limiting the time available for feedback from winds, radiation and supernovae
to open windows in the surrounding high density gas.
Recently, models which include binary stellar evolution have been shown to prolong the period of high EUV
production to $\sim10$\,Myr, and hence hold more promise for clearing of local gas, and for at least a relatively high fraction of stars
leaving the immediate environment in which they formed  \citep{Stanway2016,Ma2016}.

An alternative route, 
first proposed by \citet{Chen2007}, 
to constraining 
 \avfesc\ empirically over a broad range in redshift is 
via spectroscopy of long-duration gamma-ray burst (GRB) afterglows. 
These very bright, but short-lived, continuum sources allow detailed abundance studies of host galaxy gas along
the line of sight. Crucially, this includes calculation of the neutral hydrogen column-density, \NH, in the host from
fitting the \lya\ absorption feature when it is seen \citep[e.g.][]{Prochaska2007,Fynbo2009}. 
This column-density can be directly converted into an opacity measure for EUV ionizing radiation and hence \fesc.
For any individual host galaxy, a single sight-line does not provide a robust measure of its average escape
fraction, but since long-duration GRBs are associated with the core-collapse of
massive stars \citep[e.g.][]{Hjorth2003a,Xu2013}, a sample of GRB afterglows should be representative of
the distribution of all sight-lines specifically to the locations of young stars largely responsible for EUV production.

To date, neutral hydrogen columns reported for GRB hosts have generally been high, 
mostly classified as damped \lya\ absorbers (DLAs;  $\lognhc>20.3$),
which is usually taken as being consistent with their massive-star progenitors remaining in or close to the 
dense molecular clouds in which they formed, and/or more generally residing at the hearts of gas-rich star-forming
galaxies \citep[e.g.][]{Jakobsson2006}. 
In fact, observations of the time-variability of fine-structure transitions in  GRB afterglow spectra, 
in the (dozen or so) systems where it has been measured, has allowed the distance of the dominant absorbing clouds to be established, ranging
from $\sim50$\,pc to $\gtrsim1$\,kpc \citep[e.g.][]{Vreeswijk2013}.
These scales are comparable to the sizes of large ionized superbubbles around star-forming regions in the low redshift universe 
\citep{Oey1997,CampsFarina2017}, and so might indicate that absorption takes place due to neutral gas piled up at the
boundaries of such bubbles  for some GRBs.
In any case, this tendency towards high column-densities is potentially a problem since if the EUV escape fraction
is to fulfil the requirements for reionization, a significant proportion of
GRBs should have very low \HI\ columns ($\lognhc\ll 18$), particularly at $z>5$.

This method has the considerable advantage that GRB afterglows readily probe gas in even very faint galaxies,
for which either Lyman-continuum observations would be weakly constraining, or which may be missed completely
 in traditional galaxy surveys. 
In principle, then, with a sufficiently large sample one could trace the escape fraction both as a function of galaxy luminosity and of redshift.
The tendency of GRBs to occur preferentially in lower metallicity galaxies, $Z/Z_{\odot}\lesssim0.3$--1
\citep[e.g.][]{Perley2016b,Japelj2016,Graham2017,Vergani2017}, 
often dwarfs with high specific star formation rates \citep[e.g.][]{Svensson2010,Hunt2014}, also suggests they should be more
representative of the populations dominant during the EoR.

\citet{Chen2007} and \citet{Fynbo2009} have previously performed such analyses,
each obtaining 95\% confidence upper limits on \avfesc\ of only $\sim0.075$ based on samples
of $\sim30$ GRBs (with some overlap of their samples) with redshifts $2\lesssim z\lesssim6$.  
Here we reinvestigate this issue, using
a considerably larger sample of \sampsiz\ GRBs with \NH\ determinations, spanning a redshift range $1.6\lesssim z\lesssim6.7$.
The structure of the paper is as follows: in Section~\ref{sec:samp} we present the sample of GRBs
and describe its basic properties; in Section~\ref{sec:fesc} we outline the implications for the average escape
fraction of ionizing EUV radiation and consider evidence for evolution over cosmic history;
in Section~\ref{sec:syst} we consider a range of potential systematic uncertainties which
could bias our conclusions in either direction, and address more
fully the question of how representative GRB sight-lines are likely to be for the stellar
populations of interest for reionization; finally in Section~\ref{sec:conc} we draw our conclusions.
Further details of some individual GRBs, including  results not already reported elsewhere, are given
in an Appendix.

\section{The  H\,{\small I} column-density sample}
\label{sec:samp}

Spectra of GRB afterglows frequently exhibit strong \lya\ absorption lines which can be modelled to
constrain the column-density of neutral hydrogen responsible.
Particularly at higher redshifts, this relies largely on fitting the red wing of the line, since absorption
by the IGM \lya\ forest significantly affects the blue wing.
In the large majority of cases, the systemic redshifts are known quite precisely from metal-line detections,
which improves the precision of the \lya\ fits.
We have gathered together \HI\ column-densities towards GRBs from the literature, and combined them with
a large number of new measurements we have made using afterglow spectra from various sources.
Many of these come from the long-running Very Large Telescope (VLT) X-shooter legacy programme {\citep{Selsing2018},
but we also include data from the Nordic Optical Telescope (NOT), 
the William Herschel Telescope (WHT),
the Gran Telescopio Canarias (GTC),
the Telescopio Nazionale Galileo (TNG),
the Gemini Telescopes (both North and South),
the Asiago Copernico Telescope (CT),
and other VLT spectrographs.
The bulk of the GRBs were originally discoveries of the Neil Gehrels {\em Swift} Observatory, but there was  little consistency in terms of 
which bursts were followed-up or the kinds of observations obtained (e.g. in terms of spectral resolution, wavelength
coverage, sensitivity etc.).
The net result is an inhomogeneous sample, and potential effects of selection biases are discussed in Section~\ref{sec:syst}.

The HI column densities measured from the afterglow spectra in our sample are plotted in Figure~\ref{fig:nhz} and summarised in Table~\ref{tab:data};
corresponding primary sources for the adopted values of \lognh\ are given in the 4th column of the table, and readers
are refered also to 
Appendix~\ref{appendix} for further details regarding previously unreported fits and additional comments on some particular cases.
The total number of sight-lines is \sampsiz, which represents a more than four-fold increase over similar
previous studies \citep{Chen2007,Fynbo2009}.

The median redshift of our sample is $\tilde{z}=\medzval$.
The lower redshift cut-off, at $z\sim1.6$, occurs because the observed wavelength of
\lya\ falls in the near-UV, and begins to be strongly affected by declining atmospheric
transmission.
At the high redshift end, the sample is curtailed due to declining spectral quality; 
although bursts at $z\gtrsim6.7$ have been found, either 
the signal-to-noise has been too poor to reliably measure the red damping wing of \lya\ \citep{Tanvir2009,Salvaterra2009,Tanvir2017}
or the redshift has been inferred photometrically \citep{Cucchiara2011a}.

\begin{figure*}
\includegraphics[angle=270,width=1.7\columnwidth]{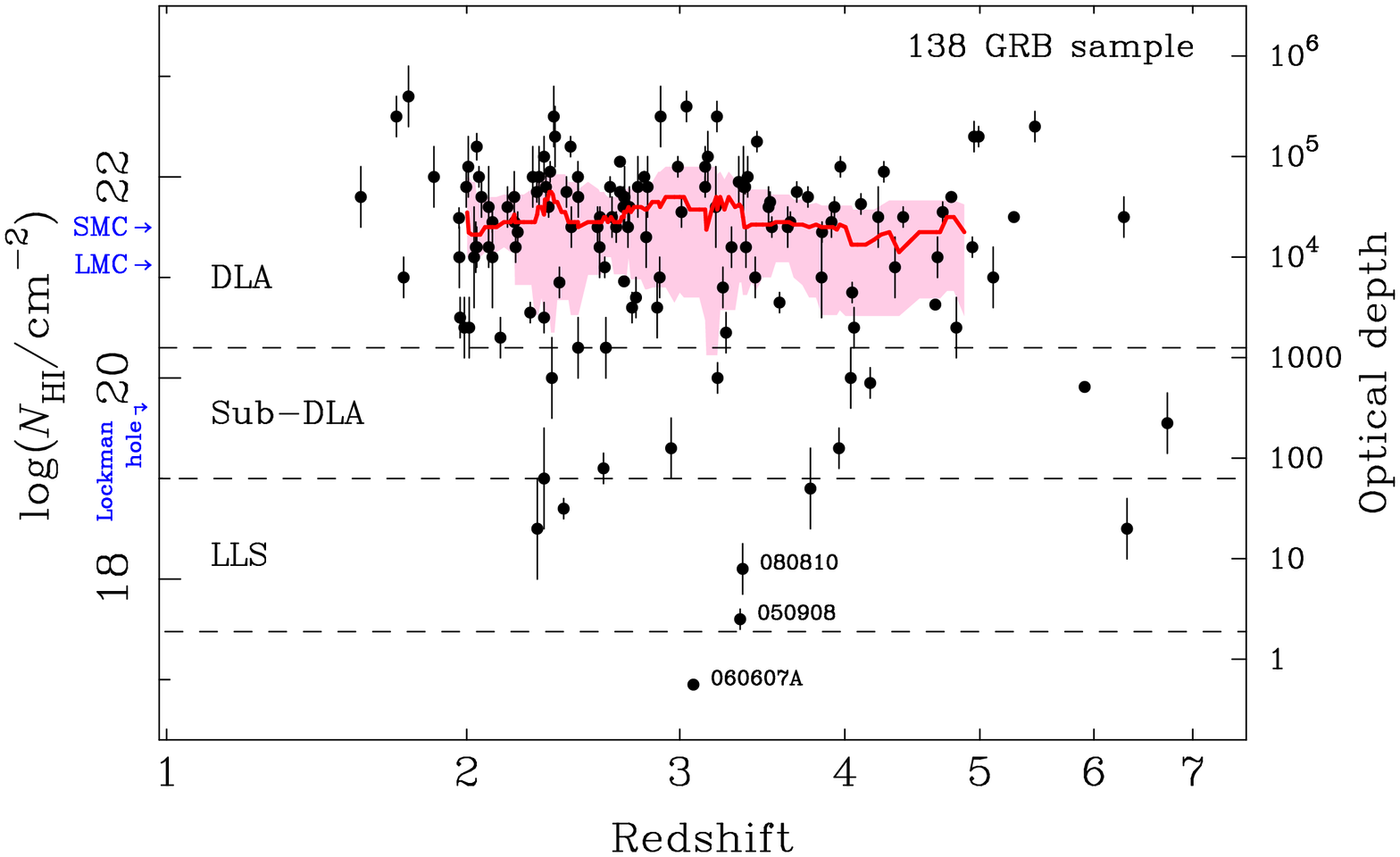}
    \caption{The values of neutral hydrogen column-density in the host plotted against redshift for the sample  of GRBs.
    The corresponding optical depth at the Lyman limit is shown on the right-hand axis.
    The large majority are DLAs, with a smaller proportion (17/\sampsiz) being classified as sub-DLAs, Lyman Limit Systems (LLS)
    and below.
    All sight-lines apart from the two with the lowest column-densities are essentially opaque ($\tau\gtrsim10$)
    to EUV ionizing radiation.
    The running median (red line) and interquartile range (pink shading) 
    of 20 points shows no evidence for significant variation with redshift.
    For comparison, on the left axis we also mark the locations of the median {\rm log}(\NH/2) values in the directions of LMC and SMC
    \HII\ regions from \citet{Pellegrini2012} (halving the measured columns is appropriate since the they include
    contributions from both the foreground and background of the \HII\ region), and the lowest column density out of the Milky Way
    from the position of the Sun,
    the ``Lockman hole" \citep{Lockman1986}.}
    \label{fig:nhz}
\end{figure*}

\begin{table*}
	\centering
	\caption{The sample of GRBs.
	References:- 
	\JensenA\ \citet{Jensen2001},
	\FynboA\ \citet{Fynbo2002},
	\VreeswijkA\ \citet{Vreeswijk2006}, \HjorthA\ \citet{Hjorth2003b}, \FynboB\ \citet{Fynbo2005}, \MollerA\ \citet{Moller2002}, 
	\ShinA\ \citet{Shin2006}, \VreeswijkB\ \citet{Vreeswijk2004}, \JakobssonA\ \citet{Jakobsson2004},
	\FynboC\ \citet{Fynbo2009},
	\BergerA\ \citet{Berger2006}, \TotaniA\ \citet{Totani2006}, \ChenA\ \citet{Chen2007}, \CharyA\ \citet{Chary2007},
	\FerreroA\ \citet{Ferrero2009}, (16) This work, 
	\WisemanA\ \citet{Wiseman2017}, \PatelA\ \citet{Patel2010}, \deUPA\ \citet{deUP2012},
	\KuinA\ \citet{Kuin2009}, \DAvanzoA\ \citet{DAvanzo2010}, \SavaglioA\ \citet{Savaglio2012}, \LevesqueA\ \citet{Levesque2010}, 
	\SelsingA\ \citet{Selsing2018}, \CucchiaraA\ \citet{Cucchiara2011b}, (26) Zafar et al. in prep.,
	\CucchiaraB\ \citet{Cucchiara2015}, \JeongA\ \citet{Jeong2014}, 
	\ChornockA\  \citet{Chornock2014},
	\MelandriA\ \citet{Melandri2015}, \PuglieseA\ Pugliese et al. in prep., \ChenB\ \citet{Chen2009}, \PerleyA\ \citet{Perley2011},
	\SchulzeA\ \citet{Schulze2015}, \GreinerA\ \citet{Greiner2015}, \McGuireA\ \citet{McGuire2016}, \TanvirA\ \citet{Tanvir2012},
	\McGuireB\ McGuire et al. in prep., \ThoeneA\ \citet{Thoene2011}, \FriisA\ \citet{Friis2015}, \PerleyC\ \citet{Perley2013},
	\LaskarA\ \citet{Laskar2011}, \PerleyB\ \citet{Perley2016b}, \MyersA\ Myers et al. in prep. \citep[in this case, the derived {\em Spitzer} photometry
	was transformed to stellar mass estimates following][]{Perley2016b}.
	}
	\label{tab:data}
	\begin{tabular}{lcclclcl} 
		\hline
GRB & $z$ & ${{\rm log}\left(\frac{\NH}{{\rm cm}^{-2}}\right)}$ &  Refs. & $M_{\rm UV,AB}$ & Refs. & log(M$_*$/M$_{\odot}$) & Refs.\\
		\hline
000301C &   2.03 &  $21.20\pm0.50$      & \JensenA   & $-16.0\pm0.5$ & \ChenB  & & \\
000926  &   2.04 &  $21.30\pm0.25$ & \FynboA  & $-20.40\pm0.07$ & \ChenB   & $9.64$ & \PerleyC \\ 
011211  &   2.14 &  $20.40\pm0.20$ & \VreeswijkA  & $-19.97\pm0.10$ & \ChenB   & $8.0$ & \PerleyC \\ 
020124  &   3.20 &  $21.70\pm0.40$ & \HjorthA & $>-15.72$ & \GreinerA   & $<9.90$ & \LaskarA \\  
021004  &   2.33 &  $19.00\pm0.50$ & \FynboB,\MollerA,\ref{app:021004}  & $-20.56\pm0.10$ & \FynboB   & $9.45$ & \PerleyC \\
030226  &   1.99 &  $20.50\pm0.30$ & \ShinA  & &   & & \\
030323  &   3.37 &  $21.90\pm0.07$ & \VreeswijkB  & $-18.47\pm0.1$ & \ChenB   & $<9.85$ & \LaskarA \\
030429  &   2.65 &  $21.60\pm0.20$ & \JakobssonA & &   & & \\
050319  &   3.24 &  $20.90\pm0.20$ &\FynboC & $>-20.01$ & \PerleyA    & $9.69$ & \PerleyB \\
050401  &   2.90 &  $22.60\pm0.30$ & \FynboC & $-19.28\pm0.31$ &  \SchulzeA    &  $9.61$ &  \PerleyB \\
050505  &   4.27 &  $22.05\pm0.10$ & \BergerA & &   & $<9.67$ & \MyersA \\ 
050730  &   3.97 &  $22.10\pm0.10$ & \FynboC & $>-17.21$ & \GreinerA    & $<9.46$ &  \PerleyB \\ 
050820A &   2.61 & $21.10\pm0.10$ & \FynboC & $-18.93\pm0.06$ & \SchulzeA    & $9.38$ &  \PerleyB \\
050904  &   6.29 &  $21.60\pm0.20$ & \TotaniA & $-19.21\pm0.2$ & \McGuireA   & $<10.07$ &   \PerleyB \\
050908  &   3.34 &  $17.60\pm0.10$ & \FynboC & $-18.18\pm0.27$ & \SchulzeA & $<9.91$ & \LaskarA \\
050922C &   2.20 & $21.55\pm0.10$ & \FynboC & $>-17.95$ & \PerleyA    & $<9.01$ &  \PerleyB \\
060115  &   3.53 &   $21.50\pm0.10$ & \FynboC &  $-18.61\pm0.27$ & \SchulzeA & $9.43$ &  \PerleyB \\   
060124  &   2.30 &  $18.50\pm0.50$ & \FynboC & &  & &  \\ 
060206  &   4.05 &  $20.85\pm0.10$ & \FynboC & $-18.47\pm0.1$ &  \GreinerA   & $<9.95$ &   \\
060210  &   3.91 &  $21.55\pm0.15$ & \FynboC & $-21.85\pm0.12$ &  \PerleyA    & $10.46$&  \PerleyB \\ 
060223A &   4.41 & $21.60\pm0.10$ & \ChenA & $-18.33\pm0.15$ &  \GreinerA  & $<10.17$ & \LaskarA \\
060510B &   4.94 & $21.30\pm0.10$ & \ChenA & $-20.51\pm0.17$ &  \GreinerA  & $9.86$ &  \MyersA \\  
060522  &   5.11 &  $20.60\pm0.30$ & \CharyA,\This & $>-18.34$ &    \TanvirA   & $<9.31$ &  \PerleyB \\  
060526  &   3.21 &  $20.00\pm0.15$ & \FynboC & $>-17.36$ &  \GreinerA    & $9.30$ &  \PerleyB \\      
060605 & 3.77 &     $18.90\pm0.40$ & \FerreroA & $-17.94\pm0.2$ & \GreinerA   & $<9.97$ & \LaskarA \\
060607A &  3.08 &  $16.95\pm0.03$ & \FynboC & $>-15.52$ &  \GreinerA   & $<9.45$ &  \PerleyB \\
060707  &   3.43 &  $21.00\pm0.20$ & \FynboC & $-20.78\pm0.06$ &  \SchulzeA    & $9.99$ &  \PerleyB \\  
060714  &   2.71 &  $21.80\pm0.10$ & \FynboC & $-18.88\pm0.28$ &  \SchulzeA   & $9.25$ &  \PerleyB \\
060906  &   3.69 &  $21.85\pm0.10$ & \FynboC & $>-20.60$ &   \GreinerA   & $<10.02$ & \LaskarA \\ 
060926  &   3.21 &  $22.60\pm0.15$ & \FynboC & $-21.59\pm0.05$ & \GreinerA    & $10.71$ & \LaskarA \\ 
060927  &   5.47 &  $22.50\pm0.15$ & \FynboC & $>-18.01$ &   \TanvirA   & $<9.63$ &  \PerleyB \\
061110B &   3.44 & $22.35\pm0.10$ & \FynboC & $-19.82\pm0.29$ & \GreinerA   & $<9.47$ &  \PerleyB\\ 
070110  &   2.35 &  $21.70\pm0.10$ & \FynboC & $-19.81\pm0.11$ &  \SchulzeA   & $<9.16$ &  \PerleyB \\
070411  &   2.95 &  $19.30\pm0.30$ & \FynboC & &   & &  \\
070506  &   2.31 &  $22.00\pm0.30$ & \FynboC & $-18.80\pm0.21$ & \SchulzeA   & &  \\
070611  &   2.04 &  $21.30\pm0.20$ & \FynboC & $>-17.52$ &  \SchulzeA   & &  \\
070721B &   3.63 & $21.50\pm0.20$ & \FynboC & $-18.39\pm0.44$ &  \GreinerA   &  $<9.42$ &  \PerleyB \\     
070802  &   2.45 &  $21.50\pm0.20$ & \FynboC & $-19.85\pm0.2$ &  \SchulzeA & $9.69$ & \PerleyC \\
070810A  &   2.17 & $21.70\pm0.20$ & \This  & &   & & \\
071031  &   2.69 &  $22.15\pm0.05$ & \FynboC & &   & &  \\
080129 & 4.35 &   $21.10\pm0.30$ & \This & &   & $<11.47$ & \MyersA \\ 
080210  &   2.64 & $21.90\pm0.10$ & \FynboC & $>-19.32$ & \PerleyA   & $<9.50$ &  \PerleyB \\
080310  &   2.43 &  $18.70\pm0.10$ & \FynboC & $>-18.45$ &   \PerleyA   & $9.78$ &  \PerleyB \\
080413A &   2.43 & $21.85\pm0.15$ & \FynboC & &   & $<9.64$ &  \PerleyB\\
080603B &   2.69 & $21.85\pm0.05$ & \FynboC & &   & $9.15$ &  \PerleyB \\
080607  &   3.04 &  $22.70\pm0.15$ & \FynboC & $-19.4\pm0.5$ &   \GreinerA    & $10.45$ &  \PerleyB \\
080721  &   2.59 &  $21.60\pm0.10$ & \FynboC & &   & $<9.63$ &  \PerleyB \\
080804  &   2.20 &  $21.30\pm0.15$ & \FynboC & &   & $9.28$ &  \PerleyB \\
080810 & 3.36 &   $18.10\pm0.25$ & \WisemanA,\ref{app:080810} & $-22.43\pm0.5$ & \GreinerA   & $10.29$ &  \MyersA \\ 
080905B & 2.37 & $22.60\pm0.30$ & \This & &   & & \\
080913  &   6.73 &  $19.55\pm0.30$ & \PatelA,\ref{app:080913} & $>-17.28$ &   \McGuireB    & & \\
081008  &   1.97 &  $21.59\pm0.10$ & \deUPA & &   & $9.18$ &  \PerleyB \\
081029 & 3.85 &   $21.45\pm0.10$ & \This & $>-19.87$ &   \GreinerA    & $<9.39$ &  \PerleyB \\
081118 & 2.58 &    $21.50\pm0.20$ & \This & &   & $9.18$ &  \PerleyB\\
081203A &  2.05 & $22.00\pm0.10$ & \KuinA & &   & & \\
		\hline
	\end{tabular}
\end{table*}

\begin{table*}
	\centering
	\contcaption{}
	\begin{tabular}{lcclclcl} 
		\hline
GRB & $z$ & ${{\rm log}\left(\frac{\NH}{{\rm cm}^{-2}}\right)}$ &  Refs. & $M_{\rm UV,AB}$ & Refs. & log(M$_*$/M$_{\odot}$) & Refs.\\
		\hline
081222 & 2.77 & $20.80\pm0.20$ & \This & &  & $9.61$ &  \PerleyB \\
090205  &   4.65 & $20.73\pm0.05$ & \DAvanzoA & $-21.26\pm0.13$ &  \GreinerA   & $<10.7$ & \DAvanzoA \\
090313  &   3.38 &  $21.30\pm0.20$ &\This & $>-21.19$ &  \GreinerA  & $<10.35$ &  \MyersA  \\ 
090323 & 3.58 & $20.75\pm0.10$ & \SavaglioA,\ref{app:090323} & $-21.60\pm0.18$ &  \GreinerA  & $10.58$ & \MyersA  \\ 
090426  &   2.61 & $19.10\pm0.15$ & \LevesqueA,\ref{app:090426} & $-20.43\pm0.08$ &  \ThoeneA  & &   \\
090516A  &   4.11 &  $21.73\pm0.10$ & \deUPA & $-20.99\pm0.4$ & \GreinerA  & $10.63$ &  \MyersA \\ 
090519 & 3.85 & $21.00\pm0.40$ & \This & $>19.09$ & \GreinerA & $<10.35$ &  \PerleyB  \\
090529 & 2.62 & $20.30\pm0.30$ & \This & &  &  &    \\
090715B & 3.01 & $21.65\pm0.15$ & \This & &  & $10.20$ &  \MyersA  \\ 
090726 & 2.71 & $21.80\pm0.30$ & \This & &  &  &    \\
090809  &   2.74 &  $21.70\pm0.20$ & {\SelsingA} & & & &   \\
090812 & 2.45 & $22.30\pm0.10$ & \deUPA  & & & $<9.35$ &  \PerleyB \\
090926A  &   2.11 &  $21.55\pm0.10$ & {\SelsingA} & & & &  \\
091029 & 2.75 & $20.70\pm0.15$ & \This & & & $<9.81$ &  \PerleyB   \\
100219A &   4.67 &  $21.20\pm0.20$ & {\SelsingA} & $-19.74\pm0.5$ & \GreinerA  & $<10.11$ & \MyersA  \\ 
100302A &   4.81 &  $20.50\pm0.30$ & \This & &  & &   \\
100316A &   3.16 &  $22.20\pm0.25$ & \This & &  & &   \\
100425A &   1.76 &  $21.00\pm0.20$ & {\SelsingA} & & & &   \\
100513A &  4.77 &   $21.80\pm0.05$ & \This & $-19.88\pm0.29$ & \GreinerA  & $<10.14$ & \MyersA  \\ 
100728B &   2.11 &  $21.20\pm0.50$ & {\SelsingA} & &  & $<9.25$ &  \PerleyB  \\
110128A &   2.34 &  $21.90\pm0.15$ & {\SelsingA} & &  & &   \\
110205A &   2.21 &  $21.45\pm0.20$ & \CucchiaraA & &  & $9.72$ &   \PerleyB \\
110731A & 2.83 &    $21.90\pm0.30$ & \This & &  & &   \\
110818A & 3.36 &    $21.90\pm0.40$ & {\SelsingA} & $-21.68\pm0.05$ &   \GreinerA & &  \\
111008A &   4.99 &  $22.40\pm0.10$ & {\SelsingA} & $>-20.80$ &  \GreinerA  & &   \\ 
111107A &   2.89 &  $21.00\pm0.20$ & {\SelsingA} & &  & &   \\
120119A &   1.73 &  $22.60\pm0.20$ & {\SelsingA} & &  & $9.91$ &  \PerleyB  \\
120327A &   2.81 &  $22.00\pm0.05$ & {\SelsingA} & &  & &   \\
120404A &   2.88 &  $20.70\pm0.30$ & {\SelsingA} & &  & &   \\
120712A &   4.17 &  $19.95\pm0.15$ & {\SelsingA} & &  & $<9.82$ & \MyersA  \\ 
120716A &   2.49 &  $22.00\pm0.15$ & {\SelsingA} & &  & &   \\
120811C & 2.67 &    $21.50\pm0.15$ & \This  & &  & &  \\
120815A &   2.36 &  $22.05\pm0.10$ & {\SelsingA} & &   & &  \\
120909A &   3.93 &  $21.70\pm0.10$ & {\SelsingA} & $-21.16\pm0.12$ & \GreinerA  & &   \\ 
121024A &   2.30 &  $21.85\pm0.10$ & {\SelsingA} & $-21.47\pm0.10$ &  \FriisA    & $9.9$ & \FriisA    \\
121027A &  1.77 &   $22.80\pm0.30$ & {\SelsingA},\ref{app:121027A} & & & &   \\
121128A &  2.20 &   $21.80\pm0.25$ & {\This} & & & &   \\
121201A &   3.39 &  $22.00\pm0.20$ & {\SelsingA} & $-20.84\pm0.21$ &    \GreinerA  & &  \\
121229A &   2.71 &  $21.70\pm0.20$ & {\SelsingA}& &   & &  \\
130408A &   3.76 &  $21.80\pm0.10$ & {\SelsingA} & $>-21.13$ & \GreinerA  & &    \\ 
130427B &   2.78 &  $21.90\pm0.30$ & {\SelsingA} & &   & &  \\
130505A &   2.27 &  $20.65\pm0.10$ & \CucchiaraB   & &  & &   \\
130518A &   2.49 &  $21.80\pm0.20$ & \This  & &  & &   \\
130606A &   5.91 &  $19.91\pm0.02$ & {\SelsingA}  & $-20.33\pm0.15$ & \McGuireA  & &   \\
130610A &   2.09 &  $21.30\pm0.20$ & \This & &  & &   \\
130612A &   2.01 &  $22.10\pm0.30$ & {\SelsingA} & &  & &   \\
131011A &   1.87 &  $22.00\pm0.30$ & {\SelsingA} & &  & &   \\
131108A &   2.40 &  $20.95\pm0.15$ & \This  & &  & &   \\
131117A &   4.04 &  $20.00\pm0.30$ & {\SelsingA}  & &  & &   \\
140206A &   2.73 &  $21.50\pm0.20$ & \This  & &  & &   \\
140226A &   1.97 &  $20.60\pm0.20$ & \CucchiaraB   & &  & &   \\
140304A &   5.28 &  $21.60$ & \JeongA  &  & & &   \\
140311A &   4.95 &  $22.40\pm0.15$ & {\SelsingA}  & & & $<10.10$ & \MyersA   \\ 
140419A &   3.96 &  $19.30\pm0.20$ & \CucchiaraB   & &  & &   \\
140423A &   3.26 &  $20.45\pm0.20$  & \CucchiaraB  & &  & &   \\
140430A & 1.60 &    $21.80\pm0.30$ & {\SelsingA}  & &  & &   \\
140515A &   6.32 &  $18.50\pm0.30$ & \ChornockA,\MelandriA,\ref{app:140515A}  & $-18.31\pm0.35$ & \McGuireA  & &   \\
140518A &   4.71 &  $21.65\pm0.10$ & \CucchiaraB   & &  & &   \\
140614A &   4.23 &  $21.60\pm0.30$ & {\SelsingA}  & &  & &   \\
140629A &   2.28 &  $22.00\pm0.30$ & \This  & &  & &   \\
140703A &   3.14 &  $21.90\pm0.10$ & \This  & &  & &   \\
140808A &   3.29 &  $21.30\pm0.20$ & \This  & &  & &   \\
141028A &   2.33 &  $20.60\pm0.15$ & {\SelsingA}  & &  & &   \\
141109A &   2.99 &  $22.10\pm0.10$ & {\SelsingA}  & &  & &   \\
		\hline
	\end{tabular}
\end{table*}

\begin{table*}
	\centering
	\contcaption{}
	\begin{tabular}{lcclclcl} 
		\hline
GRB & $z$ & ${{\rm log}\left(\frac{\NH}{{\rm cm}^{-2}}\right)}$ &  Refs. & $M_{\rm UV,AB}$ & Refs. & log(M$_*$/M$_{\odot}$) & Refs.\\
		\hline
150206A &   2.09 &  $21.70\pm0.40$ & {\SelsingA}  & &  & &   \\
150403A &   2.06 &  $21.80\pm0.20$ & {\SelsingA}  & &  & &   \\
150413A &   3.14 &  $22.10\pm0.20$ & \This & &  & &  \\
150915A &   1.97 &  $21.20\pm0.30$ & {\SelsingA}  & &  & &   \\
151021A &   2.33 &  $22.20\pm0.20$ & {\SelsingA}  & &  & &   \\
151027B &   4.06 &  $20.50\pm0.20$ & {\SelsingA}  & &  & &   \\
151215A &   2.59 &  $21.30\pm0.30$ & \This  & &  & &   \\
160203A &   3.52 &  $21.75\pm0.10$ & {\SelsingA},\PuglieseA  & &  & &   \\
160227A &   2.38 &  $22.40\pm0.30$ & \This  & &  & &   \\
160629A &   3.33 &  $21.95\pm0.25$ & \This  & &  & &   \\
161014A &   2.82 &  $21.40\pm0.30$ & {\SelsingA}  & &  & &   \\
161017A &   2.01 &  $20.50\pm0.30$ & \This  & $>-20.11$ & \This  & &  \\
161023A &   2.71 &  $20.96\pm0.05$ & {\SelsingA}  & &  & &   \\
170202A &   3.65 &  $21.55\pm0.10$ & {\SelsingA}  & &  & &   \\
170405A &   3.51 &  $21.70\pm0.20$ & \This   & &  & &   \\
170531B &   2.37 &  $20.00\pm0.40$ & \This   & &  & &   \\
180115A &   2.49 &  $20.30\pm0.30$ & \This   & &  & &   \\
180325A &   2.04 &  $22.30\pm0.14$ & {\Zafar}   & &  & &   \\
180329B &   2.00 &  $21.90\pm0.20$ & \This   & &  & &   \\
		\hline
	\end{tabular}
\end{table*}

\section{Implications for the ionizing escape fraction}
\label{sec:fesc}

Following  \citet{Chen2007}, 
we note that
the optical depth for radiation at the Lyman limit (912 \AA) along a given sight line due to absorption by neutral hydrogen
is given by $\tau=\sigma_{\rm LL}\NH$, where $\sigma_{\rm LL}=6.28\times10^{-18}$\,cm$^2$ is the photoionization
cross-section of hydrogen. Hence the average escape fraction for  $n$ sight lines is given by

\begin{equation}
    \langle{f_{\rm esc}}\rangle=\frac{1}{n}\sum\limits_{i=1}^n {{\rm exp}(-\tau_i)}.
	\label{eq:one}
\end{equation}
Considering our whole sample, we find a mean value of only $\langle f_{\rm esc}\rangle=\avfval$, well below that thought to be required in the EoR.

In Figure~\ref{fig:cum} 
we plot the cumulative distribution of \HI\ column-density measures for the whole sample.
The median value of column-density is $\lognhc=\medval$, consistent with previous studies \citep[e.g. the equivalent figure is 21.5 for the sample 
of ][]{Fynbo2009}
and also similar to the median values of \lognh\ towards \HII\ regions in the Magellanic Clouds \citep[][see Figure~\ref{fig:nhz}]{Pellegrini2012}.
We find that up to the median point the sample is well described by a simple power law distribution $P(<x)\propto x^{0.4}$,
where $x$ is the value of $\NH$, shown as an orange dashed line in the figure.
While this model is not motivated by any particular physical considerations, it does provide a smooth
representation of the data, and using it we obtain an average escape fraction 
$\langle f_{\rm esc}\rangle=0.004$, in good agreement with the value found above.

Most of the rest of this paper is concerned with the robustness of this result, 
and the statistical and potential systematic uncertainties that may affect it.

\subsection{Low column-density sight-lines}
\label{sub:low}
Only two sight-lines have $\lognhc<18$
(corresponding to $\fesc>0.002$), 
and as it happens both of these low column-density systems were already included in the \citet{Chen2007} 
and \citet{Fynbo2009} analyses.
Since the numerical result for \avfesc\ depends entirely on these two sight-lines, we review here what is known of their properties
and in particular consider whether there could be attenuation of EUV radiation by dust as well
as \HI\ absorption. We also address the effect of direct recombinations to the ground-state producing 
ionizing photons.

\subsubsection{GRB\,050908}
\label{subsub:050908}
GRB\,050908, at $z=3.34$, had a moderately bright optical afterglow, being $R\approx19$ at 15\,min post-burst \citep{Torii2005}.
Spectroscopy was obtained with Gemini/GMOS, from which
\citet{Chen2007} reported a column-density of $\lognhc=19.1$.
The lower value of $\lognhc=17.6$, corresponding to $\fesc=0.08$, used here, 
was that derived from a VLT/FORS1 spectrum \citep{Fynbo2009},  and it is preferred since
it is based on direct evidence of non-zero afterglow continuum emission below the Lyman limit.

There is no indication of excess absorption in the {\em Swift} X-ray 
observations\footnote{\url{http://www.swift.ac.uk/xrt_spectra/00154112/}}
\citep{Evans2009}, which is also consistent with a low column density and low extinction sight-line.

\subsubsection{GRB\,060607A}
\label{subsub:060607A}
GRB\,060607A, at $z=3.08$, had a very bright and well-studied early optical afterglow, which reached $r=14.3$ at 3\,min post-burst
\citep{Nysewander2009}.
This sight-line has the lowest column-density of our sample at $\lognhc=16.95$, from a VLT/UVES spectrum \citep{Fynbo2009},
corresponding $\fesc=0.57$.  The spectrum also showed evidence for emission below the Lyman limit, although only for
a small stretch of wavelength before it was cut-off by an intervening absorber, but this is consistent with a very low opacity.

The light curve and spectral energy distribution were studied in detail by \citet{Nysewander2009},
who modelled their {\it Bgri} optical data together with $H$-band photometry from
\citet{Molinari2007}.  They concluded that a rest-frame dust extinction of zero was ruled out
at the $2.6\sigma$ level.  
The shape of the extinction law is only weakly constrained by these data, so extrapolating to the
Lyman limit introduces a large systematic uncertainty, but with reasonable dust laws
their favoured extinction would correspond to a value of $A_{912}\gtrsim1$.
If this inference is correct, then it would suggest the actual escape fraction at the Lyman
limit for GRB\,060607A could be significantly diminished by dust extinction, by a factor $\sim2.5$ or
more.

We note that there is also marginal evidence of X-ray absorption\footnote{\url{http://www.swift.ac.uk/xrt_spectra/00213823/}}  
in the source-frame at a level of $N_{\rm H,X}\approx (3.1\pm1.9)\times10^{21}$\pcm\ 
\citep{Evans2009}
over the Milky Way  foreground \citep{Willingale2013}.
This would be broadly consistent with a SMC dust-to-gas ratio \citep{Bouchet1985} 
providing the hydrogen associated with this gas had largely been ionized (so it was not seen in
the optical spectrum) but  the dust had mostly not been destroyed.
On the other hand, \citet{Prochaska2008} argue that the absence of \ion{N}{V} absorption
argues for both low density and low metallicity surrounding the burst location.

\subsubsection{Direct recombinations to the ground-state}
Gas that has been ionized by massive star radiation within a host galaxy will generally recombine
quickly.
A fraction of recombining H ions will go directly to the ground-state, and so emit a photon just above the Lyman limit energy
(some higher energy photons will also be emitted by recombining He ions). 
In low column-density systems a fraction of these will escape the host without further absorption, 
and therefore re-boost the escaping ionizing flux, albeit with radiation that will soon be redshifted
to energies below 1\,Ryd.
In other words, simply translating \HI\ column-density into line-of-sight opacity is likely to lead to a 
small underestimate of the escape fraction in low column-density systems.
The net effect of this re-boost depends on various factors, but could provide an increase of up to 10--20\%
in the effective escape fraction \citep{FG2009},
thus at least partially offsetting any dust extinction.

For the remainder of 
Section~\ref{sec:fesc} we will continue to consider only the opacity
due to \HI\ absorption, but will return to the potential systematic effects of dust in Section~\ref{subsub:dust}.

\subsection{Statistical uncertainty}

Even with our considerably larger sample of sight-lines, the fact that only two have any appreciable escape
fraction means that to some extent we are still dealing with rather small number statistics.
We also lack a robust theoretical model which could be fit to the data, and so must 
explore the statistical uncertainties non-parametrically.

Again we first follow \citet{Chen2007} by performing a bootstrap exercise,
employing $10^6$  random resamples of the data with replacement.
From this we estimate a 98\% confidence upper limit of $\avfesc<\uplim$;
the result is the same whether or not we allow the resampled \NH\ values to have additional scatter based on
the error bars for each point. 

In an alternative approach, we simulated several large populations of sight-lines with higher values of average
escape fraction than found in our data (by replicating the GRB\,050908/060607A values), 
and drew $10^5$ random \sampsiz-member samples from each of these. 
For the case of the population with $\avfesc_{\rm pop}=0.02$ we found 98\% of random samples produced
$\avfesc_{\rm samp}>0.005$. 
Thus, these two methods agree on an  upper limit for \avfesc\ of $0.015$--$0.02$.
A similar analysis gives a 98\% lower limit of $\avfesc>10^{-3}$.

These are significantly tighter constraints  than found by the previous studies of \citet{Chen2007} 
and \citet{Fynbo2009} of $\avfesc<0.075$ at 95\%, due to our larger sample size
and the fact that no further very low column-density sight-lines have been identified in any of the additional GRBs. 
We note that our result is also consistent with the $\avfesc=0.020\pm0.017$ obtained by
constraining the flux below the Lyman-limit in a stacked spectrum of eleven GRB afterglows with $\bar{z}=3.66$ by \citet{Christensen2011}.

\subsection{Comparison to model predictions}
\label{subsec:model}
It is worth noting that our $\NH$ distribution is inconsistent with the predictions of \citet{Cen2014} who used high-resolution
cosmological radiation-hydro simulations of galaxies within the EoR ($z\sim7$) to explore column-densities along GRB sight-lines.
They found  a bimodal distribution 
with a peak at high column-density ($\lognhc\sim21$--22), similar to the observed distribution, 
but then another substantial peak with column-densities $\lognhc<18$ which is not seen in practice.
Part of the explanation could be that
\citet{Cen2014} assumed that the GRB rate traces the SNII rate
and found a large fraction of their low column-density GRBs occurred in super-solar metallicity environments,
whereas in reality GRB progenitors
seem to be younger at explosion \citep[e.g.][]{Larsson2007}
and crucially, unlike SNII, are rarely found in high metallicity galaxies
\citep[e.g.][]{Perley2015}.
On the other hand, their simulations do not account for the effect of GRBs in ionizing gas local to the burst (see Section~\ref{subsub:effects}), and
it seems when a high column-density is found in their models it is often due to such local gas, whereas
the low column-density cases occur for progenitors that have escaped their birth clouds.
This suggests their simulations do not capture the distributed nature of neutral hydrogen
in these star-forming galaxies, at least as it is found in $z\sim2$--5 GRB hosts.

Our results do correspond much more closely with the earlier simulations of \citet{Pontzen2010}, who similarly
investigated sight-lines to young star forming regions in model galaxies, but in this case did
specifically consider the $z\sim2$--5 range.
They found a median \lognhc\ of 21.5 and calculated an escape fraction of $\avfesc\approx0.01$, 
both close to our findings.  Indeed, their default prescription assumes that GRBs trace star formation up to
an age of 50\,Myr, whereas restricting to a perhaps more realistic 10\,Myr age reduces \avfesc\ to 0.007.
These simulations included the effect of local ionizing sources on the gas proximate to the burst, 
but again not the potential additional effect of 
ionization due to the GRB itself.
We return to these issues in Section~\ref{sec:syst}.

\subsection{Evolution with redshift}
\label{sub:evol}
As shown by the red line in Figure~\ref{fig:nhz}, there is no evidence for significant variation in the median
value of \lognh\ between redshifts $z\sim2$ to $z\sim5$.
To investigate this further, in Figure~\ref{fig:cum} we plot cumulative $\NH$ distributions for three
subsets of the whole sample cut in redshift. It is apparent that there is little difference between
the low ($z<3$) and intermediate redshift ($3<z<5$) sub-samples -- the median values are the same, 
and a two-sample Kolmogorov-Smirnov (KS) test finds them to be consistent with the null hypothesis that they 
are drawn from the same parent distribution (p-value of 0.88).
The  intermediate redshift sub-sample 
does have a somewhat longer tail to low column-density than the low redshift 
sub-sample,
and we note that low column-density systems are arguably rather more likely to
go unrecognised at lower redshifts (discussed further in Section~\ref{subsub:weak}).
However, another potential selection effect is that the proportion of dusty sight-lines appears to decline with increasing redshift
above $z\sim3$
\citep{Kann2010,Perley2016a,Perley2016b}, 
and if dusty bursts are systematically lost from the low redshift sub-sample, due to the
difficulty of locating and obtaining spectra for the afterglows,
then it could mask a more significant evolutionary trend.
The issue of biases due to dust is one we return to in subsequent sections.

There is a suggestion of a more significant decline in the typical values of \lognh\ at $z\gtrsim6$
\citep[as also pointed out by ][]{Chornock2014,Melandri2015},
which influences the final $z>5$ bin (red line).
However, 
the conclusion is still limited by small number statistics, and a KS test again finds this $z>5$ sub-sample to be consistent
with being drawn from the same distribution as the lower redshift ($z<5$) sub-sample (p-value of 0.31).

\begin{figure}
\includegraphics[angle=270,width=\columnwidth]{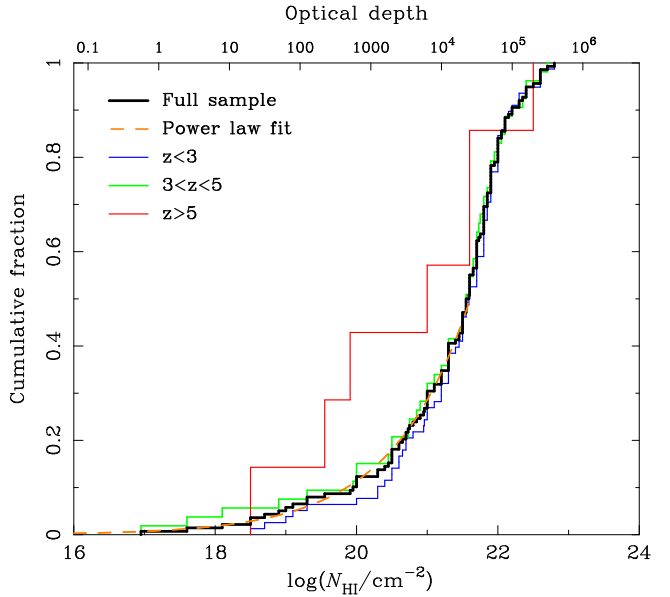}
    \caption{Cumulative distribution of \HI\ for the whole sample and split into subsets by redshift, as indicated.
The corresponding optical depth at \lya\ is shown on the top axis.
     A power-law fit to the lower column-density half of the sample is shown by the orange dashed line (see text).}
    \label{fig:cum}
\end{figure}

\subsection{Dependence on host properties}
\label{sub:hosts}

In Figure~\ref{fig:muv} we plot \lognh\ versus host UV absolute magnitude, $M_{\rm UV}$, which is a gauge of
the current (unobscured) star formation rate, for those bursts in our sample where good constraints on host luminosity are available
in the literature (37 measurements and 18 upper limits, detailed in Table~\ref{tab:data}).
The restricted number of cases for which deep host searches have been conducted, and the fact that many
are upper limits, means we cannot draw firm conclusions, but there 
is little indication of a dependence of the average \HI\ column-density on the current star formation rate.
Note, we have converted these values to a common cosmology (a flat Friedmann model with $\Omega_{\rm M}=0.3$, $H_0=70$\,\kms),
but not otherwise attempted to 
correct for differences in the procedures used by different authors (approaches to
k-corrections, for example), or small deviations from the reference 1600\,\AA\ wavelength generally adopted.
These distinctions should not be at a level that would affect the conclusion.

\begin{figure}
\center
\includegraphics[angle=270,width=0.9\columnwidth]{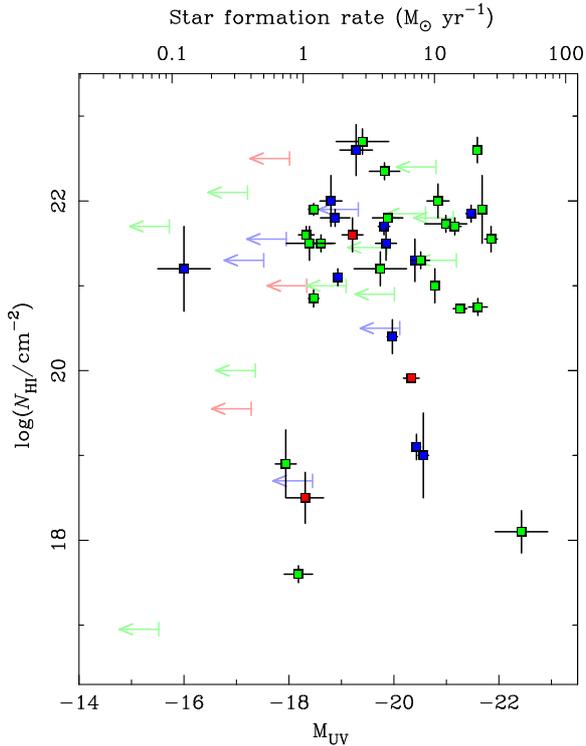}
    \caption{The values of neutral hydrogen column-density plotted against host rest-frame UV  
    absolute AB magnitude
    and corresponding star formation rate \citep[top axis, following the calibration for a $Z=Z_{\odot}/3$, constant
    star-formation rate and age $>300$\,Myr population from][]{Madau2014}, 
    for 37 detections and 19 upper limits collected from   various sources (Table~\ref{tab:data}).    
     Colour coding is as with the lines in Figure~\ref{fig:cum} and upper limits are $2\sigma$.
    }
    \label{fig:muv}
\end{figure}

Similarly, in Figure~\ref{fig:mstar} we plot \lognh\  against host stellar mass, ${\rm M_*}$, for those galaxies in our sample for which
estimates (or limits) are available in the literature (Table~\ref{tab:data}).
The bulk are based on {\em Spitzer} infrared photometry, particularly from \citet{Perley2013},  \citet{Perley2016b}, \citet{Laskar2011} and Myers et al. in prep.
Here we are restricted to
28 galaxies with ${\rm M_*}$ estimates and 31 galaxies with upper limits. Again there is little indication of any trend,
contrary to suggestions that \fesc\ may correlate strongly with galaxy size \citep[e.g][]{Anderson2017}.

\begin{figure}
\center
\includegraphics[angle=270,width=0.9\columnwidth]{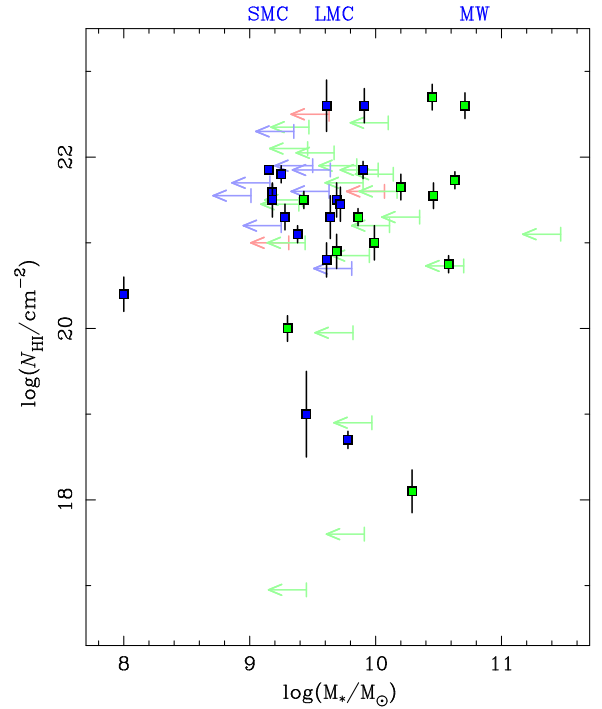}
    \caption{The values of neutral hydrogen column-density plotted against host stellar mass estimates from
    a range of sources (Table~\ref{tab:data})
    for 60 hosts, 32 of which are upper limits (shown at 2$\sigma$). Colour coding is as with the lines in Figure~\ref{fig:cum}.
    The approximate stellar masses of the Milky Way and Magellanic Clouds are indicated on the top axis for comparison.
    }
    \label{fig:mstar}
\end{figure}

One trend that is conspicuous in Figure~\ref{fig:muv}, and in particular Figure~\ref{fig:mstar}, is that
the UV brighter and more massive hosts are predominantly in the $3<z<5$ bin.  It is true that
small and faint hosts are harder to detect at higher redshifts, and that may explain some part
of this trend.  However, another factor seems to be that 
higher mass (${\rm M}_*\gtrsim10^{10}$\,\msol) hosts at $z<3$
are more often associated with heavily extinguished bursts, and hence more likely  to be highly dusty, 
than at higher redshifts \citep{Perley2016b};
these are presumably systematically under-represented in our compilation.


\section{Potential systematic uncertainties}
\label{sec:syst}

A number of systematic uncertainties may affect our analysis, potentially biasing the conclusions.  
These include observational selection effects: in order to be included in the sample
afterglows must be localised and spectra obtained which cover the \lya\ region with reasonable
signal-to-noise.

Other concerns relate to  the still-uncertain nature of the GRB progenitor itself, both in terms
of the assumption that their sight-lines are representative of dominant EUV-producing stars, and specifically whether
there could be special circumstances required to produce GRBs that necessitate an atypical environment.  
It is also pertinent to ask whether the results can be extrapolated to the EoR.

In this section we begin by investigating systematic effects which may influence our calculation of
\avfesc\ in terms of its application to the GRB progenitors. 
The majority \citep[$\sim70$\%; ][]{Gehrels2013} of {\em Swift}-discovered GRBs do not have redshifts, 
which, it may be thought, could lead to 
large biases.
With a more uniformly selected sample, some of these effects might be quantified via
simulated data-sets, but that approach would be of limited utility here.
However, by considering the nature of these selection effects, and surveying the
available data, we shall show that these biases can be understood well enough to confirm that they are unlikely to affect the main conclusions.

We then consider more carefully 
how representative the GRB progenitors are 
of the EUV-producing massive stars, and the potential
systematic uncertainties that may introduce. Finally we discuss in more detail the application of
our results to the EoR.

\subsection{Systematic over-estimation of $\avfesc$}
\label{sec:over}

Two  factors are very likely to produce an over-estimate of the escape fraction, as discussed here.  

\subsubsection{Dust extinction biases}
\label{subsub:dust}
GRB afterglow surveys are biased against  high column-density
sight-lines since they tend to  be dustier and hence harder 
to locate and obtain redshifts for in the optical band
\citep{Fynbo2009,Greiner2011,Watson2012}.  Dusty systems at higher redshift would be particularly susceptible to being
lost since even near-IR observations will be looking in the source-frame optical
or near-UV. However we expect dusty systems to be rarer at $z\gtrsim4$ \citep{Perley2016b} 
and particularly by $z\sim6$ \citep[cf.][]{Zafar2011,Schaerer2015}.
Recently \citet{Bolmer2017} have shown that the decline of obscuration in GRB afterglows with
increased redshift is likely primarily due to their hosts being less dusty rather than an observational selection effect.

Over all redshifts, \citet{Perley2016a} estimate that approximately 20\% of GRBs detected by {\em Swift} are
heavily dust obscured, and often reside in globally dusty, massive galaxies \citep{Rossi2012,Perley2013}; 
this sub-population is significantly biased against in afterglow samples.  
A similar proportion show signs of more moderate dust obscuration \citep[see also][]{Covino2013}.
Thus this effect, while important for understanding the overall GRB host population,
would likely only necessitate a comparatively minor correction to the escape fraction estimate (i.e. a $\sim20$--40\% reduction,
assuming that all very dusty sight-lines are opaque to EUV),
and the correction is likely to be greatest at $z<3$, as discussed in Section~\ref{sub:evol}.

Even in some cases where afterglows have been detected and redshifts
measured, there may be significant attenuation in the UV by dust.  
However, from the point of view of escape fraction, this is only relevant for
the two sight-lines with non-negligible \fesc, and as already discussed in Section~\ref{sub:low}
may lead to corrections (downward) of a factor $\sim2$ or more for our sample.

\subsubsection{Effects of GRB prompt emission and early afterglow on local gas and dust}
\label{subsub:effects}
The GRB prompt flash and early afterglow produces an intense radiation field that is
expected to quickly  destroy dust \citep{Waxman2000,Morgan2014} and ionize gas to distances of up to several tens of pc when the ambient
medium has a low to moderate density, $n\lesssim10$\pcc\ \citep{Perna2002,Vreeswijk2007,Krongold2013}. 
Therefore the opacity measured to the afterglow could in principle be less than the opacity that would
have been seen to the progenitor star system.
Various lines of evidence suggest the existence of an ionized gas component, likely 
reasonably local to the GRB.
In particular, it has been argued that high column-densities measured in X-ray absorption, which in some afterglows
significantly exceed the optical measures, are at least partially  due to denser gas close to the progenitor in
which the hydrogen has been ionized \citep{Watson2007,Schady2011}.
Furthermore, the observed correlation of X-ray absorption with local galaxy surface brightness (in optically
bright GRBs) may support a local origin for a significant proportion of the absorbing gas \citep{Lyman2017}.
Highly ionized species have also been seen in some afterglow spectra, which are  likely
to be of circumburst origin \citep[][Heintz et al. submitted.]{Fox2008,DeCia2011}.

On the other hand, significant  ionization may have been brought about by the stellar radiation field
prior to the burst \citep{Watson2013,Krongold2013}, rendering the ionizing effect of the GRB largely irrelevent.
This is supported by models of feedback from star formation in massive molecular clouds which show
that hot ionized bubbles can grow to tens or hundreds of pc in a few Myr, although in some circumstances, in particular if
star formation is relatively inefficient,  the outflow can stall and the cloud recollapse, leading to further star formation episodes
\citep[e.g.][]{Rahner2017}.

In principle, deep time-resolved spectroscopy may allow the GRB-driven ionization of \HI\ to be observed directly \citep{Perna2002},
although in practice, sufficiently good data-sets have rarely been acquired.
Only in one case, namely GRB\,090426, has time-variability of \lya\ been seen,
between spectra obtained at 1.1\,hr and 12\,hr post-burst, suggesting the influence of the GRB.
Here photoionization modelling placed the absorbing gas at $\sim80$\,pc \citep{Thoene2011}.  
As discussed in Appendix~\ref{app:090426}, there is uncertainty regarding this particular burst as to whether
it is of the long or short duration class, but nonetheless it confirms the local ionizing effect of GRB emission
can occur even at quite large distances.

In one other case, GRB\,080310, \citet{Vreeswijk2013} found their time-dependent photoionization model to be improved
with the addition of a cloud at $\sim10$--50\,pc from the burst, which became fully ionized  by the early afterglow emission.
The required column-density of this cloud, of $\lognhc\sim19$--20, was greater than the $\lognhc=18.7$ inferred for the 
observed neutral absorber.

In conclusion, it is likely that some fraction of GRBs exhibit a reduced \HI\ column-density due to the ionizing effect of the
burst itself.  
If windows to the IGM often occur when superbubbles puncture low-density channels out of  galactic
neutral gas \citep[e.g.][]{Dove2000,Roy2015}, then this may be largely irrelevant as far as high escape fraction sight-lines are concerned.
On the other hand, regarding our sample, if GRB\,060607A was dust extinguished (Section~\ref{subsub:060607A})
then it would be surprising if the dust was not associated with some neutral gas which was ionized by the burst.

\subsection{Systematic under-estimation of $\avfesc$}
\label{sec:under}

A reasonable question is whether some very low \NH\ systems may not have been recognised
in GRB afterglow observations.
Omitting from our sample, or over-estimating the column-density, of even a fairly small number of such bursts  could  lead
to significant under-estimation of $\avfesc$.
There are several circumstances in which this could plausibly arise that are discussed below.
The nature of such biases depends on the quality of the afterglow spectroscopy, and so we split 
our discussion into three broad categories: good S/N spectra (Sections~\ref{subsub:weak} and \ref{subsub:mismeasure}), 
poor S/N spectra that were still sufficient to provide a redshift
(Section~\ref{subsub:lowsn}), and 
instances where no redshift was obtained due to the faintness of the optical afterglow  (Section~\ref{subsub:lowdens}).
These divisions are somewhat qualitative, but this is appropriate given that we
do not have access to many individual spectra, and also noting that continuum S/N can vary significantly within a spectrum.

However, first we consider what can be said about the potential level of such selection effects, by reference to a nearly redshift-complete GRB sample.

\subsubsection{Lessons from the SHOALS sample}
\label{subsub:shoals}

Only a fraction of X-ray localised GRBs have optical/nIR afterglow identifications, and only a fraction of
those have redshifts from afterglow spectra.\footnote{By way of illustration, to the end of 2017, the database 
maintained by Jochen Greiner, \url{http://www.mpe.mpg.de/~jcg/grbgen.html}, lists 1200 GRBs with X-ray counterparts
and 720 with optical identifications.  Of these, 301 have afterglow redshifts according to the database maintained
by Daniel Perley, \url{http://www.astro.caltech.edu/grbox/grbox.php}.} 
Thus, if the chances of a redshift being obtained depend on 
the \HI\ column-density then
it could bias our results.  
On the other hand, many GRBs receive little ground-based follow-up for other reasons,
for example, because of poor weather at major observatories,
because they are badly placed for observation due to proximity to the Sun, Moon or Galactic plane,
or simply because of the limited availability of large telescopes to make the necessary rapid target-of-opportunity observations.

Since our sample is not selected or observed in a uniform way, and many values are taken from the literature,
it is hard to assess the maximum scale of these effects directly.
However, we can get a handle on them by considering the well-defined 
{\em Swift} Gamma-Ray Burst Host Galaxy Legacy Survey (SHOALS) sample \citep{Perley2016a},
which consists of 119 long-duration GRBs 
discovered by {\em Swift} up until October 2012, and excludes bursts that were poorly placed for ground observation
(whether or not they have redshifts).
SHOALS imposes a threshold on the prompt $\gamma$-ray fluence of $S_{15-150\,\rm keV} > 10^{-6}$\,erg\pcm,
which does mean it excludes some intrinsically very weak events, 
but prompt emission is thought to arise from internal processes within
the GRB jet and so not to be dependent on the nature of the ambient environment. 
Furthermore, despite many searches, there is little indication that prompt $\gamma$-ray behaviour
depends on other properties of the host, such as metallicity \citep[e.g.][]{Levesque2010b,Japelj2016}.
SHOALS also requires that bursts have identified X-ray counterparts, but since all long-bursts
are detectable in X-rays if observed sufficiently early by {\em Swift}, the selection criterion employed was
simply that a rapid autonomous slew was performed.
This is consistent with theoretical expectations that the X-ray afterglow flux should be independent of
ambient density, $n$, unless it is very low \citep[$n<10^{-3}$\pcc, e.g.][]{Hascoet2011}.

The SHOALS sample has a high degree of redshift completeness thanks in large part  to major efforts to obtain redshifts from host
galaxy observations (identified within X-ray or optical afterglow error boxes) where they had not already been obtained
from afterglow spectroscopy.
Specifically, 92\% have spectroscopic or (in a few cases) good 
photometric redshifts, and all but one have some photometric constraint on the redshift.

Here we restrict our attention to the 80 bursts for which the redshift is $z>1.6$
or for which the constraints allow the possibility of the burst being in that range.
Of these we can immediately say that 52
were very likely high column-density sight-lines, either
because \NH\ was measured directly 
(39) 
or because they were found to 
have faint afterglows with indications of
high levels of extinction (14) according to \citet[][see also Section~\ref{subsub:dust}]{Perley2016a}.
One, namely GRB\,060607A, is the same low-column system included in our sample.

Of the remainder, 
19 appear to have had at least moderately bright optical afterglows,
but either spectroscopy was obtained which  did not cover the wavelength of \lya\ (8)
and the redshifts rely on metal lines (although in no case were these lines reported as being unusually weak)
or simply no spectroscopy was  attempted to our knowledge (11). 
A further four 
had little afterglow follow-up of any kind reported; these were GRB\,050128, early in 
the {\em Swift} mission, GRB\,050726, for which real-time alerts were not sent to the ground,
GRBs\,050922B 
which occurred on the same day as several other high-priority bursts, and GRB\,070328.
We find no reason to think any of these bursts lacked \lya\ measurements due to observational
selection effects, since they all seem to be cases where only limited follow-up was attempted.

This leaves only three sources, which merit more thorough scrutiny.
One of these, GRB\,071025, was observed with Keck/HIRES, but the spectrum was low-S/N with flux
only being detected at $\lambda>7500$\,\AA. \citet{Fynbo2009} argued that this may
be due to a \lya\ break at $z\approx5.2$ (the low S/N precluding measurement of the
line strength), or alternatively that  it could indicate a highly dust reddened afterglow at lower redshift.
A photometric redshift constraint from multi-band afterglow imaging supports a high redshift ($z\sim5$)
interpretation, whilst also favouring fairly substantial dust extinction \citep{Perley2010}.
Thus it seems this is an intrinsically bright  event but, again, most likely with a high column-density sight-line.

GRB\,100305A was the target of several deep imaging observations within the first hour post-burst,
but the only candidate afterglow \citep[Gemini/GMOS observations in $riz$;][]{Cucchiara2010} was subsequently found to be outside the revised
X-ray error circle and to be present as a steady source in later imaging \citep{Perley2016a}.
We have analysed previously unpublished early UKIRT data, and also find no afterglow down
to a 2$\sigma$ limit of $K_{\rm AB}=20.7$ at 40\,min after the trigger.  
In fact the {\em Swift}/XRT spectrum \citep[see \url{http://www.swift.ac.uk/xrt_spectra/00414905/};][]{Evans2009}
does show significant X-ray absorption above the Galactic value, suggesting a high column sightline, possibly
combined with moderately high redshift making the optical/nIR afterglow faint.

Finally we have GRB\,070223
is known to be at redshift $z=1.63$ from the host galaxy \citep{Perley2016a}.
Here the afterglow was faint in both the optical and near-infrared, despite early follow-up,
meaning that no spectroscopy was attempted.
The host galaxy was detected in {\em Spitzer} 3.6\,$\mu$m imaging, but
the implied stellar mass is a relatively modest ${\rm{log(M_*/M_{\odot})}\approx9.5}$
\citep{Perley2016b}.
We have reanalysed the early imaging obtained at the Liverpool Telescope and the WHT
at $\approx3$\,hr post-burst (details are given in Appendix~\ref{app:070223}),
finding AB magnitudes of $r=23.8\pm0.3$ and $K=22.0\pm0.3$ for a faint source at
the X-ray afterglow position.
However, we have also analysed the SDSS and PanSTARRS imaging of the same region,
and in both cases find a persistent source, presumably the host, at the same location, with a
magnitude $r=23.5\pm0.3$.
Thus it seems clear that the optical source seen by the LT \citep[and also the MDM 1.3m; ][]{Mirabal2007}
was actually host dominated, and hence the optical afterglow must have been substantially fainter.
By contrast, the $K$-band source faded  by 0.7\,mag by the following week, confirming
an afterglow detection in the near-IR \citep{Rol2007}.
Thus it seems likely that, despite not being in a massive dusty host, this event too was heavily
extinguished, which is consistent with the high column-density inferred
from the X-ray spectrum of ${\rm log}(N_{\ion{H}{I}\rm,X}/{\rm cm}^{-2})\approx{22.6}$
\citep[see \url{http://www.swift.ac.uk/xrt_spectra/00261664/};][]{Evans2009}.

In summary, from our analysis of the SHOALS sample,  of 80 bursts that may be at $z>1.6$, 55 have evidence of high-\NH\ and/or
high extinction and 1 has low column ($\lognh<18$).
In all other cases, limited follow-up seems to be the primary reason for a lack of a constraint on \NH.
This is worth emphasising: even amongst bursts that were chosen as being well-placed for follow-up and which had at least moderately bright afterglows, 
a significant number of events ($\sim25$\%) lack spectroscopic constraints on \lya\ absorption 
for reasons that seem to be unrelated to the afterglow properties.
Thus, it seems that the large majority of optically faint
bursts are dust extinguished, with a smaller number at high redshift and hence optical ``drop-outs".
The predominant selection effect, then, leads to high-\NH\ bursts being lost from the sample (already discussed in Section~\ref{subsub:dust}).
This suggests that any bursts that are lost from our sample due to selection against low column-density systems
must be few in number.

\subsubsection{Featureless or very weak-lined GRB afterglow spectra}
\label{subsub:weak}
In rare cases, like GRB\,071025 discussed above,
afterglow spectra are acquired in which no absorption features can be seen at a reasonable confidence level,
or that exhibit only marginal features that cannot be unambiguously identified.
This could be  due to foreground gas in the host having very low column-density such that it
produces neither a clear \lya\ feature nor detectable metal lines,
with the net result that no redshift is obtained.
However, in our experience,  such apparently featureless spectra are nearly always cases where 
either the continuum level has very
low signal-to-noise (S/N) ratio (as was the case for GRB\,071025), thus not necessitating an especially low column-density, and/or
the spectrum only covers a relatively short wavelength range and so may easily miss prominent absorption features.

Problems associated with low-S/N spectra are discussed in Section~\ref{subsub:lowsn}.
The possibility that intrinsically fainter afterglows, which typically result in no afterglow redshift being determined, 
may on average have low column-density absorbers, we return to  in Section~\ref{subsub:lowdens}. 
Here we restrict attention to whether weak absorption features could have led to no redshift being found
despite the spectra being of moderate to good S/N and spanning a wide wavelength range.

Again, our experience suggests such circumstances are very rare: we are not aware of any 
compelling examples, published or unpublished.
A much discussed near-miss was GRB\,070125, for which the absorption lines were very weak, but ultimately the redshift 
was found to be $z=1.55$ from  \ion{Mg}{ii} absorption seen in a Gemini/GMOS spectrum \citep{Cenko2008}.
A later Keck LRIS spectrum of the afterglow, which extended to shorter wavelengths, showed marginal evidence for \lya\ absorption, but this was
only sufficient to conclude $\lognhc<20.3$ in the host \citep{Updike2008}.
Based on the weakness of the metal lines,   \citet{DeCia2011} argued that the neutral hydrogen column-density
was probably low, likely in the LLS range, but that this could have been substantially diminished by the particularly intense
afterglow radiation ionizing gas to a considerable distance.
Given that \lya\ was so far into the near-UV
in this case,  around 3100\,\AA, which is hard to calibrate in ground-based data, we did not include it in
our sample.
The  unusual nature of this system is illustrated by the fact that GRB\,070125 had the lowest ``line-strength-parameter"
(an index based on the strength of absorption lines compared to the average over the sample) out of 69 spectra studied by \citet{deUP2012}.

Another instructive case is GRB\,140928A, for which spectroscopy was obtained with Gemini/GMOS-S 
\citep{Cucchiara2014}.
Here the afterglow continuum was clearly detected, but no unambiguous lines were seen, 
despite the S/N being moderately good (S/N\,$\approx$\,8 per spectral resolution element at 6500\,\AA).
In this case the spectral range was 5680\,\AA--10250\,\AA, with two 80\,\AA\ chip gaps, thus it is plausible
that simply  no intrinsically strong lines happened to lie within this window.
What we can say, though, is that for \lya\ to fall within the spectrum would have required $z>3.7$, which would mean
we would have expected a clear break due to the onset of the \lya-forest, irrespective of the host column-density.
This  is not seen, so we can conclude that \lya\ very likely was not within the spectral window in this case.

Thus this example highlights  an important point regarding weak-lined spectra, namely
that at least above redshift $z\sim3$  strong attenuation due to the \lya\ forest would normally
be expected to be clearly seen in reasonable S/N optical spectra covering the relevant wavelength range, 
giving good indications of the redshift, even in the absence of any host absorption.

Finally we note that, while there have been occasional instances when host galaxy follow-up has revealed an earlier
claimed afterglow redshift (based on a low-S/N spectrum) to be mistaken \citep[e.g.][]{Jakobsson2012},
to our knowledge none of these have indicated a case where the afterglow spectrum 
should have revealed \lya\ absorption which was not seen.
All these considerations suggest that any bias introduced by the effect of low host column-density going unrecognised
despite good afterglow spectroscopy should be minor compared to the other effects we consider.

\subsubsection{Mis-measuring low column-density systems}
\label{subsub:mismeasure}

A more subtle question is whether \NH\ values may be over-estimated simply due
to the measurement process, particularly for low-S/N spectra.  
This should not be a major concern in the majority of cases, where damping wings are clearly
seen and fitted, confirming the high column-density.
For cases with rest-frame equivalent width of \lya\ less than $W_0\sim5$\,\AA\ (roughly $\lognhc\sim19.5$), 
especially when observed at  low spectral resolution (typically $R\lesssim2000$),
uncertainty in the velocity structure of the absorbing gas leads to relatively high uncertainty in the 
inferred \HI\ column-density.  If the range in velocity of the absorbing gas is under-estimated,
for example if due to several clouds with different velocities, then it would lead
to an over-estimate of the column-density.

Of our sample, three bursts both fall into this category and lack direct evidence of emission or
otherwise below the Lyman limit, namely GRBs 060124, 060605 and 090426.
The last of these was unusual in exhibiting apparent variability of \lya\ absorption (Appendix~\ref{app:090426}),
suggesting absorption dominated by a single absorber.
The other two are more difficult cases, although the spectral resolution is sufficient to
rule out  a high spread in velocity (cf. GRB\,021004, Appendix~\ref{app:021004}), and the inferred
\HI\ columns (and error bars)
appear to have considered a fairly conservative range of Doppler parameters, making a significant over-estimate unlikely.

\subsubsection{Misidentification of the host absorber}
\label{subsub:misid}
A similar possible scenario involving very low column-density would be where
 the host absorption lines were not identified in the spectrum at all, 
 but instead chance alignment with a stronger intervening absorption system led
to the incorrect assignation of its redshift as the redshift of the burst, along with an erroneous column-density.
Again, this is likely to be a rare circumstance since the incidence of strong intervening absorbers is not high
and one would normally expect to see the \lya\ forest from the IGM,
particularly above $z\sim3$, which would allow identification of an unassociated \lya\ absorber  as
being due to an intervening system.
We also note that in some spectra we detect metal fine-structure lines, which 
are thought to be the result of excitation by the burst itself of gas within its host galaxy,
confirming the association \citep[e.g.][]{Vreeswijk2006}.

A particular example that highlighted this concern was GRB\,071003,
in which it was found that the highest redshift system, a detection of \ion{Mg}{ii}, 
presumed to be from the host, was notably weaker than some intervening
absorbers  \citep{Perley2008}.
Similarly, GRB\,060605 exhibited weak \lya\ from the host, but stronger ($\lognhc=20.9$) from
an intervening system at slightly lower redshift \citep{Ferrero2009}.

Another pertinent case is  GRB\,141026A, the afterglow of which was observed by GTC, with a spectrum covering
wavelength range 5100--9800\,\AA.  The S/N was rather poor, but an absorption line was seen close to the blue end 
of the spectrum that if interpreted as \lya\ would imply $z=3.35$ and a low column-density of
$\lognhc\lesssim20$ \citep{deUgarte2014}. In this instance there were no other features seen to confirm
the line identification, and no evidence of a decrement that could be ascribed to the \lya\ forest, for which reasons
we chose not to include this burst in our sample.
Thus, this example illustrates that misidentification of redshift might in some circumstances result in a
bias in the opposite direction, namely toward lower column-density.

Once again, we conclude that whilst it is hard to rule out completely, the rate 
of strong intervening absorbers being falsely identified as host systems, providing good 
spectra are obtained, must be very low.

\subsubsection{Low signal-to-noise spectra}
\label{subsub:lowsn}

Some afterglow spectra are sufficient to provide redshifts, but the signal-to-noise, at least around the \lya\ region, is poor.
This may lead to  \NH\ being undetermined, particularly if it is low, thus creating a bias in favour of including higher column-density systems.
Amongst our sample, only eleven bursts lack clear metal line detections, and of these
five have tentative metal line detections (GRBs 020124, 060927, 080129, 080913, 121229A),
and five have no metal line detections but do show an unambiguous continuum break at \lya\  that
is sufficiently well defined to constrain the wing profile
(GRBs 060522, 081203, 090519, 100316A, 130427B, 140515A).
The latter subset all have low S/N, and  the search for metal lines was complicated by low spectral resolution and/or
being in a difficult region of the spectrum,
but reassuringly they span a wide range of \NH\ values, which is not suggestive of any particular bias.
This gives confidence that our sample derives predominantly from high-S/N spectra, and
contains few bursts which are only included  because they had a particularly high value of \HI\ column-density.

Several other bursts have a redshift determined from the \lya\ break, but the S/N proved insufficient to 
estimate the \NH\ value.  These cases are few in number:
apart from several at $z\gtrsim6$, from a search of spectra we have ourselves and the literature
we have only identified GRBs 071025 ($z\approx5$; Section~\ref{subsub:shoals}), 
140428A \citep[$z\approx4.7$;][]{Perley2014} and 160327A \citep[$z\approx5$;][]{deUP2016}.
Thus we believe that these cases, while they may be below the median \NH\ for all bursts,
are not likely to be unusually low-\NH.  A small bias could partially offset the
bias against dusty sight-lines discussed previously.

It is notable that redshifts can be obtained from low-S/N spectra when the redshift is comparatively high,
which can be understood because the strength of the \lya-break increases with redshift.
At redshifts below $z\sim3$ such spectra likely will not yield secure redshifts, a category that is discussed in the next section.

\subsubsection{When redshifts are not obtained: could GRBs in low density environments have intrinsically fainter afterglows?}
\label{subsub:lowdens}

We have argued in the preceding sections that bursts are unlikely to have been lost from our sample
due to weak absorption lines providing that good spectra were obtained.
However, bursts with very faint optical afterglows will be under-represented 
due to the increased difficulty of arcsecond localisation and redshift determination (either because spectroscopy
was not attempted, or because spectra had too low S/N to give a conclusive redshift or \NH\ measure).
Thus, if bursts occurring in low density environments had weaker lines and also on average fainter afterglows, 
then that potentially may lead us to systematically lose bursts with high \fesc.
One way a GRB progenitor could find itself in a lower density environment, would be if it was formed by a so-called
``runaway" star.  We consider this particular issue in Section~\ref{subsec:env}, but here focus on the potential
effect of low density on the brightness of afterglows and the likelihood that such systems have been missed.

As discussed above,
 the majority of optically faint afterglows are dust extinguished,
and have high EUV opacities,
while a smaller number are high redshift optical drop-outs.
We should also remember that some afterglows were faint when observed simply due to the delay in acquiring
spectroscopy.
This suggests that the fraction of
systems that are faint due to low density circumburst media is low. 
On the other hand, basic synchrotron afterglow theory provides some motivation for thinking
such a trend might occur.
In particular, for a relativistic jet shocking a medium of uniform density, $n$, in typical circumstances 
the  optical afterglow flux should scale with $n^{1/2}$ \citep{Granot2002}.
In fact, with sufficiently good wide-band monitoring of the afterglow, the ambient density of the medium in which
the jet is travelling (i.e. sub-pc scales) can be calculated.
The range of circumburst medium densities inferred from such modelling is quite wide,
from $10^{-4}$ to $10^{3}$\pcc\ \citep[e.g.][]{Laskar2014}, but equally is subject to model assumptions and
large uncertainties in many cases.

However, the crucial point is that the density
structure of the immediate circumburst environment is likely determined  by the recent mass loss history of the
progenitor system, and potentially that of any companions \citep{vanMarle2006,vanMarle2008}.
This is very unlikely to be  correlated with the density of neutral gas  producing the \lya\ absorption, which
is generally situated at significant distances of at least tens and often hundreds of pc from the burst site \citep[e.g.][]{Prochaska2006,Vreeswijk2013}.

From an empirical point of view, there are few indications of any correlations of afterglow
intrinsic luminosity with other host properties, including column-density.
For example, \citet{deUP2012} found no evidence of a correlation of afterglow luminosity at 
the time of observation with the observed line strength in a sample of 69 bursts.
In fact, some low column-density systems actually have notably bright optical afterglows,
including, as mentioned in Section~\ref{subsub:060607A}, GRB\,060607A, which has the lowest
value of \NH\ in our sample.
Other low column-density GRBs with bright afterglows (given the time post-burst
they were first observed) were GRBs 070125 \citep[$V=18.5$ at 13\,hr post-burst][]{Updike2008}, 
071003 \citep[$R=12.8$ at 42\,s post-burst][]{Perley2008} 
and  140928A \citep[$r=20.8$ at 22\,hr post-burst][]{Varela2014},  
all discussed above.
To some extent this could be regarded as a selection effect, since weak lines can only be
detected, or even searched for, in high-S/N spectra. It is also the case that one expects GRBs with highly luminous
optical flashes to be more effective at ionizing gas to larger distances (Section~\ref{subsub:effects}).
However, this does at least indicate a large scatter in any putative correlation of intrinsic afterglow luminosity and
\lya\ strength, and combined with the comparative dearth of featureless afterglow spectra (Section~\ref{subsub:weak}),
leads us to conclude that any such bias must be small.

\subsection{How representative are GRB progenitors of the dominant stellar sources of EUV radiation?}
\label{subsec:howrep}

An essential assumption in our analysis is that GRBs are good tracers of the locations of populations
of massive stars likely to be responsible for the bulk of  EUV radiation production.
In this section we discuss the extent to which this is true, and consider
the potential implications for our results.

\subsubsection{Metallicity effects}
GRBs preferentially occur in low (sub-solar) metallicity environments \citep{Kruehler2015,Japelj2016,Perley2016b,Graham2017,Vergani2017}, 
which are typically (but not solely) in
less dusty and smaller galaxies \citep[e.g.][]{Schulze2015,Blanchard2016}, 
and therefore might be expected to have lower neutral gas column-densities
\citep[although see e.g.][for counter arguments]{Gnedin2008,Sharma2016}.
Lower metallicity populations also produce more EUV for a given star formation rate \citep{Stanway2016}.
These factors may result in an over-estimate of the escape fraction averaged over all galaxies
at lower redshifts, but at higher redshifts
(above $z\sim4$) we would expect low metallicity to be the dominant mode of star formation,
and for it to be increasingly occurring in small galaxies (we return to this issue in Section~\ref{sec:eor}).

\subsubsection{Timescales of EUV emission compared to GRB progenitor lifetimes}
\label{subsub:time1}
GRB positions are well correlated spatially with regions of high mid- and near-UV emission
in their hosts, and specifically more highly correlated than are most (type II) core-collapse
supernovae \citep{Fruchter2006,Svensson2010,Blanchard2016}.
This  has previously been used to argue that if the GRB progenitor is a single star
it is likely to have initial mass $M>20$--25\,M$_{\odot}$ \citep{Larsson2007,Anderson2012},
but in any case, whether single or binary, it suggests average lifetimes less than those of more common core-collapse supernovae.

We can investigate this question more quantitatively using stellar population synthesis models.
Specifically, we consider the BPASS models of \citet{Eldridge2009}.
These include prescriptions for the contribution of binary stars, which is essential given the importance
of binary interactions to the evolution of massive stars \citep{Sana2012}. 
In fact binaries both enhance the total EUV output for a given stellar mass
and extend the emission in time. This will  increase the total number of ionizing photons
produced and also the effectiveness of feedback, tipping the balance of  ionization versus recombination 
in the environs of newly formed stars
and allowing a greater period for the gas to be cleared.
The net result is an increase of up to factors of several in the predicted escape fraction \citep{Stanway2016,Ma2016}.

In Figure~\ref{fig:bpass} we show the cumulative EUV production for a single burst population
as a function of time for a range of metallicities. A single burst of star formation represents one extreme: if, as is quite plausible,
star formation is more continuous in a region of a galaxy \citep[c.f.][]{Ochsendorf2017}, then GRBs will also be spread over time
and their locations will naturally be more representative of the EUV production sites.
The figure shows that the large bulk ($\gtrsim80$\%) of such radiation is still produced in the first 10\,Myr, in other
words during the lifetimes of all but the most massive stars.  
Comparatively little EUV is produced after an age of $\sim10$\,Myr \citep[cf.][]{Ma2016}, 
so even if older stars are less deeply embedded in their nascent gas clouds (e.g. through  having
moved from their birth sites and/or there being more time for stellar feedback, and in particular the accumulated action of supernovae,
to carve low density channels through the neutral gas in their vicinity)
they can contribute rather little to the total ionizing radiation output.
A caveat is that binary interaction, particularly at low metallicity, may result in envelope stripping of rather
less massive stars ($\sim12$\,M$_\odot$), which may then emit ionizing radiation over a longer period of
time ($\sim20$\,Myr), so could make a significant contribution to the total EUV output that has generally
been neglected to-date \citep{Gotberg2017}.

\begin{figure}
\includegraphics[angle=270,width=\columnwidth]{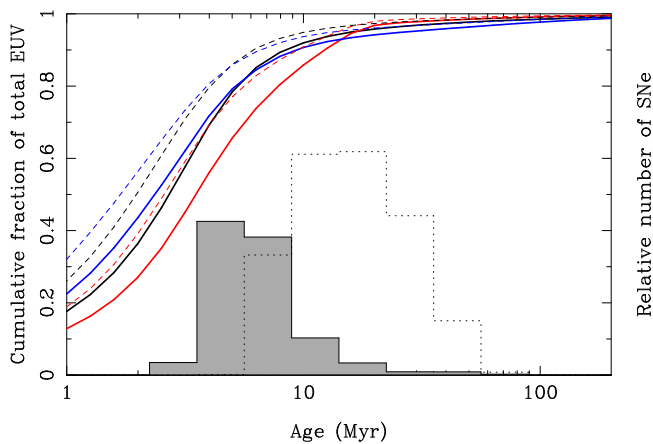}
    \caption{Output of EUV radiation for a single burst of star formation as a function of age from
    the BPASS models, which include prescriptions for binary stellar evolution \citep{Eldridge2009}.
    Shown are three models, $0.1\,Z_{\odot}$ (red), $0.3\,Z_{\odot}$ (black), $Z_{\odot}$ (blue), for a
    stellar initial mass function of power-law slope $\alpha=-1.3$ between 0.1 and 0.5\,\msol\ and slope $\alpha=-2.35$ from 0.5 to 100\,\msol.
    Dashed lines are the equivalent models with an IMF going up to 300\,\msol.
    Also shown for the $0.3\,Z_{\odot}$ model 
    are the relative numbers of type Ic supernovae (shaded histogram), and type II supernovae
    (dotted histogram) in each logarithmic age bin.}
    \label{fig:bpass}
\end{figure}

Although the nature of GRB progenitors remains
uncertain, most scenarios suggest they are indeed likely to have lifetimes of order 5--10\,Myr.
For example, this is indicated by their correlation with regions of high UV emission \citep{Larsson2007},
and is roughly the range spanned by the  viable single star chemically homogeneous evolution 
models studied by \citet{Yoon2006}.
To place this in context, the relative numbers of type Ic supernovae in logarithmic bins are also shown on
Figure~\ref{fig:bpass}, and are predominantly in the range of 3\,Myr to 15\,Myr.  
Since the supernovae accompanying long-duration GRBs are also stripped-envelope events \citep{HjorthBloom2012}, and 
correlate with the UV light of their hosts in a similar way \citep{Kelly2008},
it is reasonable to suppose that they span a comparable range of ages at explosion.

\subsection{How representative are GRB progenitors of sources of escaping EUV radiation?}
\label{subsec:env}

Even if GRB locations are good tracers of the dominant sources of stellar EUV radiation, it might be that they
under-represent the stars from which the bulk of the {\em escaping} radiation is produced.
Here we consider several such scenarios.

\subsubsection{Timescales revisited - might GRBs not sample peak periods of transparency?}
\label{subsub:time2}
In Section~\ref{subsub:time1} we argued that GRBs likely do explode on time-scales relevant for a significant
proportion of EUV emission from a single age stellar population.
However, if the peak episodes of EUV escape generally occur before the first GRBs explode following
a burst of star formation, it is plausible we may not sample the relevant periods (again, for
star formation of a more continuous nature, this will not be a concern).

Modelling in detail the escape of ionizing radiation from young massive stars in a range of
realistic scenarios is highly challenging.  On the scales of individual clouds 
outward pressure created by winds, radiation and supernovae competes with gravitational infall,  
while ionization competes with recombination.
These clouds also vary in size and shape, and have complex turbulent internal structure and magnetic fields.
It is also essential to incorporate the  processes of star formation and evolution, which introduces further uncertainties.
Finally the star forming regions exist in larger scale galactic environments.
In order to escape a galaxy, radiation must first escape its local environment, likely
the molecular cloud in which the stars formed, and subsequently leak out through the larger scale
neutral gas distribution.
Here we consider the lessons from recent state-of-the-art models which focus on different
aspects of the problem.

\citet{Howard2018} examined the escape of EUV radiation from  a range of massive ($10^4$--$10^6$\,\msol) giant molecular clouds (GMCs) 
containing multiple massive star clusters, based on 3D simulations with ongoing star formation.
They found  the EUV escape from the clouds themselves to be variable in time, with 
occasional peaks above 10\% from 2\,Myr after the onset of star formation.
The lower mass clouds tend to achieve high escape fractions of 20--100\% by $\sim5$\,Myr, due to near complete ionization of the clouds.
This is  within the time-frame that some GRBs likely occur, even if the bulk of progenitors have
longer lifetimes (Section~\ref{subsub:time1}).
The intermittency here is partly the result of small-scale  density structure due to the turbulent nature of the 
cloud, which produces time-changing local density field around the clusters.
This results in a very anisotropic directional distribution of escaping radiation \citep[something also seen in 3D models of small molecular clouds
dominated by a single O star, by][]{Walch2012}.
Placing these clouds into a galactic context produces galaxies with rapid (10--20\,Myr) fluctuations
in SFR and \fesc, particularly in dwarf galaxies, with a general trend of stronger episodes of
star formation being associated with lower \fesc.
However, these simulations did not include the effects of winds or supernovae,
and also use single star stellar population prescriptions.  As noted previously, inclusion of binaries is likely to
increase the effectiveness of feedback, and extend the time-scales.

\citet{Rahner2017} performed 1D spherically symmetric calculations, including winds, radiative transfer and
the effects of supernovae, covering a range of cloud masses, densities, star-formation efficiencies and metallicities.
Here absorption is dominated by neutral gas in the swept up shell of material surrounding
the central low density ionized bubble.
In those models that show any appreciable EUV escape from the birth cloud at all,
they also generally find a high proportion  
occurs during the first 2--6\,Myr.
Of course, these calculations are not able to include anisotropies in the shell structure, which may be
key to understanding the escape fraction and,
again, presumably star formation and/or EUV production more extended in time would modify these results.

Simulations that place star formation in a more cosmological context, while reliant on less sophisticated prescriptions for the 
feedback physics,
have tended to find episodes of high escape fraction, at least in early galaxies, to have scale times of 10-20\,Myr 
\citep[e.g.][]{Wise2014,Kimm2014,Ma2016,Trebitsch2017}. This reflects timescales of star formation activity
and consequent supernova feedback which has the dominant effect on the galactic scale gas distribution.
Interestingly, \citet{Toy2016} suggested that GRB hosts likely have had episodic star formation, based on comparison
of enrichment timescales with observed metallicities.

For comparison, star formation  in the 30 Doradus \HII\ region (the Tarantula Nebula) in the Large Magellanic Cloud,
often regarded as a local prototype of low-metallicity star-forming regions that may have been highly abundant in the early universe,
has been occurring in different clumps and clusters for at least $\sim10$\,Myr \citep{Sabbi2016}.
Inferring the EUV escape fraction from the Tarantula Nebula  region is subject to large uncertainties, but a recent
detailed study of its massive star population  constrained it to be in the range 0--0.6, with a preferred value of 0.06 \citep{Doran2013}.

Clearly this is a field where much work remains to be done.  It is plausible that in some circumstances,
a burst of star formation in a molecular cloud towards the edge of its galaxy may lead to a brief period 
of high EUV escape to the IGM before the first supernovae explode.  However, it does not seem likely this could 
be a common occurrence, and that to produce high average escape fractions would require 
the combined feedback effects of radiation, winds and supernovae  to disperse and ionize both local and global
gas, on timescales comparable to GRB progenitor lifetimes.

\subsubsection{Could GRB progenitors preferentially form in higher density environments?}
It has  been suggested that GRBs  may favour not only low metallicity, but possibly also high density
sites \citep[e.g.][]{Kelly2014,Perley2015}, for example due to dynamical processes in young dense stellar clusters being important in the
formation of their progenitors \citep{vandenHeuvel2013}. 

However, even if this is true, it is not obvious that it would significantly affect our conclusions.
Very massive and dense clouds are likely to recollapse without dispersal \citep{Rahner2017},
and would also be much less affected by the GRB event itself, so GRBs forming preferentially
in such environments seems to contradict the observation that only a minority of bursts are
heavily dust obscured, and that absorption often is predominantly at large distances.
Cases of massive GMCs where feedback does drive a strong outflow might in fact provide the best chances
of creating windows of low density ionized gas to the IGM, thus favouring low column-density systems.
In any event, it remains the case that GRBs occur in a range of environments, based on their galactic locations
and the evidence we have of the local density, which all suggests little bias compared to the stars
we expect to dominate the escaping EUV radiation.

\subsubsection{Could GRB progenitors preferentially remain in higher density environments?}
A sizeable fraction ($\sim30$\%) of OB stars in the Milky Way are found to have sufficiently high space velocities (several 10s of km\,s$^{-1}$),
presumably as a result of dynamical interactions, that they will end their lives well outside their nascent birth clouds \citep[e.g.][]{Tetzlaff2011}.
Such runaway stars may sometimes spend much of their lives
in relatively low density regions, and so could have a higher \avfesc\ than stars that
remain close to their birth sites.
If for some reason, such as a requirement to be a binary system, GRB progenitors were less likely to be runaways,
then, on the face of it, sight-lines to GRBs would not sample that population. 
On the other hand, since GRBs themselves ionize gas in their locality to significant distance (see Section~\ref{subsub:effects}),
and given that including runaways in hydrodynamic simulations only results in a modest 
increase of \avfesc\ of $\sim20$\% \citep{Kimm2014},
missing runaways are unlikely to have a large effect on our conclusions.

\subsubsection{Could GRB progenitors create higher density sight-lines?}
\label{subsub:credens}
We can ask whether the special nature of the GRB progenitor may influence the column-densities we measure. 
In particular, to allow a GRB jet to reach highly relativistic velocities it is thought that their 
progenitors must have no extended envelope. Indeed the lack of hydrogen and helium in the spectra
of supernovae accompanying GRBs confirms this picture.
Expelling the envelope without also losing significant angular momentum is a potential problem
for the collapsar scenario for GRB production \citep[e.g.][]{Detmers2008}. 
One possibility is that high rotation could lead to chemically homogeneous evolution, essentially consuming the envelope \citep{Yoon2005}.  
Alternatively, the hydrogen and helium layers might be lost, for example
through explosive common-envelope ejection in a tight binary \citep{Podsiadlowski2010}.
In such cases it is plausible that the expelled material, which could amount to $\sim10$\,\msol,
might provide enhanced absorption if it remains close enough to provide a significant column-density,
but far enough not to be ionized by the ambient UV radiation field prior to the burst.
Thus, even though large column-densities generally seem to be produced by gas at relatively large distance
from the GRB site, it could be that a modest contribution from gas at 10--20\,pc expelled by the progenitor
sets an effective floor to the distribution of $\lognhc\sim17.5$--18 in the GRB sample.
Other massive stars that did not produce such high mass loss would therefore have higher escape fractions.

However, this ignores the ionizing flux of the optical flash and early afterglow of the GRBs themselves.  In cases
such as the two lowest column-density systems in our sample, GRBs\,050908 and 060607A (Section~\ref{sub:low}), the bright
afterglows exceeded the flux required to ionize this mass of local gas by orders of magnitude.
Only if the peak UV luminosity of the burst was at least as faint as $M_{\rm AB}\approx-22$ could a
proportion of the expelled gas remain neutral, and even at $z\approx2$ this translates to 
a peak apparent magnitude of $m_{\rm V}\approx23$, which is effectively unfeasible for follow-up spectroscopy
with reasonable S/N given current technology.
Thus, excess absorption from gas expelled by GRB progenitors can have no effect on our sample
of \NH\ measurements.

\subsection{Applicability to the era of reionization}
\label{sec:eor}

As already discussed, GRBs favour low metallicity environments, and in particular appear to have
a roughly constant efficiency of occurrence up to around 0.3--1 times Solar metallicity, with a rapid drop
above that threshold \citep{Cucchiara2015,Perley2016b,Graham2017,Vergani2017}.
This certainly suggests that GRBs should form during the EoR when few galaxies were highly metal enriched.
Furthermore, known GRB hosts span a wide range in stellar mass \citep{Perley2016b}, including 
small, low metallicity galaxies likely representative of the galaxy populations predominant in the EoR.
In particular, while three GRB hosts have now been detected by the {\em Hubble Space Telescope} ({\em HST}) at $z\sim6$ with properties similar
to those of Lyman-break galaxies at the same redshift \citep{McGuire2016}, it remains the case that the
majority of GRB hosts at $z>6$ appear to be undetected to {\em HST} limits, consistent with them
being drawn from a galaxy luminosity function in which the faint end dominates \citep{Tanvir2012,Trenti2015}.

Further evidence that GRBs effectively sample stellar populations predominant in the EoR is that
the rate of GRBs relative to total star formation appears to rise with redshift even faster than expected
due to the metallicity sensitivity already discussed \citep{Kistler2009,Robertson2012,Perley2016b},
although number statistics are small at the highest redshifts.

Thus to the extent to which analogues of EoR galaxies exist at lower redshifts, we would expect them to
appear in the GRB host samples, and to be included in our column-density sample.
As discussed in Section~\ref{subsec:model}, the high-resolution cosmological simulations of galaxies at $z\sim7$ 
of \citet{Cen2014} find
that roughly half of GRBs should have $\lognhc<18$, which we certainly do not see at lower
redshifts. If the prediction is correct -- and limited statistics rule out
a strong test at this stage -- then it would require that there is a large sub-population of high-\fesc\ hosts at $z>6$
that barely exist in the  $z<5$ host sample.

\section{Discussion and Conclusions}
\label{sec:conc}

We have compiled a large sample of \HI\ column-density measurements obtained
from GRB afterglow spectroscopy.  Because GRBs select hosts over a very broad range of
luminosity, we consider that they provide a good representation of the dominant populations of (not highly dusty) star 
forming galaxies at $z\gtrsim2$.
Despite uncertainties about the exact nature of the GRB progenitor, we also argue
 that the life-times and locations of GRB progenitors likely make them well suited to
sampling the periods of high EUV production, and in particular the peak episodes for its escape.

Out of \sampsiz, only two sight-lines have sufficiently low column-density to allow any significant EUV to emerge.
Assuming this sample is representative of the sight-lines  to the massive stars dominating ionizing radiation production, we
conclude an average escape fraction at the Lyman limit of $\avfesc\approx\avfval$, with a 98\% confidence
upper limit of $\avfesc\approx\uplim$.
This value is in reasonable agreement with the $\avfesc\sim0.01$ predicted for GRB sight-lines
based on the hydrodynamic simulations of \citet{Pontzen2010}. 
It suggests that only in rare cases does stellar feedback puncture holes out of the dense
ISM, providing clear windows to the IGM.
This is a more stringent limit than was obtained from the recent direct search for escaping
EUV at $z\approx3.3$ by \citet{Grazian2017}, who found $\avfesc<0.017$ (67\% confidence) for
galaxies brighter than $L^*$ and $\avfesc\lesssim0.1$ for a sample brighter than $0.2L^*$.

If we account for the additional opacity due to dust for the two sight-lines for which there is
any appreciable escape fraction, and the loss of highly dusty bursts from the sample
altogether, then these numbers could reduce by factors of $2$--$3$.  If radiation from the GRBs
themselves ionized some gas in these two high-\fesc\ cases, then the estimate of \avfesc\ should 
be reduced further.

On the other side of the balance sheet, the difficulty of finding low column-density 
systems in  spectra with poor signal-to-noise does provide a modest 
selection effect in the opposite direction.  However, inspection of more complete
sub-samples suggests that even missing one or two cases would be surprising,
so the net effect is unlikely to be more than a factor $\sim2$ increase in our
estimate of \avfesc.

The bulk of the events are at redshifts $z=2$--5, which, even accounting for these systematic
uncertainties, indicates that stellar EUV falls short of
providing the radiation field needed to explain the properties of the \lya\ forest in this redshift
range \citep{Becker2013,Stanway2016}.
More crucially, if their properties apply to galaxies during the EoR, 
these limits on \avfesc\ are at least an order of magnitude below what
is required to maintain an ionized IGM in that era  \citep{Robertson2015,Stanway2016}.

There is a weak suggestion of a 
decline in average \HI\ column-density at  the highest redshifts ($z\gtrsim6$), but statistics remain too poor
for definitive conclusions, and none of the $z>5$ sight-lines have an appreciable escape fraction. 
Furthermore, there is no indication of a strong correlation of neutral hydrogen column-density 
with either galaxy UV luminosity or stellar mass, for the subsets of our sample for
which these measures are available.
Thus we find no evidence to support the suggestion that ionising escape fraction may be much higher for the small galaxies
that likely dominated star formation in the EoR.

Avoiding this conclusion would seem to require either that GRBs are not good tracers of the
primary sources of escaping EUV, for instance because the large majority occurs before
GRBs explode or because there are classes of older and less-massive stars that produce substantially 
more EUV than has  hitherto been appreciated  \citep[e.g. as a result of binary interactions;][]{Gotberg2017},
or that there is very marked evolution in the properties of GRB sight-lines between $z\sim7$ and $z\sim2$--5.

Overall this work shows the power of GRBs to address the difficult question of the Lyman continuum
escape fraction averaged over the dominant populations of high-redshift star forming galaxies, and demonstrates the benefits
of long-term campaigns to obtain GRB afterglow spectroscopy.
The evidence that GRBs occur preferentially in low metallicity systems, and that their
rate relative to the star formation rate increases with redshift, all suggest that they are likely
good tracers of star formation in the EoR.
Future samples of larger numbers of $z>6$ GRBs with good \HI\ column-density determinations may benefit
soon from the availability of {\em JWST}, and 
in the 2020s through the {\em SVOM} satellite working in conjunction with follow-up spectroscopy on 30-m class
ground-based telescopes. 
Hence we may hope to obtain much tighter and more direct
constraints on the contribution of stars to reionization \citep{Yuan2016}.

\section*{Acknowledgements}

{\em This work has benefitted significantly from
the leading contributions to our field over many years
of three sadly departed colleagues: Neil Gehrels, Javier Gorosabel and Peter Curran.}

The authors acknowledge useful discussions with Avi Loeb, Andrew Pontzen, Martin Haehnelt and Alex de Koter.

Partly based on observations made with ESO Telescopes at the La Silla Paranal Observatory under programme IDs
280.D-5059, 081.A-0856, 082.A-0301, 083.A-0644, 091.A-0442, 100.D-0648.

Partly based on observations made with the Nordic Optical Telescope,
operated by the Nordic Optical Telescope Scientific Association at the
Observatorio del Roque de los Muchachos, La Palma, Spain, of the
Instituto de Astrof\'isica de Canarias, under programs 31-014, 32-010,
39-023, 48-005 and 51-504.

Partly based on observations made with the Gran Telescopio Canarias (GTC), installed in the Spanish Observatorio del Roque de los Muchachos of the Instituto de Astrof'sica de Canarias, in the island of La Palma, Spain.

Partly based on observations made with the Italian Telescopio Nazionale
Galileo (TNG) operated on the island of La Palma by the Fundaci\'on
Galileo Galilei of the INAF (Istituto Nazionale di Astrofisica) at the
Spanish Observatorio del Roque de los Muchachos of the Instituto de
Astrof\'isica de Canarias, under programs A26TAC\_63 and A32TAC\_5. 

The WHT and its override programme (for 090715B: programme W09AN001,
P.I. Curran; for 161017A: programme W/2017A/23, P.I. Levan) are operated on the island of La Palma by the Isaac Newton
Group in the Spanish Observatorio del Roque de los Muchachos of the
Instituto de Astrof\'isica de Canarias. We thank A. Kamble, R. Starling
and P. Curran for their help with the 090715B observations, and Marie Hrudkova for executing
the 161017A observations.

Partly based on observations obtained at the Gemini Observatory, which is operated by the Association of Universities for Research in Astronomy, Inc., under a cooperative agreement with the NSF on behalf of the Gemini partnership: the National Science Foundation (United States), the National Research Council (Canada), CONICYT (Chile), Ministerio de Ciencia, Tecnolog\'{i}a e Innovaci\'{o}n Productiva (Argentina), and Minist\'{e}rio da Ci\^{e}ncia, Tecnologia e Inova\c{c}\~{a}o (Brazil).

Partially based on data from the GTC Public Archive at CAB (INTA-CSIC).

This work made use of the GRBspec database \url{http://grbspec.iaa.es}.

JJ acknowledges support from NOVA and NWO-FAPESP grant for advanced instrumentation in astronomy.

KEH acknowledges support by a Project Grant (162948--051) from The Icelandic Research Fund.

DAK acknowledges support from the Spanish research project AYA
2014-58381-P and Juan de la Cierva Incorporaci\'on IJCI-2015-26153.

AJL and ERS acknowledge STFC consolidated grant ST/L000733/1.

NRT and KW acknowledge STFC consolidated grant ST/N000757/1.

AC acknowledges NASA grant NNX15AP95A.

AdUP acknowledges support from a Ram\'on y Cajal fellowship (RyC-2012-09975), a 
2016 BBVA Foundation Grant for Researchers and Cultural Creators, and from the Spanish research project AYA 2014-58381-P.

This publication makes use of data products from the Two Micron All Sky Survey, which is a joint project of the University of Massachusetts and the Infrared Processing and Analysis Center/California Institute of Technology, funded by the National Aeronautics and Space Administration and the National Science Foundation.

The Pan-STARRS1 Surveys (PS1) and the PS1 public science archive have been made possible through contributions by the Institute for Astronomy, the University of Hawaii, the Pan-STARRS Project Office, the Max-Planck Society and its participating institutes, the Max Planck Institute for Astronomy, Heidelberg and the Max Planck Institute for Extraterrestrial Physics, Garching, The Johns Hopkins University, Durham University, the University of Edinburgh, the Queen's University Belfast, the Harvard-Smithsonian Center for Astrophysics, the Las Cumbres Observatory Global Telescope Network Incorporated, the National Central University of Taiwan, the Space Telescope Science Institute, the National Aeronautics and Space Administration under Grant No. NNX08AR22G issued through the Planetary Science Division of the NASA Science Mission Directorate, the National Science Foundation Grant No. AST-1238877, the University of Maryland, Eotvos Lorand University (ELTE), the Los Alamos National Laboratory, and the Gordon and Betty Moore Foundation.

Funding for SDSS-III has been provided by the Alfred P. Sloan Foundation, the Participating Institutions, the National Science Foundation, and the U.S. Department of Energy Office of Science. The SDSS-III web site is http://www.sdss3.org/.

SDSS-III is managed by the Astrophysical Research Consortium for the Participating Institutions of the SDSS-III Collaboration including the University of Arizona, the Brazilian Participation Group, Brookhaven National Laboratory, Carnegie Mellon University, University of Florida, the French Participation Group, the German Participation Group, Harvard University, the Instituto de Astrofisica de Canarias, the Michigan State/Notre Dame/JINA Participation Group, Johns Hopkins University, Lawrence Berkeley National Laboratory, Max Planck Institute for Astrophysics, Max Planck Institute for Extraterrestrial Physics, New Mexico State University, New York University, Ohio State University, Pennsylvania State University, University of Portsmouth, Princeton University, the Spanish Participation Group, University of Tokyo, University of Utah, Vanderbilt University, University of Virginia, University of Washington, and Yale University. 

This project has received funding from the European Research Council (ERC) under the European Union's Horizon 2020 research and innovation programme (grant agreement \Num~725246). \\
\\
Affiliations\\
$^{1}$University of Leicester, Department of Physics \& Astronomy and Leicester Institute of Space \& Earth Observation, University Road, Leicester, LE1 7RH, UK\\
$^{2}$Dark Cosmology Centre, Niels Bohr Institute, University of Copenhagen, Juliane Maries Vej 30, 2100 Copenhagen \O, Denmark\\
$^{3}$Instituto de Astrof\'{\i}sica de Andaluc\'{\i}a (IAA-CSIC), Glorieta de la Astronom\'{\i}a s/n, E-18008, Granada, Spain\\
$^{4}$Astronomical Institute Anton Pannekoek, University of Amsterdam,  PO Box 94249, 1090 GE Amsterdam, the Netherlands \\
$^{5}$Astrophysics Research Institute, Liverpool John Moores University, IC2, Liverpool Science Park, 146 Brownlow Hill,  Liverpool L3 5RF, UK\\
$^{6}$Department of Physics, University of Warwick, Coventry, CV4 7AL, UK\\
$^{7}$Harvard-Smithsonian Center for Astrophysics, 60 Garden Street, Cambridge, MA 02138, USA \\
$^{8}$NASA's Goddard Space Flight Center, Greenbelt, MD 20771, USA \\
$^{9}$Joint Space-Science Institute, University of Maryland, College Park, MD 20742, USA \\
$^{10}$Astrophysical Institute, Department of Physics and Astronomy, Ohio University, Athens, OH 45701, USA\\
$^{11}$INAF-Osservatorio Astronomico di Brera, Via Bianchi 46,  23807 Merate, Italy \\
$^{12}$University of the Virgin Islands, College of Science and Mathematics, \#2 Brewers Bay Road, Charlotte Amalie, USVI 00802\\
$^{13}$INAF - Osservatorio Astronomico di Roma, Via Frascati 33, I-00040 Monte Porzio Catone (RM), 00078, Italy \\
$^{14}$ASI-Science Data Centre, Via del Politecnico snc, I-00133 Rome, Italy\\
$^{15}$APC, Astroparticule et Cosmologie, Universit\'e Paris Diderot, CNRS/IN2P3, CEA/Irfu, Observatoire de Paris,  Sorbonne Paris Cit\'e, 10, Rue Alice Domon et L\'eonie Duquet, 75205, Paris Cedex 13, France \\
$^{16}$Centre for Astrophysics and Cosmology, University of Nova Gorica, Vipavska 11c, 5270 Ajdov\v s\v cina, Slovenia\\
$^{17}$Centre for Astrophysics and Cosmology, Science Institute, University of Iceland, Dunhagi 5, 107 Reykjav\'ik, Iceland\\
$^{18}$Max-Planck-Institut f\"ur Extraterrestrische Physik, Giessenbachstrasse, 85748 Garching, Germany\\
$^{19}$National Radio Astronomy Observatory, 520 Edgemont Road, Charlottesville, VA 22903, USA
$^{20}$Department of Astronomy, University of California, 501 Campbell Hall, Berkeley, CA 94720-3411, USA\\
$^{21}$Department of Particle Physics and Astrophysics, Weizmann Institute of Science, Rehovot 7610001, Israel \\
$^{22}$Heidelberger Institut f{\"u}r Theoretische Studien, Schloss-Wolfsbrunnenweg 35, 69118 Heidelberg, Germany\\
$^{23}$GEPI, Observatoire de Paris, PSL Research University, CNRS, Univ. Paris Diderot, Sorbonne Paris Cit\'e,  Place Jules Janssen, 92195, Meudon, France\\
$^{24}$CAS Key Laboratory of Space Astronomy and Technology, National Astronomical Observatories, Chinese Academy of Sciences,  Beijing 100012,
China




\bibliographystyle{mnras}
\bibliography{HI_paper} 

\begin{thebibliography}{}
\makeatletter
\relax
\def\mn@urlcharsother{\let\do\@makeother \do\$\do\&\do\#\do\^\do\_\do\%\do\~}
\def\mn@doi{\begingroup\mn@urlcharsother \@ifnextchar [ {\mn@doi@}
  {\mn@doi@[]}}
\def\mn@doi@[#1]#2{\def\@tempa{#1}\ifx\@tempa\@empty \href
  {http://dx.doi.org/#2} {doi:#2}\else \href {http://dx.doi.org/#2} {#1}\fi
  \endgroup}
\def\mn@eprint#1#2{\mn@eprint@#1:#2::\@nil}
\def\mn@eprint@arXiv#1{\href {http://arxiv.org/abs/#1} {{\tt arXiv:#1}}}
\def\mn@eprint@dblp#1{\href {http://dblp.uni-trier.de/rec/bibtex/#1.xml}
  {dblp:#1}}
\def\mn@eprint@#1:#2:#3:#4\@nil{\def\@tempa {#1}\def\@tempb {#2}\def\@tempc
  {#3}\ifx \@tempc \@empty \let \@tempc \@tempb \let \@tempb \@tempa \fi \ifx
  \@tempb \@empty \def\@tempb {arXiv}\fi \@ifundefined
  {mn@eprint@\@tempb}{\@tempb:\@tempc}{\expandafter \expandafter \csname
  mn@eprint@\@tempb\endcsname \expandafter{\@tempc}}}

\bibitem[\protect\citeauthoryear{{Anderson}, {Habergham}, {James}  \&
  {Hamuy}}{{Anderson} et~al.}{2012}]{Anderson2012}
{Anderson} J.~P.,  {Habergham} S.~M.,  {James} P.~A.,   {Hamuy} M.,  2012,
  \mn@doi [\mnras] {10.1111/j.1365-2966.2012.21324.x}, \href
  {http://adsabs.harvard.edu/abs/2012MNRAS.424.1372A} {424, 1372}

\bibitem[\protect\citeauthoryear{{Anderson}, {Governato}, {Karcher}, {Quinn}
  \& {Wadsley}}{{Anderson} et~al.}{2017}]{Anderson2017}
{Anderson} L.,  {Governato} F.,  {Karcher} M.,  {Quinn} T.,   {Wadsley} J.,
  2017, \mn@doi [\mnras] {10.1093/mnras/stx709}, \href
  {http://adsabs.harvard.edu/abs/2017MNRAS.468.4077A} {468, 4077}

\bibitem[\protect\citeauthoryear{{Becker} \& {Bolton}}{{Becker} \&
  {Bolton}}{2013}]{Becker2013}
{Becker} G.~D.,  {Bolton} J.~S.,  2013, \mn@doi [\mnras]
  {10.1093/mnras/stt1610}, \href
  {http://adsabs.harvard.edu/abs/2013MNRAS.436.1023B} {436, 1023}

\bibitem[\protect\citeauthoryear{{Berger}, {Penprase}, {Cenko}, {Kulkarni},
  {Fox}, {Steidel}  \& {Reddy}}{{Berger} et~al.}{2006}]{Berger2006}
{Berger} E.,  {Penprase} B.~E.,  {Cenko} S.~B.,  {Kulkarni} S.~R.,  {Fox}
  D.~B.,  {Steidel} C.~C.,   {Reddy} N.~A.,  2006, \mn@doi [\apj]
  {10.1086/501162}, \href {http://adsabs.harvard.edu/abs/2006ApJ...642..979B}
  {642, 979}

\bibitem[\protect\citeauthoryear{{Blanchard}, {Berger}  \& {Fong}}{{Blanchard}
  et~al.}{2016}]{Blanchard2016}
{Blanchard} P.~K.,  {Berger} E.,   {Fong} W.-f.,  2016, \mn@doi [\apj]
  {10.3847/0004-637X/817/2/144}, \href
  {http://adsabs.harvard.edu/abs/2016ApJ...817..144B} {817, 144}

\bibitem[\protect\citeauthoryear{{Bolmer}, {Greiner}, {Kr{\"u}hler}, {Schady},
  {Ledoux}, {Tanvir}  \& {Levan}}{{Bolmer} et~al.}{2018}]{Bolmer2017}
{Bolmer} J.,  {Greiner} J.,  {Kr{\"u}hler} T.,  {Schady} P.,  {Ledoux} C.,
  {Tanvir} N.~R.,   {Levan} A.~J.,  2018, \mn@doi [\aap]
  {10.1051/0004-6361/201731255}, \href
  {http://adsabs.harvard.edu/abs/2018A%26A...609A..62B} {609, A62}

\bibitem[\protect\citeauthoryear{{Bouchet}, {Lequeux}, {Maurice}, {Prevot}  \&
  {Prevot-Burnichon}}{{Bouchet} et~al.}{1985}]{Bouchet1985}
{Bouchet} P.,  {Lequeux} J.,  {Maurice} E.,  {Prevot} L.,   {Prevot-Burnichon}
  M.~L.,  1985, \aap, \href
  {http://adsabs.harvard.edu/abs/1985A%26A...149..330B} {149, 330}

\bibitem[\protect\citeauthoryear{{Bouwens} et~al.,}{{Bouwens}
  et~al.}{2012}]{Bouwens2012}
{Bouwens} R.~J.,  et~al., 2012, \mn@doi [\apjl] {10.1088/2041-8205/752/1/L5},
  \href {http://adsabs.harvard.edu/abs/2012ApJ...752L...5B} {752, L5}

\bibitem[\protect\citeauthoryear{{Camps-Fari{\~n}a}, {Zaragoza-Cardiel},
  {Beckman}, {Font}, {Vel{\'a}zquez}, {Rodr{\'{\i}}guez-Gonz{\'a}lez}  \&
  {Rosado}}{{Camps-Fari{\~n}a} et~al.}{2017}]{CampsFarina2017}
{Camps-Fari{\~n}a} A.,  {Zaragoza-Cardiel} J.,  {Beckman} J.~E.,  {Font} J.,
  {Vel{\'a}zquez} P.~F.,  {Rodr{\'{\i}}guez-Gonz{\'a}lez} A.,   {Rosado} M.,
  2017, \mn@doi [\mnras] {10.1093/mnras/stx551}, \href
  {http://adsabs.harvard.edu/abs/2017MNRAS.468.4134C} {468, 4134}

\bibitem[\protect\citeauthoryear{{Castro-Tirado} et~al.,}{{Castro-Tirado}
  et~al.}{2010}]{CastroTirado2010}
{Castro-Tirado} A.~J.,  et~al., 2010, \mn@doi [\aap]
  {10.1051/0004-6361/200913966}, \href
  {http://adsabs.harvard.edu/abs/2010A%26A...517A..61C} {517, A61}

\bibitem[\protect\citeauthoryear{{Castro-Tirado} et~al.,}{{Castro-Tirado}
  et~al.}{2014}]{CastroTirado2014}
{Castro-Tirado} A.~J.,  et~al., 2014, GRB Coordinates Network, \href
  {http://adsabs.harvard.edu/abs/2014GCN..16505...1C} {16505}

\bibitem[\protect\citeauthoryear{{Castro-Tirado} et~al.,}{{Castro-Tirado}
  et~al.}{2016}]{CastroTirado2016}
{Castro-Tirado} A.~J.,  et~al., 2016, GRB Coordinates Network, \href
  {http://adsabs.harvard.edu/abs/2016GCN..19632...1C} {19632}

\bibitem[\protect\citeauthoryear{{Cen} \& {Kimm}}{{Cen} \&
  {Kimm}}{2014}]{Cen2014}
{Cen} R.,  {Kimm} T.,  2014, \mn@doi [\apj] {10.1088/0004-637X/794/1/50}, \href
  {http://adsabs.harvard.edu/abs/2014ApJ...794...50C} {794, 50}

\bibitem[\protect\citeauthoryear{{Cenko}, {Berger}, {Djorgovski}, {Mahabal}  \&
  {Fox}}{{Cenko} et~al.}{2006}]{Cenko2006}
{Cenko} S.~B.,  {Berger} E.,  {Djorgovski} S.~G.,  {Mahabal} A.~A.,   {Fox}
  D.~B.,  2006, GRB Coordinates Network, \href
  {http://adsabs.harvard.edu/abs/2006GCN..5155....1C} {5155}

\bibitem[\protect\citeauthoryear{{Cenko} et~al.,}{{Cenko}
  et~al.}{2008}]{Cenko2008}
{Cenko} S.~B.,  et~al., 2008, \mn@doi [\apj] {10.1086/526491}, \href
  {http://adsabs.harvard.edu/abs/2008ApJ...677..441C} {677, 441}

\bibitem[\protect\citeauthoryear{{Cenko}, {Perley}, {Morgan}, {Klein}, {Bloom},
  {Butler}  \& {Cobb}}{{Cenko} et~al.}{2010}]{Cenko2010}
{Cenko} S.~B.,  {Perley} D.~A.,  {Morgan} A.~N.,  {Klein} C.~R.,  {Bloom}
  J.~S.,  {Butler} N.~R.,   {Cobb} B.~E.,  2010, GRB Coordinates Network, \href
  {http://adsabs.harvard.edu/abs/2010GCN..10752...1C} {10752}

\bibitem[\protect\citeauthoryear{{Chary}, {Berger}  \& {Cowie}}{{Chary}
  et~al.}{2007}]{Chary2007}
{Chary} R.,  {Berger} E.,   {Cowie} L.,  2007, \mn@doi [\apj] {10.1086/522692},
  \href {http://adsabs.harvard.edu/abs/2007ApJ...671..272C} {671, 272}

\bibitem[\protect\citeauthoryear{{Chen}, {Prochaska}  \& {Gnedin}}{{Chen}
  et~al.}{2007}]{Chen2007}
{Chen} H.-W.,  {Prochaska} J.~X.,   {Gnedin} N.~Y.,  2007, \mn@doi [\apjl]
  {10.1086/522306}, \href {http://adsabs.harvard.edu/abs/2007ApJ...667L.125C}
  {667, L125}

\bibitem[\protect\citeauthoryear{{Chen} et~al.,}{{Chen}
  et~al.}{2009}]{Chen2009}
{Chen} H.-W.,  et~al., 2009, \mn@doi [\apj] {10.1088/0004-637X/691/1/152},
  \href {http://adsabs.harvard.edu/abs/2009ApJ...691..152C} {691, 152}

\bibitem[\protect\citeauthoryear{{Chornock}, {Perley}, {Cenko}, {Bloom}, {Cobb}
   \& {Prochaska}}{{Chornock} et~al.}{2009a}]{Chornock2009}
{Chornock} R.,  {Perley} D.~A.,  {Cenko} S.~B.,  {Bloom} J.~S.,  {Cobb} B.,
  {Prochaska} J.~X.,  2009a, GRB Coordinates Network, \href
  {http://adsabs.harvard.edu/abs/2009GCN..8994....1C} {8994}

\bibitem[\protect\citeauthoryear{{Chornock}, {Perley}  \& {Cobb}}{{Chornock}
  et~al.}{2009b}]{Chornock2009b}
{Chornock} R.,  {Perley} D.~A.,   {Cobb} B.~E.,  2009b, GRB Coordinates
  Network, \href {http://adsabs.harvard.edu/abs/2009GCN..10100...1C} {10100}

\bibitem[\protect\citeauthoryear{{Chornock}, {Cucchiara}, {Fox}  \&
  {Berger}}{{Chornock} et~al.}{2010}]{Chornock2010}
{Chornock} R.,  {Cucchiara} A.,  {Fox} D.,   {Berger} E.,  2010, GRB
  Coordinates Network, \href
  {http://adsabs.harvard.edu/abs/2010GCN..10466...1C} {10466}

\bibitem[\protect\citeauthoryear{{Chornock}, {Berger}, {Fox}, {Fong}, {Laskar}
  \& {Roth}}{{Chornock} et~al.}{2014}]{Chornock2014}
{Chornock} R.,  {Berger} E.,  {Fox} D.~B.,  {Fong} W.,  {Laskar} T.,   {Roth}
  K.~C.,  2014, preprint, \href
  {http://adsabs.harvard.edu/abs/2014arXiv1405.7400C} {} (\mn@eprint {arXiv}
  {1405.7400})

\bibitem[\protect\citeauthoryear{{Christensen}, {Fynbo}, {Prochaska},
  {Th{\"o}ne}, {de Ugarte Postigo}  \& {Jakobsson}}{{Christensen}
  et~al.}{2011}]{Christensen2011}
{Christensen} L.,  {Fynbo} J.~P.~U.,  {Prochaska} J.~X.,  {Th{\"o}ne} C.~C.,
  {de Ugarte Postigo} A.,   {Jakobsson} P.,  2011, \mn@doi [\apj]
  {10.1088/0004-637X/727/2/73}, \href
  {http://adsabs.harvard.edu/abs/2011ApJ...727...73C} {727, 73}

\bibitem[\protect\citeauthoryear{{Ciardi}, {Bolton}, {Maselli}  \&
  {Graziani}}{{Ciardi} et~al.}{2012}]{Ciardi2012}
{Ciardi} B.,  {Bolton} J.~S.,  {Maselli} A.,   {Graziani} L.,  2012, \mn@doi
  [\mnras] {10.1111/j.1365-2966.2012.20902.x}, \href
  {http://adsabs.harvard.edu/abs/2012MNRAS.423..558C} {423, 558}

\bibitem[\protect\citeauthoryear{{Covino} et~al.,}{{Covino}
  et~al.}{2013}]{Covino2013}
{Covino} S.,  et~al., 2013, \mn@doi [\mnras] {10.1093/mnras/stt540}, \href
  {http://adsabs.harvard.edu/abs/2013MNRAS.432.1231C} {432, 1231}

\bibitem[\protect\citeauthoryear{{Cucchiara}}{{Cucchiara}}{2010}]{Cucchiara2010}
{Cucchiara} A.,  2010, GRB Coordinates Network, Circular Service, No.~10478,
  \#1 (2010), \href {http://adsabs.harvard.edu/abs/2010GCN.10478....1C} {10478}

\bibitem[\protect\citeauthoryear{{Cucchiara}, {Fox}, {Cenko}  \&
  {Berger}}{{Cucchiara} et~al.}{2008a}]{Cucchiara2008b}
{Cucchiara} A.,  {Fox} D.~B.,  {Cenko} S.~B.,   {Berger} E.,  2008a, GRB
  Coordinates Network, \href
  {http://adsabs.harvard.edu/abs/2008GCN..8448....1C} {8448}

\bibitem[\protect\citeauthoryear{{Cucchiara}, {Fox}, {Cenko}  \&
  {Berger}}{{Cucchiara} et~al.}{2008b}]{Cucchiara2008}
{Cucchiara} A.,  {Fox} D.~B.,  {Cenko} S.~B.,   {Berger} E.,  2008b, GRB
  Coordinates Network, \href
  {http://adsabs.harvard.edu/abs/2008GCN..8713....1C} {8713}

\bibitem[\protect\citeauthoryear{{Cucchiara} et~al.,}{{Cucchiara}
  et~al.}{2011a}]{Cucchiara2011a}
{Cucchiara} A.,  et~al., 2011a, \mn@doi [\apj] {10.1088/0004-637X/736/1/7},
  \href {http://adsabs.harvard.edu/abs/2011ApJ...736....7C} {736, 7}

\bibitem[\protect\citeauthoryear{{Cucchiara} et~al.,}{{Cucchiara}
  et~al.}{2011b}]{Cucchiara2011b}
{Cucchiara} A.,  et~al., 2011b, \mn@doi [\apj] {10.1088/0004-637X/743/2/154},
  \href {http://adsabs.harvard.edu/abs/2011ApJ...743..154C} {743, 154}

\bibitem[\protect\citeauthoryear{{Cucchiara}, {Cenko}  \& {Perley}}{{Cucchiara}
  et~al.}{2014}]{Cucchiara2014}
{Cucchiara} A.,  {Cenko} S.~B.,   {Perley} D.~A.,  2014, GRB Coordinates
  Network, \href {http://adsabs.harvard.edu/abs/2014GCN..16856...1C} {16856}

\bibitem[\protect\citeauthoryear{{Cucchiara}, {Fumagalli}, {Rafelski},
  {Kocevski}, {Prochaska}, {Cooke}  \& {Becker}}{{Cucchiara}
  et~al.}{2015}]{Cucchiara2015}
{Cucchiara} A.,  {Fumagalli} M.,  {Rafelski} M.,  {Kocevski} D.,  {Prochaska}
  J.~X.,  {Cooke} R.~J.,   {Becker} G.~D.,  2015, \mn@doi [\apj]
  {10.1088/0004-637X/804/1/51}, \href
  {http://adsabs.harvard.edu/abs/2015ApJ...804...51C} {804, 51}

\bibitem[\protect\citeauthoryear{{D'Avanzo} et~al.,}{{D'Avanzo}
  et~al.}{2010}]{DAvanzo2010}
{D'Avanzo} P.,  et~al., 2010, \mn@doi [\aap] {10.1051/0004-6361/201014801},
  \href {http://adsabs.harvard.edu/abs/2010A%26A...522A..20D} {522, A20}

\bibitem[\protect\citeauthoryear{{D'Avanzo}, {Malesani}, {D'Elia}, {Antonelli},
  {Tagliaferri}, {Vergani}, {Fiorenzano}  \& {Mainella}}{{D'Avanzo}
  et~al.}{2014}]{DAvanzo2014}
{D'Avanzo} P.,  {Malesani} D.,  {D'Elia} V.,  {Antonelli} L.~A.,  {Tagliaferri}
  G.,  {Vergani} S.~D.,  {Fiorenzano} A.,   {Mainella} G.,  2014, GRB
  Coordinates Network, \href
  {http://adsabs.harvard.edu/abs/2014GCN..16493...1D} {16493}

\bibitem[\protect\citeauthoryear{{D'Avanzo}, {Malesani}, {D'Elia}, {Melandri},
  {Lorenzi}  \& {Stoev}}{{D'Avanzo} et~al.}{2016}]{DAvanzo2016}
{D'Avanzo} P.,  {Malesani} D.,  {D'Elia} V.,  {Melandri} A.,  {Lorenzi} V.,
  {Stoev} H.,  2016, GRB Coordinates Network, \href
  {http://ukads.nottingham.ac.uk/abs/2016GCN..20078...1D} {20078}

\bibitem[\protect\citeauthoryear{{D'Elia}, {Thoene}, {de Ugarte Postigo},
  {D'Avanzo}, {Covino}, {Piranomonte}, {Salvaterra}  \& {Chincarini}}{{D'Elia}
  et~al.}{2008}]{DElia2008}
{D'Elia} V.,  {Thoene} C.~C.,  {de Ugarte Postigo} A.,  {D'Avanzo} P.,
  {Covino} S.,  {Piranomonte} S.,  {Salvaterra} R.,   {Chincarini} G.,  2008,
  GRB Coordinates Network, \href
  {http://adsabs.harvard.edu/abs/2008GCN..8531....1D} {8531}

\bibitem[\protect\citeauthoryear{{D'Elia}, {D'Avanzo}, {Covino}, {Melandri},
  {Vergani}  \& {di Fabrizio}}{{D'Elia} et~al.}{2014}]{DElia2014}
{D'Elia} V.,  {D'Avanzo} P.,  {Covino} S.,  {Melandri} A.,  {Vergani} S.~D.,
  {di Fabrizio} L.,  2014, GRB Coordinates Network, Circular Service,
  No.~15802, \#1 (2014), \href
  {http://adsabs.harvard.edu/abs/2014GCN.15802....1D} {15802}

\bibitem[\protect\citeauthoryear{{De Cia} et~al.,}{{De Cia}
  et~al.}{2011}]{DeCia2011}
{De Cia} A.,  et~al., 2011, \mn@doi [\mnras]
  {10.1111/j.1365-2966.2011.19471.x}, \href
  {http://adsabs.harvard.edu/abs/2011MNRAS.418..129D} {418, 129}

\bibitem[\protect\citeauthoryear{{Detmers}, {Langer}, {Podsiadlowski}  \&
  {Izzard}}{{Detmers} et~al.}{2008}]{Detmers2008}
{Detmers} R.~G.,  {Langer} N.,  {Podsiadlowski} P.,   {Izzard} R.~G.,  2008,
  \mn@doi [\aap] {10.1051/0004-6361:200809371}, \href
  {http://adsabs.harvard.edu/abs/2008A%26A...484..831D} {484, 831}

\bibitem[\protect\citeauthoryear{{Doran} et~al.,}{{Doran}
  et~al.}{2013}]{Doran2013}
{Doran} E.~I.,  et~al., 2013, \mn@doi [\aap] {10.1051/0004-6361/201321824},
  \href {http://adsabs.harvard.edu/abs/2013A%26A...558A.134D} {558, A134}

\bibitem[\protect\citeauthoryear{{Dove}, {Shull}  \& {Ferrara}}{{Dove}
  et~al.}{2000}]{Dove2000}
{Dove} J.~B.,  {Shull} J.~M.,   {Ferrara} A.,  2000, \mn@doi [\apj]
  {10.1086/308481}, \href {http://adsabs.harvard.edu/abs/2000ApJ...531..846D}
  {531, 846}

\bibitem[\protect\citeauthoryear{{Eldridge} \& {Stanway}}{{Eldridge} \&
  {Stanway}}{2009}]{Eldridge2009}
{Eldridge} J.~J.,  {Stanway} E.~R.,  2009, \mn@doi [\mnras]
  {10.1111/j.1365-2966.2009.15514.x}, \href
  {http://adsabs.harvard.edu/abs/2009MNRAS.400.1019E} {400, 1019}

\bibitem[\protect\citeauthoryear{{Evans} et~al.,}{{Evans}
  et~al.}{2009}]{Evans2009}
{Evans} P.~A.,  et~al., 2009, \mn@doi [\mnras]
  {10.1111/j.1365-2966.2009.14913.x}, \href
  {http://adsabs.harvard.edu/abs/2009MNRAS.397.1177E} {397, 1177}

\bibitem[\protect\citeauthoryear{{Faisst}}{{Faisst}}{2016}]{Faisst2016}
{Faisst} A.~L.,  2016, \mn@doi [\apj] {10.3847/0004-637X/829/2/99}, \href
  {http://adsabs.harvard.edu/abs/2016ApJ...829...99F} {829, 99}

\bibitem[\protect\citeauthoryear{{Fatkhullin} et~al.,}{{Fatkhullin}
  et~al.}{2009}]{Fatkhullin2009}
{Fatkhullin} T.,  et~al., 2009, GRB Coordinates Network, \href
  {http://adsabs.harvard.edu/abs/2009GCN..9712....1F} {9712}

\bibitem[\protect\citeauthoryear{{Faucher-Gigu{\`e}re}, {Lidz}, {Zaldarriaga}
  \& {Hernquist}}{{Faucher-Gigu{\`e}re} et~al.}{2009}]{FG2009}
{Faucher-Gigu{\`e}re} C.-A.,  {Lidz} A.,  {Zaldarriaga} M.,   {Hernquist} L.,
  2009, \mn@doi [\apj] {10.1088/0004-637X/703/2/1416}, \href
  {http://adsabs.harvard.edu/abs/2009ApJ...703.1416F} {703, 1416}

\bibitem[\protect\citeauthoryear{{Ferrero} et~al.,}{{Ferrero}
  et~al.}{2009}]{Ferrero2009}
{Ferrero} P.,  et~al., 2009, \mn@doi [\aap] {10.1051/0004-6361/200809980},
  \href {http://adsabs.harvard.edu/abs/2009A%26A...497..729F} {497, 729}

\bibitem[\protect\citeauthoryear{{Finkelstein} et~al.,}{{Finkelstein}
  et~al.}{2012}]{Finkelstein2012}
{Finkelstein} S.~L.,  et~al., 2012, \mn@doi [\apj]
  {10.1088/0004-637X/758/2/93}, \href
  {http://adsabs.harvard.edu/abs/2012ApJ...758...93F} {758, 93}

\bibitem[\protect\citeauthoryear{{Fiore} et~al.,}{{Fiore}
  et~al.}{2005}]{Fiore2005}
{Fiore} F.,  et~al., 2005, \mn@doi [\apj] {10.1086/429385}, \href
  {http://adsabs.harvard.edu/abs/2005ApJ...624..853F} {624, 853}

\bibitem[\protect\citeauthoryear{{Fontanot}, {Cristiani}, {Pfrommer}, {Cupani}
  \& {Vanzella}}{{Fontanot} et~al.}{2014}]{Fontanot2014}
{Fontanot} F.,  {Cristiani} S.,  {Pfrommer} C.,  {Cupani} G.,   {Vanzella} E.,
  2014, \mn@doi [\mnras] {10.1093/mnras/stt2332}, \href
  {http://adsabs.harvard.edu/abs/2014MNRAS.438.2097F} {438, 2097}

\bibitem[\protect\citeauthoryear{{Fox}, {Ledoux}, {Vreeswijk}, {Smette}  \&
  {Jaunsen}}{{Fox} et~al.}{2008}]{Fox2008}
{Fox} A.~J.,  {Ledoux} C.,  {Vreeswijk} P.~M.,  {Smette} A.,   {Jaunsen} A.~O.,
   2008, \mn@doi [\aap] {10.1051/0004-6361:200810286}, \href
  {http://adsabs.harvard.edu/abs/2008A%26A...491..189F} {491, 189}

\bibitem[\protect\citeauthoryear{{Fragos}, {Lehmer}, {Naoz}, {Zezas}  \&
  {Basu-Zych}}{{Fragos} et~al.}{2013}]{Fragos2013}
{Fragos} T.,  {Lehmer} B.~D.,  {Naoz} S.,  {Zezas} A.,   {Basu-Zych} A.,  2013,
  \mn@doi [\apjl] {10.1088/2041-8205/776/2/L31}, \href
  {http://adsabs.harvard.edu/abs/2013ApJ...776L..31F} {776, L31}

\bibitem[\protect\citeauthoryear{{Friis} et~al.,}{{Friis}
  et~al.}{2015}]{Friis2015}
{Friis} M.,  et~al., 2015, \mn@doi [\mnras] {10.1093/mnras/stv960}, \href
  {http://adsabs.harvard.edu/abs/2015MNRAS.451..167F} {451, 167}

\bibitem[\protect\citeauthoryear{{Fruchter} et~al.,}{{Fruchter}
  et~al.}{2006}]{Fruchter2006}
{Fruchter} A.~S.,  et~al., 2006, \mn@doi [\nat] {10.1038/nature04787}, \href
  {http://adsabs.harvard.edu/abs/2006Natur.441..463F} {441, 463}

\bibitem[\protect\citeauthoryear{{Fynbo} et~al.,}{{Fynbo}
  et~al.}{2002}]{Fynbo2002}
{Fynbo} J.~P.~U.,  et~al., 2002, in {Gilfanov} M.,  {Sunyeav} R.,   {Churazov}
  E.,  eds, Lighthouses of the Universe: The Most Luminous Celestial Objects
  and Their Use for Cosmology. p.~187 (\mn@eprint {} {astro-ph/0110603}),
  \mn@doi{10.1007/10856495_24}

\bibitem[\protect\citeauthoryear{{Fynbo} et~al.,}{{Fynbo}
  et~al.}{2005}]{Fynbo2005}
{Fynbo} J.~P.~U.,  et~al., 2005, \mn@doi [\apj] {10.1086/432633}, \href
  {http://adsabs.harvard.edu/abs/2005ApJ...633..317F} {633, 317}

\bibitem[\protect\citeauthoryear{{Fynbo} et~al.,}{{Fynbo}
  et~al.}{2009}]{Fynbo2009}
{Fynbo} J.~P.~U.,  et~al., 2009, \mn@doi [\apjs] {10.1088/0067-0049/185/2/526},
  \href {http://adsabs.harvard.edu/abs/2009ApJS..185..526F} {185, 526}

\bibitem[\protect\citeauthoryear{{Gehrels} \& {Razzaque}}{{Gehrels} \&
  {Razzaque}}{2013}]{Gehrels2013}
{Gehrels} N.,  {Razzaque} S.,  2013, \mn@doi [Frontiers of Physics]
  {10.1007/s11467-013-0282-3}, \href
  {http://adsabs.harvard.edu/abs/2013FrPhy...8..661G} {8, 661}

\bibitem[\protect\citeauthoryear{{Gnedin}, {Kravtsov}  \& {Chen}}{{Gnedin}
  et~al.}{2008}]{Gnedin2008}
{Gnedin} N.~Y.,  {Kravtsov} A.~V.,   {Chen} H.-W.,  2008, \mn@doi [\apj]
  {10.1086/524007}, \href
  {http://ukads.nottingham.ac.uk/abs/2008ApJ...672..765G} {672, 765}

\bibitem[\protect\citeauthoryear{{Gorosabel}, {de Ugarte Postigo}, {Th\"one},
  {Perley}  \& {Garcia Rodriguez}}{{Gorosabel} et~al.}{2014}]{Gorosabel2014}
{Gorosabel} J.,  {de Ugarte Postigo} A.,  {Th\"one} C.,  {Perley} D.,   {Garcia
  Rodriguez} A.,  2014, GRB Coordinates Network, 16671

\bibitem[\protect\citeauthoryear{{G{\"o}tberg}, {de Mink}  \&
  {Groh}}{{G{\"o}tberg} et~al.}{2017}]{Gotberg2017}
{G{\"o}tberg} Y.,  {de Mink} S.~E.,   {Groh} J.~H.,  2017, \mn@doi [\aap]
  {10.1051/0004-6361/201730472}, \href
  {http://adsabs.harvard.edu/abs/2017A%26A...608A..11G} {608, A11}

\bibitem[\protect\citeauthoryear{{Graham} \& {Fruchter}}{{Graham} \&
  {Fruchter}}{2017}]{Graham2017}
{Graham} J.~F.,  {Fruchter} A.~S.,  2017, \mn@doi [\apj]
  {10.3847/1538-4357/834/2/170}, \href
  {http://adsabs.harvard.edu/abs/2017ApJ...834..170G} {834, 170}

\bibitem[\protect\citeauthoryear{{Granot} \& {Sari}}{{Granot} \&
  {Sari}}{2002}]{Granot2002}
{Granot} J.,  {Sari} R.,  2002, \mn@doi [\apj] {10.1086/338966}, \href
  {http://adsabs.harvard.edu/abs/2002ApJ...568..820G} {568, 820}

\bibitem[\protect\citeauthoryear{{Grazian} et~al.,}{{Grazian}
  et~al.}{2017}]{Grazian2017}
{Grazian} A.,  et~al., 2017, \mn@doi [\aap] {10.1051/0004-6361/201730447},
  \href {http://adsabs.harvard.edu/abs/2017A%26A...602A..18G} {602, A18}

\bibitem[\protect\citeauthoryear{{Greiner} et~al.,}{{Greiner}
  et~al.}{2009a}]{Greiner2009b}
{Greiner} J.,  et~al., 2009a, \mn@doi [\apj] {10.1088/0004-637X/693/2/1610},
  \href {http://adsabs.harvard.edu/abs/2009ApJ...693.1610G} {693, 1610}

\bibitem[\protect\citeauthoryear{{Greiner} et~al.,}{{Greiner}
  et~al.}{2009b}]{Greiner2009}
{Greiner} J.,  et~al., 2009b, \mn@doi [\apj] {10.1088/0004-637X/693/2/1912},
  \href {http://adsabs.harvard.edu/abs/2009ApJ...693.1912G} {693, 1912}

\bibitem[\protect\citeauthoryear{{Greiner} et~al.,}{{Greiner}
  et~al.}{2011}]{Greiner2011}
{Greiner} J.,  et~al., 2011, \mn@doi [\aap] {10.1051/0004-6361/201015458},
  \href {http://adsabs.harvard.edu/abs/2011A%26A...526A..30G} {526, A30}

\bibitem[\protect\citeauthoryear{{Greiner} et~al.,}{{Greiner}
  et~al.}{2015a}]{Greiner2015b}
{Greiner} J.,  et~al., 2015a, \mn@doi [\nat] {10.1038/nature14579}, \href
  {http://adsabs.harvard.edu/abs/2015Natur.523..189G} {523, 189}

\bibitem[\protect\citeauthoryear{{Greiner} et~al.,}{{Greiner}
  et~al.}{2015b}]{Greiner2015}
{Greiner} J.,  et~al., 2015b, \mn@doi [\apj] {10.1088/0004-637X/809/1/76},
  \href {http://adsabs.harvard.edu/abs/2015ApJ...809...76G} {809, 76}

\bibitem[\protect\citeauthoryear{{Gunn} \& {Peterson}}{{Gunn} \&
  {Peterson}}{1965}]{Gunn1965}
{Gunn} J.~E.,  {Peterson} B.~A.,  1965, \mn@doi [\apj] {10.1086/148444}, \href
  {http://adsabs.harvard.edu/abs/1965ApJ...142.1633G} {142, 1633}

\bibitem[\protect\citeauthoryear{{Hansen} \& {Haiman}}{{Hansen} \&
  {Haiman}}{2004}]{Hansen2004}
{Hansen} S.~H.,  {Haiman} Z.,  2004, \mn@doi [\apj] {10.1086/379636}, \href
  {http://adsabs.harvard.edu/abs/2004ApJ...600...26H} {600, 26}

\bibitem[\protect\citeauthoryear{{Hasco{\"e}t}, {Uhm}, {Mochkovitch}  \&
  {Daigne}}{{Hasco{\"e}t} et~al.}{2011}]{Hascoet2011}
{Hasco{\"e}t} R.,  {Uhm} Z.~L.,  {Mochkovitch} R.,   {Daigne} F.,  2011,
  \mn@doi [\aap] {10.1051/0004-6361/201117404}, \href
  {http://ukads.nottingham.ac.uk/abs/2011A%26A...534A.104H} {534, A104}

\bibitem[\protect\citeauthoryear{{Hassan}, {Dav{\'e}}, {Mitra}, {Finlator},
  {Ciardi}  \& {Santos}}{{Hassan} et~al.}{2018}]{Hassan2018}
{Hassan} S.,  {Dav{\'e}} R.,  {Mitra} S.,  {Finlator} K.,  {Ciardi} B.,
  {Santos} M.~G.,  2018, \mn@doi [\mnras] {10.1093/mnras/stx2194}, \href
  {http://adsabs.harvard.edu/abs/2018MNRAS.473..227H} {473, 227}

\bibitem[\protect\citeauthoryear{{Hjorth} \& {Bloom}}{{Hjorth} \&
  {Bloom}}{2012}]{HjorthBloom2012}
{Hjorth} J.,  {Bloom} J.~S.,  2012, in Chapter 9 in ``Gamma-Ray Bursts'',
  Cambridge Astrophysics Series 51, eds.~C.~Kouveliotou, R.~A.~M.~J.~Wijers and
  S.~Woosley, Cambridge University Press (Cambridge). pp 169--190

\bibitem[\protect\citeauthoryear{{Hjorth} et~al.,}{{Hjorth}
  et~al.}{2003a}]{Hjorth2003a}
{Hjorth} J.,  et~al., 2003a, \mn@doi [\nat] {10.1038/nature01750}, \href
  {http://adsabs.harvard.edu/abs/2003Natur.423..847H} {423, 847}

\bibitem[\protect\citeauthoryear{{Hjorth} et~al.,}{{Hjorth}
  et~al.}{2003b}]{Hjorth2003b}
{Hjorth} J.,  et~al., 2003b, \mn@doi [\apj] {10.1086/378493}, \href
  {http://adsabs.harvard.edu/abs/2003ApJ...597..699H} {597, 699}

\bibitem[\protect\citeauthoryear{{Howard}, {Pudritz}, {Harris}  \&
  {Klessen}}{{Howard} et~al.}{2018}]{Howard2018}
{Howard} C.~S.,  {Pudritz} R.~E.,  {Harris} W.~E.,   {Klessen} R.~S.,  2018,
  \mn@doi [\mnras] {10.1093/mnras/stx3276}, \href
  {http://adsabs.harvard.edu/abs/2018MNRAS.475.3121H} {475, 3121}

\bibitem[\protect\citeauthoryear{{Hunt} et~al.,}{{Hunt}
  et~al.}{2014}]{Hunt2014}
{Hunt} L.~K.,  et~al., 2014, \mn@doi [\aap] {10.1051/0004-6361/201323340},
  \href {http://adsabs.harvard.edu/abs/2014A%26A...565A.112H} {565, A112}

\bibitem[\protect\citeauthoryear{{Izzo} et~al.,}{{Izzo}
  et~al.}{2018}]{Izzo2018}
{Izzo} L.,  et~al., 2018, GRB Coordinates Network, Circular Service, No.~22567,
  \#1 (2018), \href {http://adsabs.harvard.edu/abs/2018GCN.22567....1I} {22567}

\bibitem[\protect\citeauthoryear{{Jakobsson} et~al.,}{{Jakobsson}
  et~al.}{2004}]{Jakobsson2004}
{Jakobsson} P.,  et~al., 2004, \mn@doi [\aap] {10.1051/0004-6361:20041233},
  \href {http://adsabs.harvard.edu/abs/2004A%26A...427..785J} {427, 785}

\bibitem[\protect\citeauthoryear{{Jakobsson} et~al.,}{{Jakobsson}
  et~al.}{2006}]{Jakobsson2006}
{Jakobsson} P.,  et~al., 2006, \mn@doi [\aap] {10.1051/0004-6361:20066405},
  \href {http://adsabs.harvard.edu/abs/2006A%26A...460L..13J} {460, L13}

\bibitem[\protect\citeauthoryear{{Jakobsson} et~al.,}{{Jakobsson}
  et~al.}{2012}]{Jakobsson2012}
{Jakobsson} P.,  et~al., 2012, \mn@doi [\apj] {10.1088/0004-637X/752/1/62},
  \href {http://adsabs.harvard.edu/abs/2012ApJ...752...62J} {752, 62}

\bibitem[\protect\citeauthoryear{{Japelj} et~al.,}{{Japelj}
  et~al.}{2016}]{Japelj2016}
{Japelj} J.,  et~al., 2016, \mn@doi [\aap] {10.1051/0004-6361/201628314}, \href
  {http://adsabs.harvard.edu/abs/2016A%26A...590A.129J} {590, A129}

\bibitem[\protect\citeauthoryear{{Japelj} et~al.,}{{Japelj}
  et~al.}{2017}]{Japelj2017}
{Japelj} J.,  et~al., 2017, \mn@doi [\mnras] {10.1093/mnras/stx477}, \href
  {http://adsabs.harvard.edu/abs/2017MNRAS.468..389J} {468, 389}

\bibitem[\protect\citeauthoryear{{Jensen} et~al.,}{{Jensen}
  et~al.}{2001}]{Jensen2001}
{Jensen} B.~L.,  et~al., 2001, \mn@doi [\aap] {10.1051/0004-6361:20010291},
  \href {http://adsabs.harvard.edu/abs/2001A%26A...370..909J} {370, 909}

\bibitem[\protect\citeauthoryear{{Jeong}, {Sanchez-Ramirez}, {Gorosabel}  \&
  {Castro-Tirado}}{{Jeong} et~al.}{2014}]{Jeong2014}
{Jeong} S.,  {Sanchez-Ramirez} R.,  {Gorosabel} J.,   {Castro-Tirado} A.~J.,
  2014, GRB Coordinates Network, \href
  {http://adsabs.harvard.edu/abs/2014GCN..15936...1J} {15936}

\bibitem[\protect\citeauthoryear{{Kann} et~al.,}{{Kann}
  et~al.}{2010}]{Kann2010}
{Kann} D.~A.,  et~al., 2010, \mn@doi [\apj] {10.1088/0004-637X/720/2/1513},
  \href {http://adsabs.harvard.edu/abs/2010ApJ...720.1513K} {720, 1513}

\bibitem[\protect\citeauthoryear{{Kann} et~al.,}{{Kann}
  et~al.}{2017}]{Kann2017}
{Kann} D.~A.,  et~al., 2017, preprint, \href
  {http://adsabs.harvard.edu/abs/2017arXiv170600601K} {} (\mn@eprint {arXiv}
  {1706.00601})

\bibitem[\protect\citeauthoryear{{Kelly}, {Kirshner}  \& {Pahre}}{{Kelly}
  et~al.}{2008}]{Kelly2008}
{Kelly} P.~L.,  {Kirshner} R.~P.,   {Pahre} M.,  2008, \mn@doi [\apj]
  {10.1086/591925}, \href {http://adsabs.harvard.edu/abs/2008ApJ...687.1201K}
  {687, 1201}

\bibitem[\protect\citeauthoryear{{Kelly}, {Filippenko}, {Modjaz}  \&
  {Kocevski}}{{Kelly} et~al.}{2014}]{Kelly2014}
{Kelly} P.~L.,  {Filippenko} A.~V.,  {Modjaz} M.,   {Kocevski} D.,  2014,
  \mn@doi [\apj] {10.1088/0004-637X/789/1/23}, \href
  {http://adsabs.harvard.edu/abs/2014ApJ...789...23K} {789, 23}

\bibitem[\protect\citeauthoryear{{Khaire}, {Srianand}, {Choudhury}  \&
  {Gaikwad}}{{Khaire} et~al.}{2016}]{Khaire2016}
{Khaire} V.,  {Srianand} R.,  {Choudhury} T.~R.,   {Gaikwad} P.,  2016, \mn@doi
  [\mnras] {10.1093/mnras/stw192}, \href
  {http://adsabs.harvard.edu/abs/2016MNRAS.457.4051K} {457, 4051}

\bibitem[\protect\citeauthoryear{{Kimm} \& {Cen}}{{Kimm} \&
  {Cen}}{2014}]{Kimm2014}
{Kimm} T.,  {Cen} R.,  2014, \mn@doi [\apj] {10.1088/0004-637X/788/2/121},
  \href {http://adsabs.harvard.edu/abs/2014ApJ...788..121K} {788, 121}

\bibitem[\protect\citeauthoryear{{Kistler}, {Y{\"u}ksel}, {Beacom}, {Hopkins}
  \& {Wyithe}}{{Kistler} et~al.}{2009}]{Kistler2009}
{Kistler} M.~D.,  {Y{\"u}ksel} H.,  {Beacom} J.~F.,  {Hopkins} A.~M.,
  {Wyithe} J.~S.~B.,  2009, \mn@doi [\apjl] {10.1088/0004-637X/705/2/L104},
  \href {http://adsabs.harvard.edu/abs/2009ApJ...705L.104K} {705, L104}

\bibitem[\protect\citeauthoryear{{Knevitt}, {Wynn}, {Power}  \&
  {Bolton}}{{Knevitt} et~al.}{2014}]{Knevitt2014}
{Knevitt} G.,  {Wynn} G.~A.,  {Power} C.,   {Bolton} J.~S.,  2014, \mn@doi
  [\mnras] {10.1093/mnras/stu1803}, \href
  {http://adsabs.harvard.edu/abs/2014MNRAS.445.2034K} {445, 2034}

\bibitem[\protect\citeauthoryear{{Krongold} \& {Prochaska}}{{Krongold} \&
  {Prochaska}}{2013}]{Krongold2013}
{Krongold} Y.,  {Prochaska} J.~X.,  2013, \mn@doi [\apj]
  {10.1088/0004-637X/774/2/115}, \href
  {http://adsabs.harvard.edu/abs/2013ApJ...774..115K} {774, 115}

\bibitem[\protect\citeauthoryear{{Kr{\"u}hler} et~al.,}{{Kr{\"u}hler}
  et~al.}{2015}]{Kruehler2015}
{Kr{\"u}hler} T.,  et~al., 2015, \mn@doi [\aap] {10.1051/0004-6361/201425561},
  \href {http://adsabs.harvard.edu/abs/2015A%26A...581A.125K} {581, A125}

\bibitem[\protect\citeauthoryear{{Kuhlen} \& {Faucher-Gigu{\`e}re}}{{Kuhlen} \&
  {Faucher-Gigu{\`e}re}}{2012}]{Kuhlen2012}
{Kuhlen} M.,  {Faucher-Gigu{\`e}re} C.-A.,  2012, \mn@doi [\mnras]
  {10.1111/j.1365-2966.2012.20924.x}, \href
  {http://adsabs.harvard.edu/abs/2012MNRAS.423..862K} {423, 862}

\bibitem[\protect\citeauthoryear{{Kuin} et~al.,}{{Kuin}
  et~al.}{2009}]{Kuin2009}
{Kuin} N.~P.~M.,  et~al., 2009, \mn@doi [\mnras]
  {10.1111/j.1745-3933.2009.00632.x}, \href
  {http://adsabs.harvard.edu/abs/2009MNRAS.395L..21K} {395, L21}

\bibitem[\protect\citeauthoryear{{Larsson}, {Levan}, {Davies}  \&
  {Fruchter}}{{Larsson} et~al.}{2007}]{Larsson2007}
{Larsson} J.,  {Levan} A.~J.,  {Davies} M.~B.,   {Fruchter} A.~S.,  2007,
  \mn@doi [\mnras] {10.1111/j.1365-2966.2007.11523.x}, \href
  {http://adsabs.harvard.edu/abs/2007MNRAS.376.1285L} {376, 1285}

\bibitem[\protect\citeauthoryear{{Laskar}, {Berger}  \& {Chary}}{{Laskar}
  et~al.}{2011}]{Laskar2011}
{Laskar} T.,  {Berger} E.,   {Chary} R.-R.,  2011, \mn@doi [\apj]
  {10.1088/0004-637X/739/1/1}, \href
  {http://adsabs.harvard.edu/abs/2011ApJ...739....1L} {739, 1}

\bibitem[\protect\citeauthoryear{{Laskar} et~al.,}{{Laskar}
  et~al.}{2014}]{Laskar2014}
{Laskar} T.,  et~al., 2014, \mn@doi [\apj] {10.1088/0004-637X/781/1/1}, \href
  {http://adsabs.harvard.edu/abs/2014ApJ...781....1L} {781, 1}

\bibitem[\protect\citeauthoryear{{Levan} et~al.,}{{Levan}
  et~al.}{2014}]{Levan2014}
{Levan} A.~J.,  et~al., 2014, \mn@doi [\apj] {10.1088/0004-637X/781/1/13},
  \href {http://adsabs.harvard.edu/abs/2014ApJ...781...13L} {781, 13}

\bibitem[\protect\citeauthoryear{{Levesque} et~al.,}{{Levesque}
  et~al.}{2010a}]{Levesque2010}
{Levesque} E.~M.,  et~al., 2010a, \mn@doi [\mnras]
  {10.1111/j.1365-2966.2009.15733.x}, \href
  {http://adsabs.harvard.edu/abs/2010MNRAS.401..963L} {401, 963}

\bibitem[\protect\citeauthoryear{{Levesque}, {Soderberg}, {Kewley}  \&
  {Berger}}{{Levesque} et~al.}{2010b}]{Levesque2010b}
{Levesque} E.~M.,  {Soderberg} A.~M.,  {Kewley} L.~J.,   {Berger} E.,  2010b,
  \mn@doi [\apj] {10.1088/0004-637X/725/1/1337}, \href
  {http://adsabs.harvard.edu/abs/2010ApJ...725.1337L} {725, 1337}

\bibitem[\protect\citeauthoryear{{Lockman}, {Jahoda}  \& {McCammon}}{{Lockman}
  et~al.}{1986}]{Lockman1986}
{Lockman} F.~J.,  {Jahoda} K.,   {McCammon} D.,  1986, \mn@doi [\apj]
  {10.1086/164002}, \href {http://adsabs.harvard.edu/abs/1986ApJ...302..432L}
  {302, 432}

\bibitem[\protect\citeauthoryear{{Lyman} et~al.,}{{Lyman}
  et~al.}{2017}]{Lyman2017}
{Lyman} J.~D.,  et~al., 2017, \mn@doi [\mnras] {10.1093/mnras/stx220}, \href
  {http://adsabs.harvard.edu/abs/2017MNRAS.467.1795L} {467, 1795}

\bibitem[\protect\citeauthoryear{{Ma}, {Hopkins}, {Kasen}, {Quataert},
  {Faucher-Gigu{\`e}re}, {Kere{\v s}}, {Murray}  \& {Strom}}{{Ma}
  et~al.}{2016}]{Ma2016}
{Ma} X.,  {Hopkins} P.~F.,  {Kasen} D.,  {Quataert} E.,  {Faucher-Gigu{\`e}re}
  C.-A.,  {Kere{\v s}} D.,  {Murray} N.,   {Strom} A.,  2016, \mn@doi [\mnras]
  {10.1093/mnras/stw941}, \href
  {http://adsabs.harvard.edu/abs/2016MNRAS.459.3614M} {459, 3614}

\bibitem[\protect\citeauthoryear{{Madau} \& {Dickinson}}{{Madau} \&
  {Dickinson}}{2014}]{Madau2014}
{Madau} P.,  {Dickinson} M.,  2014, \mn@doi [\araa]
  {10.1146/annurev-astro-081811-125615}, \href
  {http://adsabs.harvard.edu/abs/2014ARA%26A..52..415M} {52, 415}

\bibitem[\protect\citeauthoryear{{Madau} \& {Fragos}}{{Madau} \&
  {Fragos}}{2017}]{Madau2017}
{Madau} P.,  {Fragos} T.,  2017, \mn@doi [\apj] {10.3847/1538-4357/aa6af9},
  \href {http://adsabs.harvard.edu/abs/2017ApJ...840...39M} {840, 39}

\bibitem[\protect\citeauthoryear{{Madau} \& {Haardt}}{{Madau} \&
  {Haardt}}{2015}]{Madau2015}
{Madau} P.,  {Haardt} F.,  2015, \mn@doi [\apjl] {10.1088/2041-8205/813/1/L8},
  \href {http://adsabs.harvard.edu/abs/2015ApJ...813L...8M} {813, L8}

\bibitem[\protect\citeauthoryear{{Malesani}, {Fynbo}, {D'Elia}, {de Ugarte
  Postigo}, {Jakobsson}  \& {Thoene}}{{Malesani} et~al.}{2009}]{Malesani2009}
{Malesani} D.,  {Fynbo} J.~P.~U.,  {D'Elia} V.,  {de Ugarte Postigo} A.,
  {Jakobsson} P.,   {Thoene} C.~C.,  2009, GRB Coordinates Network, \href
  {http://adsabs.harvard.edu/abs/2009GCN..9457....1M} {9457}

\bibitem[\protect\citeauthoryear{{Malesani} et~al.,}{{Malesani}
  et~al.}{2014}]{Malesani2014}
{Malesani} D.,  et~al., 2014, GRB Coordinates Network, \href
  {http://adsabs.harvard.edu/abs/2014GCN..15800...1M} {15800}

\bibitem[\protect\citeauthoryear{{Marchi} et~al.,}{{Marchi}
  et~al.}{2017}]{Marchi2017}
{Marchi} F.,  et~al., 2017, \mn@doi [\aap] {10.1051/0004-6361/201630054}, \href
  {http://adsabs.harvard.edu/abs/2017A%26A...601A..73M} {601, A73}

\bibitem[\protect\citeauthoryear{{McGuire} et~al.,}{{McGuire}
  et~al.}{2016}]{McGuire2016}
{McGuire} J.~T.~W.,  et~al., 2016, \mn@doi [\apj]
  {10.3847/0004-637X/825/2/135}, \href
  {http://adsabs.harvard.edu/abs/2016ApJ...825..135M} {825, 135}

\bibitem[\protect\citeauthoryear{{Melandri} et~al.,}{{Melandri}
  et~al.}{2008}]{Melandri2008}
{Melandri} A.,  et~al., 2008, \mn@doi [\apj] {10.1086/591243}, \href
  {http://adsabs.harvard.edu/abs/2008ApJ...686.1209M} {686, 1209}

\bibitem[\protect\citeauthoryear{{Melandri} et~al.,}{{Melandri}
  et~al.}{2015}]{Melandri2015}
{Melandri} A.,  et~al., 2015, \mn@doi [\aap] {10.1051/0004-6361/201526660},
  \href {http://adsabs.harvard.edu/abs/2015A%26A...581A..86M} {581, A86}

\bibitem[\protect\citeauthoryear{{Mirabal}, {Melandri}  \& {Halpern}}{{Mirabal}
  et~al.}{2007}]{Mirabal2007}
{Mirabal} N.,  {Melandri} A.,   {Halpern} J.~P.,  2007, GRB Coordinates
  Network, \href {http://adsabs.harvard.edu/abs/2007GCN..6162....1M} {6162}

\bibitem[\protect\citeauthoryear{{Mirabel}, {Dijkstra}, {Laurent}, {Loeb}  \&
  {Pritchard}}{{Mirabel} et~al.}{2011}]{Mirabel2011}
{Mirabel} I.~F.,  {Dijkstra} M.,  {Laurent} P.,  {Loeb} A.,   {Pritchard}
  J.~R.,  2011, \mn@doi [\aap] {10.1051/0004-6361/201016357}, \href
  {http://adsabs.harvard.edu/abs/2011A%26A...528A.149M} {528, A149}

\bibitem[\protect\citeauthoryear{{Molinari} et~al.,}{{Molinari}
  et~al.}{2007}]{Molinari2007}
{Molinari} E.,  et~al., 2007, \mn@doi [\aap] {10.1051/0004-6361:20077388},
  \href {http://adsabs.harvard.edu/abs/2007A%26A...469L..13M} {469, L13}

\bibitem[\protect\citeauthoryear{{M{\o}ller} et~al.,}{{M{\o}ller}
  et~al.}{2002}]{Moller2002}
{M{\o}ller} P.,  et~al., 2002, \mn@doi [\aap] {10.1051/0004-6361:20021612},
  \href {http://adsabs.harvard.edu/abs/2002A%26A...396L..21M} {396, L21}

\bibitem[\protect\citeauthoryear{{Morgan} et~al.,}{{Morgan}
  et~al.}{2014}]{Morgan2014}
{Morgan} A.~N.,  et~al., 2014, \mn@doi [\mnras] {10.1093/mnras/stu344}, \href
  {http://adsabs.harvard.edu/abs/2014MNRAS.440.1810M} {440, 1810}

\bibitem[\protect\citeauthoryear{{Mostardi}, {Shapley}, {Nestor}, {Steidel},
  {Reddy}  \& {Trainor}}{{Mostardi} et~al.}{2013}]{Mostardi2013}
{Mostardi} R.~E.,  {Shapley} A.~E.,  {Nestor} D.~B.,  {Steidel} C.~C.,  {Reddy}
  N.~A.,   {Trainor} R.~F.,  2013, \mn@doi [\apj] {10.1088/0004-637X/779/1/65},
  \href {http://adsabs.harvard.edu/abs/2013ApJ...779...65M} {779, 65}

\bibitem[\protect\citeauthoryear{{Nestor}, {Shapley}, {Kornei}, {Steidel}  \&
  {Siana}}{{Nestor} et~al.}{2013}]{Nestor2013}
{Nestor} D.~B.,  {Shapley} A.~E.,  {Kornei} K.~A.,  {Steidel} C.~C.,   {Siana}
  B.,  2013, \mn@doi [\apj] {10.1088/0004-637X/765/1/47}, \href
  {http://adsabs.harvard.edu/abs/2013ApJ...765...47N} {765, 47}

\bibitem[\protect\citeauthoryear{{Nicuesa Guelbenzu} et~al.,}{{Nicuesa
  Guelbenzu} et~al.}{2011}]{Nicuesa2011}
{Nicuesa Guelbenzu} A.,  et~al., 2011, \mn@doi [\aap]
  {10.1051/0004-6361/201116657}, \href
  {http://adsabs.harvard.edu/abs/2011A%26A...531L...6N} {531, L6}

\bibitem[\protect\citeauthoryear{{Nysewander}, {Reichart}, {Crain}, {Foster},
  {Haislip}, {Ivarsen}, {Lacluyze}  \& {Trotter}}{{Nysewander}
  et~al.}{2009}]{Nysewander2009}
{Nysewander} M.,  {Reichart} D.~E.,  {Crain} J.~A.,  {Foster} A.,  {Haislip}
  J.,  {Ivarsen} K.,  {Lacluyze} A.,   {Trotter} A.,  2009, \mn@doi [\apj]
  {10.1088/0004-637X/693/2/1417}, \href
  {http://adsabs.harvard.edu/abs/2009ApJ...693.1417N} {693, 1417}

\bibitem[\protect\citeauthoryear{{Ochsendorf}, {Meixner}, {Roman-Duval},
  {Rahman}  \& {Evans}}{{Ochsendorf} et~al.}{2017}]{Ochsendorf2017}
{Ochsendorf} B.~B.,  {Meixner} M.,  {Roman-Duval} J.,  {Rahman} M.,   {Evans}
  II N.~J.,  2017, \mn@doi [\apj] {10.3847/1538-4357/aa704a}, \href
  {http://adsabs.harvard.edu/abs/2017ApJ...841..109O} {841, 109}

\bibitem[\protect\citeauthoryear{{Oey} \& {Clarke}}{{Oey} \&
  {Clarke}}{1997}]{Oey1997}
{Oey} M.~S.,  {Clarke} C.~J.,  1997, \mn@doi [\mnras]
  {10.1093/mnras/289.3.570}, \href
  {http://adsabs.harvard.edu/abs/1997MNRAS.289..570O} {289}

\bibitem[\protect\citeauthoryear{{Ouchi} et~al.,}{{Ouchi}
  et~al.}{2009}]{Ouchi2009}
{Ouchi} M.,  et~al., 2009, \mn@doi [\apj] {10.1088/0004-637X/706/2/1136}, \href
  {http://adsabs.harvard.edu/abs/2009ApJ...706.1136O} {706, 1136}

\bibitem[\protect\citeauthoryear{{Page} et~al.,}{{Page}
  et~al.}{2009}]{Page2009}
{Page} K.~L.,  et~al., 2009, \mn@doi [\mnras]
  {10.1111/j.1365-2966.2009.15462.x}, \href
  {http://adsabs.harvard.edu/abs/2009MNRAS.400..134P} {400, 134}

\bibitem[\protect\citeauthoryear{{Patel}, {Warren}, {Mortlock}  \&
  {Fynbo}}{{Patel} et~al.}{2010}]{Patel2010}
{Patel} M.,  {Warren} S.~J.,  {Mortlock} D.~J.,   {Fynbo} J.~P.~U.,  2010,
  \mn@doi [\aap] {10.1051/0004-6361/200913876}, \href
  {http://adsabs.harvard.edu/abs/2010A%26A...512L...3P} {512, L3}

\bibitem[\protect\citeauthoryear{{Pellegrini}, {Oey}, {Winkler}, {Points},
  {Smith}, {Jaskot}  \& {Zastrow}}{{Pellegrini} et~al.}{2012}]{Pellegrini2012}
{Pellegrini} E.~W.,  {Oey} M.~S.,  {Winkler} P.~F.,  {Points} S.~D.,  {Smith}
  R.~C.,  {Jaskot} A.~E.,   {Zastrow} J.,  2012, \mn@doi [\apj]
  {10.1088/0004-637X/755/1/40}, \href
  {http://adsabs.harvard.edu/abs/2012ApJ...755...40P} {755, 40}

\bibitem[\protect\citeauthoryear{{Perley}}{{Perley}}{2011}]{Perley2011}
{Perley} D.~A.,  2011, PhD thesis, University of California, Berkeley

\bibitem[\protect\citeauthoryear{{Perley}}{{Perley}}{2014}]{Perley2014}
{Perley} D.~A.,  2014, GRB Coordinates Network, \href
  {http://adsabs.harvard.edu/abs/2014GCN..16181...1P} {16181}

\bibitem[\protect\citeauthoryear{{Perley} et~al.,}{{Perley}
  et~al.}{2008}]{Perley2008}
{Perley} D.~A.,  et~al., 2008, \mn@doi [\apj] {10.1086/591961}, \href
  {http://adsabs.harvard.edu/abs/2008ApJ...688..470P} {688, 470}

\bibitem[\protect\citeauthoryear{{Perley} et~al.,}{{Perley}
  et~al.}{2010}]{Perley2010}
{Perley} D.~A.,  et~al., 2010, \mn@doi [\mnras]
  {10.1111/j.1365-2966.2010.16772.x}, \href
  {http://adsabs.harvard.edu/abs/2010MNRAS.406.2473P} {406, 2473}

\bibitem[\protect\citeauthoryear{{Perley} et~al.,}{{Perley}
  et~al.}{2013}]{Perley2013}
{Perley} D.~A.,  et~al., 2013, \mn@doi [\apj] {10.1088/0004-637X/778/2/128},
  \href {http://adsabs.harvard.edu/abs/2013ApJ...778..128P} {778, 128}

\bibitem[\protect\citeauthoryear{{Perley} et~al.,}{{Perley}
  et~al.}{2015}]{Perley2015}
{Perley} D.~A.,  et~al., 2015, \mn@doi [\apj] {10.1088/0004-637X/801/2/102},
  \href {http://adsabs.harvard.edu/abs/2015ApJ...801..102P} {801, 102}

\bibitem[\protect\citeauthoryear{{Perley} et~al.,}{{Perley}
  et~al.}{2016a}]{Perley2016a}
{Perley} D.~A.,  et~al., 2016a, \mn@doi [\apj] {10.3847/0004-637X/817/1/7},
  \href {http://adsabs.harvard.edu/abs/2016ApJ...817....7P} {817, 7}

\bibitem[\protect\citeauthoryear{{Perley} et~al.,}{{Perley}
  et~al.}{2016b}]{Perley2016b}
{Perley} D.~A.,  et~al., 2016b, \mn@doi [\apj] {10.3847/0004-637X/817/1/8},
  \href {http://adsabs.harvard.edu/abs/2016ApJ...817....8P} {817, 8}

\bibitem[\protect\citeauthoryear{{Perna} \& {Lazzati}}{{Perna} \&
  {Lazzati}}{2002}]{Perna2002}
{Perna} R.,  {Lazzati} D.,  2002, \mn@doi [\apj] {10.1086/343081}, \href
  {http://adsabs.harvard.edu/abs/2002ApJ...580..261P} {580, 261}

\bibitem[\protect\citeauthoryear{{Planck Collaboration} et~al.,}{{Planck
  Collaboration} et~al.}{2016}]{Planck2016}
{Planck Collaboration} et~al., 2016, \mn@doi [\aap]
  {10.1051/0004-6361/201628897}, \href
  {http://adsabs.harvard.edu/abs/2016A%26A...596A.108P} {596, A108}

\bibitem[\protect\citeauthoryear{{Podsiadlowski}, {Ivanova}, {Justham}  \&
  {Rappaport}}{{Podsiadlowski} et~al.}{2010}]{Podsiadlowski2010}
{Podsiadlowski} P.,  {Ivanova} N.,  {Justham} S.,   {Rappaport} S.,  2010,
  \mn@doi [\mnras] {10.1111/j.1365-2966.2010.16751.x}, \href
  {http://adsabs.harvard.edu/abs/2010MNRAS.406..840P} {406, 840}

\bibitem[\protect\citeauthoryear{{Pontzen} et~al.,}{{Pontzen}
  et~al.}{2010}]{Pontzen2010}
{Pontzen} A.,  et~al., 2010, \mn@doi [\mnras]
  {10.1111/j.1365-2966.2009.16017.x}, \href
  {http://adsabs.harvard.edu/abs/2010MNRAS.402.1523P} {402, 1523}

\bibitem[\protect\citeauthoryear{{Prochaska}, {Chen}  \& {Bloom}}{{Prochaska}
  et~al.}{2006}]{Prochaska2006}
{Prochaska} J.~X.,  {Chen} H.-W.,   {Bloom} J.~S.,  2006, \mn@doi [\apj]
  {10.1086/505737}, \href {http://adsabs.harvard.edu/abs/2006ApJ...648...95P}
  {648, 95}

\bibitem[\protect\citeauthoryear{{Prochaska}, {Chen}, {Dessauges-Zavadsky}  \&
  {Bloom}}{{Prochaska} et~al.}{2007}]{Prochaska2007}
{Prochaska} J.~X.,  {Chen} H.-W.,  {Dessauges-Zavadsky} M.,   {Bloom} J.~S.,
  2007, \mn@doi [\apj] {10.1086/520042}, \href
  {http://adsabs.harvard.edu/abs/2007ApJ...666..267P} {666, 267}

\bibitem[\protect\citeauthoryear{{Prochaska}, {Perley}, {Howard}, {Chen},
  {Marcy}, {Fischer}  \& {Wilburn}}{{Prochaska} et~al.}{2008}]{Prochaska2008}
{Prochaska} J.~X.,  {Perley} D.,  {Howard} A.,  {Chen} H.-W.,  {Marcy} G.,
  {Fischer} D.,   {Wilburn} C.,  2008, GRB Coordinates Network, \href
  {http://adsabs.harvard.edu/abs/2008GCN..8083....1P} {8083}

\bibitem[\protect\citeauthoryear{{Rahner}, {Pellegrini}, {Glover}  \&
  {Klessen}}{{Rahner} et~al.}{2017}]{Rahner2017}
{Rahner} D.,  {Pellegrini} E.~W.,  {Glover} S.~C.~O.,   {Klessen} R.~S.,  2017,
  \mn@doi [\mnras] {10.1093/mnras/stx1532}, \href
  {http://adsabs.harvard.edu/abs/2017MNRAS.470.4453R} {470, 4453}

\bibitem[\protect\citeauthoryear{{Razoumov} \& {Sommer-Larsen}}{{Razoumov} \&
  {Sommer-Larsen}}{2010}]{Razoumov2010}
{Razoumov} A.~O.,  {Sommer-Larsen} J.,  2010, \mn@doi [\apj]
  {10.1088/0004-637X/710/2/1239}, \href
  {http://adsabs.harvard.edu/abs/2010ApJ...710.1239R} {710, 1239}

\bibitem[\protect\citeauthoryear{{Robertson} \& {Ellis}}{{Robertson} \&
  {Ellis}}{2012}]{Robertson2012}
{Robertson} B.~E.,  {Ellis} R.~S.,  2012, \mn@doi [\apj]
  {10.1088/0004-637X/744/2/95}, \href
  {http://adsabs.harvard.edu/abs/2012ApJ...744...95R} {744, 95}

\bibitem[\protect\citeauthoryear{{Robertson}, {Ellis}, {Furlanetto}  \&
  {Dunlop}}{{Robertson} et~al.}{2015}]{Robertson2015}
{Robertson} B.~E.,  {Ellis} R.~S.,  {Furlanetto} S.~R.,   {Dunlop} J.~S.,
  2015, \mn@doi [\apjl] {10.1088/2041-8205/802/2/L19}, \href
  {http://adsabs.harvard.edu/abs/2015ApJ...802L..19R} {802, L19}

\bibitem[\protect\citeauthoryear{{Rol} et~al.,}{{Rol} et~al.}{2007}]{Rol2007}
{Rol} E.,  et~al., 2007, GRB Coordinates Network, \href
  {http://adsabs.harvard.edu/abs/2007GCN..6221....1R} {6221}

\bibitem[\protect\citeauthoryear{{Rossi} et~al.,}{{Rossi}
  et~al.}{2012}]{Rossi2012}
{Rossi} A.,  et~al., 2012, \mn@doi [\aap] {10.1051/0004-6361/201117201}, \href
  {http://adsabs.harvard.edu/abs/2012A%26A...545A..77R} {545, A77}

\bibitem[\protect\citeauthoryear{{Roy}, {Nath}  \& {Sharma}}{{Roy}
  et~al.}{2015}]{Roy2015}
{Roy} A.,  {Nath} B.~B.,   {Sharma} P.,  2015, \mn@doi [\mnras]
  {10.1093/mnras/stv1006}, \href
  {http://adsabs.harvard.edu/abs/2015MNRAS.451.1939R} {451, 1939}

\bibitem[\protect\citeauthoryear{{Rutkowski} et~al.,}{{Rutkowski}
  et~al.}{2017}]{Rutkowski2017}
{Rutkowski} M.~J.,  et~al., 2017, \mn@doi [\apjl] {10.3847/2041-8213/aa733b},
  \href {http://adsabs.harvard.edu/abs/2017ApJ...841L..27R} {841, L27}

\bibitem[\protect\citeauthoryear{{Sabbi} et~al.,}{{Sabbi}
  et~al.}{2016}]{Sabbi2016}
{Sabbi} E.,  et~al., 2016, \mn@doi [\apjs] {10.3847/0067-0049/222/1/11}, \href
  {http://adsabs.harvard.edu/abs/2016ApJS..222...11S} {222, 11}

\bibitem[\protect\citeauthoryear{{Salvaterra} et~al.,}{{Salvaterra}
  et~al.}{2009}]{Salvaterra2009}
{Salvaterra} R.,  et~al., 2009, \mn@doi [\nat] {10.1038/nature08445}, \href
  {http://adsabs.harvard.edu/abs/2009Natur.461.1258S} {461, 1258}

\bibitem[\protect\citeauthoryear{{Sana} et~al.,}{{Sana}
  et~al.}{2012}]{Sana2012}
{Sana} H.,  et~al., 2012, \mn@doi [Science] {10.1126/science.1223344}, \href
  {http://adsabs.harvard.edu/abs/2012Sci...337..444S} {337, 444}

\bibitem[\protect\citeauthoryear{{S{\'a}nchez-Ram{\'{\i}}rez}
  et~al.,}{{S{\'a}nchez-Ram{\'{\i}}rez} et~al.}{2013a}]{SanchezRamirez2013b}
{S{\'a}nchez-Ram{\'{\i}}rez} R.,  et~al., 2013a, in Revista Mexicana de
  Astronomia y Astrofisica Conference Series. pp 113--113

\bibitem[\protect\citeauthoryear{{Sanchez-Ramirez}, {Gorosabel},
  {Castro-Tirado}, {Cepa}  \& {Gomez-Velarde}}{{Sanchez-Ramirez}
  et~al.}{2013b}]{SanchezRamirez2013}
{Sanchez-Ramirez} R.,  {Gorosabel} J.,  {Castro-Tirado} A.~J.,  {Cepa} J.,
  {Gomez-Velarde} G.,  2013b, GRB Coordinates Network, \href
  {http://adsabs.harvard.edu/abs/2013GCN..14685...1S} {14685}

\bibitem[\protect\citeauthoryear{{Savaglio} et~al.,}{{Savaglio}
  et~al.}{2012}]{Savaglio2012}
{Savaglio} S.,  et~al., 2012, \mn@doi [\mnras]
  {10.1111/j.1365-2966.2011.20074.x}, \href
  {http://adsabs.harvard.edu/abs/2012MNRAS.420..627S} {420, 627}

\bibitem[\protect\citeauthoryear{{Schady}, {Savaglio}, {Kr{\"u}hler}, {Greiner}
   \& {Rau}}{{Schady} et~al.}{2011}]{Schady2011}
{Schady} P.,  {Savaglio} S.,  {Kr{\"u}hler} T.,  {Greiner} J.,   {Rau} A.,
  2011, \mn@doi [\aap] {10.1051/0004-6361/201015608}, \href
  {http://adsabs.harvard.edu/abs/2011A%26A...525A.113S} {525, A113}

\bibitem[\protect\citeauthoryear{{Schaerer}, {Boone}, {Zamojski}, {Staguhn},
  {Dessauges-Zavadsky}, {Finkelstein}  \& {Combes}}{{Schaerer}
  et~al.}{2015}]{Schaerer2015}
{Schaerer} D.,  {Boone} F.,  {Zamojski} M.,  {Staguhn} J.,
  {Dessauges-Zavadsky} M.,  {Finkelstein} S.,   {Combes} F.,  2015, \mn@doi
  [\aap] {10.1051/0004-6361/201424649}, \href
  {http://adsabs.harvard.edu/abs/2015A%26A...574A..19S} {574, A19}

\bibitem[\protect\citeauthoryear{{Schlafly} \& {Finkbeiner}}{{Schlafly} \&
  {Finkbeiner}}{2011}]{Schlafly2011}
{Schlafly} E.~F.,  {Finkbeiner} D.~P.,  2011, \mn@doi [\apj]
  {10.1088/0004-637X/737/2/103}, \href
  {http://adsabs.harvard.edu/abs/2011ApJ...737..103S} {737, 103}

\bibitem[\protect\citeauthoryear{{Schulze} et~al.,}{{Schulze}
  et~al.}{2015}]{Schulze2015}
{Schulze} S.,  et~al., 2015, \mn@doi [\apj] {10.1088/0004-637X/808/1/73}, \href
  {http://adsabs.harvard.edu/abs/2015ApJ...808...73S} {808, 73}

\bibitem[\protect\citeauthoryear{{Sciama}}{{Sciama}}{1982}]{Sciama1982}
{Sciama} D.~W.,  1982, \mn@doi [\mnras] {10.1093/mnras/198.1.1P}, \href
  {http://adsabs.harvard.edu/abs/1982MNRAS.198P...1S} {198, 1P}

\bibitem[\protect\citeauthoryear{{Selsing} et~al.,}{{Selsing}
  et~al.}{2018}]{Selsing2018}
{Selsing} J.,  et~al., 2018, preprint, \href
  {http://adsabs.harvard.edu/abs/2018arXiv180207727S} {} (\mn@eprint {arXiv}
  {1802.07727})

\bibitem[\protect\citeauthoryear{{Shapley}, {Steidel}, {Pettini}, {Adelberger}
  \& {Erb}}{{Shapley} et~al.}{2006}]{Shapley2006}
{Shapley} A.~E.,  {Steidel} C.~C.,  {Pettini} M.,  {Adelberger} K.~L.,   {Erb}
  D.~K.,  2006, \mn@doi [\apj] {10.1086/507511}, \href
  {http://adsabs.harvard.edu/abs/2006ApJ...651..688S} {651, 688}

\bibitem[\protect\citeauthoryear{{Shapley}, {Steidel}, {Strom},
  {Bogosavljevi{\'c}}, {Reddy}, {Siana}, {Mostardi}  \& {Rudie}}{{Shapley}
  et~al.}{2016}]{Shapley2016}
{Shapley} A.~E.,  {Steidel} C.~C.,  {Strom} A.~L.,  {Bogosavljevi{\'c}} M.,
  {Reddy} N.~A.,  {Siana} B.,  {Mostardi} R.~E.,   {Rudie} G.~C.,  2016,
  \mn@doi [\apjl] {10.3847/2041-8205/826/2/L24}, \href
  {http://adsabs.harvard.edu/abs/2016ApJ...826L..24S} {826, L24}

\bibitem[\protect\citeauthoryear{{Sharma}, {Theuns}, {Frenk}, {Bower}, {Crain},
  {Schaller}  \& {Schaye}}{{Sharma} et~al.}{2016}]{Sharma2016}
{Sharma} M.,  {Theuns} T.,  {Frenk} C.,  {Bower} R.,  {Crain} R.,  {Schaller}
  M.,   {Schaye} J.,  2016, \mn@doi [\mnras] {10.1093/mnrasl/slw021}, \href
  {http://adsabs.harvard.edu/abs/2016MNRAS.458L..94S} {458, L94}

\bibitem[\protect\citeauthoryear{{Shin} et~al.,}{{Shin}
  et~al.}{2006}]{Shin2006}
{Shin} M.-S.,  et~al., 2006, preprint, \href
  {http://adsabs.harvard.edu/abs/2006astro.ph..8327S} {} (\mn@eprint {}
  {astro-ph/0608327})

\bibitem[\protect\citeauthoryear{{Siana} et~al.,}{{Siana}
  et~al.}{2015}]{Siana2015}
{Siana} B.,  et~al., 2015, \mn@doi [\apj] {10.1088/0004-637X/804/1/17}, \href
  {http://adsabs.harvard.edu/abs/2015ApJ...804...17S} {804, 17}

\bibitem[\protect\citeauthoryear{{Smette}, {Ledoux}, {Vreeswijk}, {De Cia},
  {Petitjean}, {Fynbo}, {Malesani}  \& {Fox}}{{Smette}
  et~al.}{2013}]{Smette2013}
{Smette} A.,  {Ledoux} C.,  {Vreeswijk} P.,  {De Cia} A.,  {Petitjean} P.,
  {Fynbo} J.,  {Malesani} D.,   {Fox} A.,  2013, GRB Coordinates Network, \href
  {http://adsabs.harvard.edu/abs/2013GCN..14848...1S} {14848}

\bibitem[\protect\citeauthoryear{{Stanway}, {Eldridge}  \& {Becker}}{{Stanway}
  et~al.}{2016}]{Stanway2016}
{Stanway} E.~R.,  {Eldridge} J.~J.,   {Becker} G.~D.,  2016, \mn@doi [\mnras]
  {10.1093/mnras/stv2661}, \href
  {http://adsabs.harvard.edu/abs/2016MNRAS.456..485S} {456, 485}

\bibitem[\protect\citeauthoryear{{Starling}, {Wijers}, {Hughes}, {Tanvir},
  {Vreeswijk}, {Rol}  \& {Salamanca}}{{Starling} et~al.}{2005}]{Starling2005}
{Starling} R.~L.~C.,  {Wijers} R.~A.~M.~J.,  {Hughes} M.~A.,  {Tanvir} N.~R.,
  {Vreeswijk} P.~M.,  {Rol} E.,   {Salamanca} I.,  2005, \mn@doi [\mnras]
  {10.1111/j.1365-2966.2005.09042.x}, \href
  {http://adsabs.harvard.edu/abs/2005MNRAS.360..305S} {360, 305}

\bibitem[\protect\citeauthoryear{{Steidel}, {Pettini}  \&
  {Adelberger}}{{Steidel} et~al.}{2001}]{Steidel2001}
{Steidel} C.~C.,  {Pettini} M.,   {Adelberger} K.~L.,  2001, \mn@doi [\apj]
  {10.1086/318323}, \href {http://adsabs.harvard.edu/abs/2001ApJ...546..665S}
  {546, 665}

\bibitem[\protect\citeauthoryear{{Svensson}, {Levan}, {Tanvir}, {Fruchter}  \&
  {Strolger}}{{Svensson} et~al.}{2010}]{Svensson2010}
{Svensson} K.~M.,  {Levan} A.~J.,  {Tanvir} N.~R.,  {Fruchter} A.~S.,
  {Strolger} L.-G.,  2010, \mn@doi [\mnras] {10.1111/j.1365-2966.2010.16442.x},
  \href {http://adsabs.harvard.edu/abs/2010MNRAS.405...57S} {405, 57}

\bibitem[\protect\citeauthoryear{{Tanvir} et~al.,}{{Tanvir}
  et~al.}{2009}]{Tanvir2009}
{Tanvir} N.~R.,  et~al., 2009, \mn@doi [\nat] {10.1038/nature08459}, \href
  {http://adsabs.harvard.edu/abs/2009Natur.461.1254T} {461, 1254}

\bibitem[\protect\citeauthoryear{{Tanvir}, {Wiersema}, {Levan}, {Cenko}  \&
  {Geballe}}{{Tanvir} et~al.}{2011}]{Tanvir2011}
{Tanvir} N.~R.,  {Wiersema} K.,  {Levan} A.~J.,  {Cenko} S.~B.,   {Geballe} T.,
   2011, GRB Coordinates Network, \href
  {http://adsabs.harvard.edu/abs/2011GCN..12225...1T} {12225}

\bibitem[\protect\citeauthoryear{{Tanvir} et~al.,}{{Tanvir}
  et~al.}{2012a}]{Tanvir2012}
{Tanvir} N.~R.,  et~al., 2012a, \mn@doi [\apj] {10.1088/0004-637X/754/1/46},
  \href {http://adsabs.harvard.edu/abs/2012ApJ...754...46T} {754, 46}

\bibitem[\protect\citeauthoryear{{Tanvir}, {Levan}  \& {Matulonis}}{{Tanvir}
  et~al.}{2012b}]{Tanvir2012b}
{Tanvir} N.~R.,  {Levan} A.~J.,   {Matulonis} T.,  2012b, GRB Coordinates
  Network, \href {http://adsabs.harvard.edu/abs/2012GCN..14009...1T} {14009}

\bibitem[\protect\citeauthoryear{{Tanvir} et~al.,}{{Tanvir}
  et~al.}{2017}]{Tanvir2017}
{Tanvir} N.~R.,  et~al., 2017, preprint, \href
  {http://adsabs.harvard.edu/abs/2017arXiv170309052T} {} (\mn@eprint {arXiv}
  {1703.09052})

\bibitem[\protect\citeauthoryear{{Tetzlaff}, {Neuh{\"a}user}  \&
  {Hohle}}{{Tetzlaff} et~al.}{2011}]{Tetzlaff2011}
{Tetzlaff} N.,  {Neuh{\"a}user} R.,   {Hohle} M.~M.,  2011, \mn@doi [\mnras]
  {10.1111/j.1365-2966.2010.17434.x}, \href
  {http://adsabs.harvard.edu/abs/2011MNRAS.410..190T} {410, 190}

\bibitem[\protect\citeauthoryear{{Thoene} et~al.,}{{Thoene}
  et~al.}{2009}]{Thoene2009}
{Thoene} C.~C.,  et~al., 2009, GRB Coordinates Network, \href
  {http://adsabs.harvard.edu/abs/2009GCN..9409....1T} {9409}

\bibitem[\protect\citeauthoryear{{Th\"one}, {Perley}, {Cooke}, {Bloom}, {Chen}
  \& {Barton}}{{Th\"one} et~al.}{2007}]{Thoene2007}
{Th\"one} C.~C.,  {Perley} D.~A.,  {Cooke} J.,  {Bloom} J.~S.,  {Chen} H.-W.,
  {Barton} E.,  2007, GRB Coordinates Network, \href
  {http://adsabs.harvard.edu/abs/2007GCN..6741....1T} {6741}

\bibitem[\protect\citeauthoryear{{Th{\"o}ne} et~al.,}{{Th{\"o}ne}
  et~al.}{2011}]{Thoene2011}
{Th{\"o}ne} C.~C.,  et~al., 2011, \mn@doi [\mnras]
  {10.1111/j.1365-2966.2011.18408.x}, \href
  {http://adsabs.harvard.edu/abs/2011MNRAS.414..479T} {414, 479}

\bibitem[\protect\citeauthoryear{{Th\"one}, {de Ugarte Postigo}, {Gorosabel},
  {Sanchez-Ramirez}, {Fynbo}  \& {Gomez Velarde}}{{Th\"one}
  et~al.}{2012}]{Thoene2012}
{Th\"one} C.~C.,  {de Ugarte Postigo} A.,  {Gorosabel} J.,  {Sanchez-Ramirez}
  R.,  {Fynbo} J.~P.~U.,   {Gomez Velarde} G.,  2012, GRB Coordinates Network,
  \href {http://adsabs.harvard.edu/abs/2012GCN..13628...1T} {13628}

\bibitem[\protect\citeauthoryear{{Torii}}{{Torii}}{2005}]{Torii2005}
{Torii} K.,  2005, GRB Coordinates Network, \href
  {http://adsabs.harvard.edu/abs/2005GCN..3943....1T} {3943}

\bibitem[\protect\citeauthoryear{{Totani}, {Kawai}, {Kosugi}, {Aoki}, {Yamada},
  {Iye}, {Ohta}  \& {Hattori}}{{Totani} et~al.}{2006}]{Totani2006}
{Totani} T.,  {Kawai} N.,  {Kosugi} G.,  {Aoki} K.,  {Yamada} T.,  {Iye} M.,
  {Ohta} K.,   {Hattori} T.,  2006, \mn@doi [\pasj] {10.1093/pasj/58.3.485},
  \href {http://adsabs.harvard.edu/abs/2006PASJ...58..485T} {58, 485}

\bibitem[\protect\citeauthoryear{{Toy} et~al.,}{{Toy} et~al.}{2016}]{Toy2016}
{Toy} V.~L.,  et~al., 2016, \mn@doi [\apj] {10.3847/0004-637X/832/2/175}, \href
  {http://adsabs.harvard.edu/abs/2016ApJ...832..175T} {832, 175}

\bibitem[\protect\citeauthoryear{{Trebitsch}, {Blaizot}, {Rosdahl}, {Devriendt}
   \& {Slyz}}{{Trebitsch} et~al.}{2017}]{Trebitsch2017}
{Trebitsch} M.,  {Blaizot} J.,  {Rosdahl} J.,  {Devriendt} J.,   {Slyz} A.,
  2017, \mn@doi [\mnras] {10.1093/mnras/stx1060}, \href
  {http://adsabs.harvard.edu/abs/2017MNRAS.470..224T} {470, 224}

\bibitem[\protect\citeauthoryear{{Trenti}, {Perna}  \& {Jimenez}}{{Trenti}
  et~al.}{2015}]{Trenti2015}
{Trenti} M.,  {Perna} R.,   {Jimenez} R.,  2015, \mn@doi [\apj]
  {10.1088/0004-637X/802/2/103}, \href
  {http://adsabs.harvard.edu/abs/2015ApJ...802..103T} {802, 103}

\bibitem[\protect\citeauthoryear{{Updike} et~al.,}{{Updike}
  et~al.}{2008}]{Updike2008}
{Updike} A.~C.,  et~al., 2008, \mn@doi [\apj] {10.1086/590236}, \href
  {http://adsabs.harvard.edu/abs/2008ApJ...685..361U} {685, 361}

\bibitem[\protect\citeauthoryear{{Vanzella} et~al.,}{{Vanzella}
  et~al.}{2010}]{Vanzella2010}
{Vanzella} E.,  et~al., 2010, \mn@doi [\apj] {10.1088/0004-637X/725/1/1011},
  \href {http://adsabs.harvard.edu/abs/2010ApJ...725.1011V} {725, 1011}

\bibitem[\protect\citeauthoryear{{Vanzella} et~al.,}{{Vanzella}
  et~al.}{2012}]{Vanzella2012}
{Vanzella} E.,  et~al., 2012, \mn@doi [\apj] {10.1088/0004-637X/751/1/70},
  \href {http://adsabs.harvard.edu/abs/2012ApJ...751...70V} {751, 70}

\bibitem[\protect\citeauthoryear{{Vanzella} et~al.,}{{Vanzella}
  et~al.}{2015}]{Vanzella2015}
{Vanzella} E.,  et~al., 2015, \mn@doi [\aap] {10.1051/0004-6361/201525651},
  \href {http://adsabs.harvard.edu/abs/2015A%26A...576A.116V} {576, A116}

\bibitem[\protect\citeauthoryear{{Vanzella} et~al.,}{{Vanzella}
  et~al.}{2016}]{Vanzella2016}
{Vanzella} E.,  et~al., 2016, \mn@doi [\apj] {10.3847/0004-637X/825/1/41},
  \href {http://adsabs.harvard.edu/abs/2016ApJ...825...41V} {825, 41}

\bibitem[\protect\citeauthoryear{{Varela}, {Kann}, {Klose}  \&
  {Greiner}}{{Varela} et~al.}{2014}]{Varela2014}
{Varela} K.,  {Kann} D.~A.,  {Klose} S.,   {Greiner} J.,  2014, GRB Coordinates
  Network, \href {http://adsabs.harvard.edu/abs/2014GCN..16849...1V} {16849}

\bibitem[\protect\citeauthoryear{{Vergani} et~al.,}{{Vergani}
  et~al.}{2017}]{Vergani2017}
{Vergani} S.~D.,  et~al., 2017, \mn@doi [\aap] {10.1051/0004-6361/201629759},
  \href {http://adsabs.harvard.edu/abs/2017A%26A...599A.120V} {599, A120}

\bibitem[\protect\citeauthoryear{{Vreeswijk} et~al.,}{{Vreeswijk}
  et~al.}{2004}]{Vreeswijk2004}
{Vreeswijk} P.~M.,  et~al., 2004, \mn@doi [\aap] {10.1051/0004-6361:20040086},
  \href {http://adsabs.harvard.edu/abs/2004A%26A...419..927V} {419, 927}

\bibitem[\protect\citeauthoryear{{Vreeswijk} et~al.,}{{Vreeswijk}
  et~al.}{2006}]{Vreeswijk2006}
{Vreeswijk} P.~M.,  et~al., 2006, \mn@doi [\aap] {10.1051/0004-6361:20053795},
  \href {http://adsabs.harvard.edu/abs/2006A%26A...447..145V} {447, 145}

\bibitem[\protect\citeauthoryear{{Vreeswijk} et~al.,}{{Vreeswijk}
  et~al.}{2007}]{Vreeswijk2007}
{Vreeswijk} P.~M.,  et~al., 2007, \mn@doi [\aap] {10.1051/0004-6361:20066780},
  \href {http://adsabs.harvard.edu/abs/2007A%26A...468...83V} {468, 83}

\bibitem[\protect\citeauthoryear{{Vreeswijk}, {Fynbo}, {Malesani}, {Hjorth}  \&
  {de Ugarte Postigo}}{{Vreeswijk} et~al.}{2008}]{Vreeswijk2008}
{Vreeswijk} P.~M.,  {Fynbo} J.~P.~U.,  {Malesani} D.,  {Hjorth} J.,   {de
  Ugarte Postigo} A.,  2008, GRB Coordinates Network, \href
  {http://adsabs.harvard.edu/abs/2008GCN..8191....1V} {8191}

\bibitem[\protect\citeauthoryear{{Vreeswijk} et~al.,}{{Vreeswijk}
  et~al.}{2013}]{Vreeswijk2013}
{Vreeswijk} P.~M.,  et~al., 2013, \mn@doi [\aap] {10.1051/0004-6361/201219652},
  \href {http://adsabs.harvard.edu/abs/2013A%26A...549A..22V} {549, A22}

\bibitem[\protect\citeauthoryear{{Walch}, {Whitworth}, {Bisbas}, {W{\"u}nsch}
  \& {Hubber}}{{Walch} et~al.}{2012}]{Walch2012}
{Walch} S.~K.,  {Whitworth} A.~P.,  {Bisbas} T.,  {W{\"u}nsch} R.,   {Hubber}
  D.,  2012, \mn@doi [\mnras] {10.1111/j.1365-2966.2012.21767.x}, \href
  {http://adsabs.harvard.edu/abs/2012MNRAS.427..625W} {427, 625}

\bibitem[\protect\citeauthoryear{{Watson} \& {Jakobsson}}{{Watson} \&
  {Jakobsson}}{2012}]{Watson2012}
{Watson} D.,  {Jakobsson} P.,  2012, \mn@doi [\apj]
  {10.1088/0004-637X/754/2/89}, \href
  {http://adsabs.harvard.edu/abs/2012ApJ...754...89W} {754, 89}

\bibitem[\protect\citeauthoryear{{Watson}, {Hjorth}, {Fynbo}, {Jakobsson},
  {Foley}, {Sollerman}  \& {Wijers}}{{Watson} et~al.}{2007}]{Watson2007}
{Watson} D.,  {Hjorth} J.,  {Fynbo} J.~P.~U.,  {Jakobsson} P.,  {Foley} S.,
  {Sollerman} J.,   {Wijers} R.~A.~M.~J.,  2007, \mn@doi [\apjl]
  {10.1086/518310}, \href {http://adsabs.harvard.edu/abs/2007ApJ...660L.101W}
  {660, L101}

\bibitem[\protect\citeauthoryear{{Watson} et~al.,}{{Watson}
  et~al.}{2013}]{Watson2013}
{Watson} D.,  et~al., 2013, \mn@doi [\apj] {10.1088/0004-637X/768/1/23}, \href
  {http://adsabs.harvard.edu/abs/2013ApJ...768...23W} {768, 23}

\bibitem[\protect\citeauthoryear{{Waxman} \& {Draine}}{{Waxman} \&
  {Draine}}{2000}]{Waxman2000}
{Waxman} E.,  {Draine} B.~T.,  2000, \mn@doi [\apj] {10.1086/309053}, \href
  {http://adsabs.harvard.edu/abs/2000ApJ...537..796W} {537, 796}

\bibitem[\protect\citeauthoryear{{Wiersema}, {Levan}, {Kamble}, {Tanvir}  \&
  {Malesani}}{{Wiersema} et~al.}{2009}]{Wiersema2009}
{Wiersema} K.,  {Levan} A.,  {Kamble} A.,  {Tanvir} N.,   {Malesani} D.,  2009,
  GRB Coordinates Network, \href
  {http://adsabs.harvard.edu/abs/2009GCN..9673....1W} {9673}

\bibitem[\protect\citeauthoryear{{Willingale}, {Starling}, {Beardmore},
  {Tanvir}  \& {O'Brien}}{{Willingale} et~al.}{2013}]{Willingale2013}
{Willingale} R.,  {Starling} R.~L.~C.,  {Beardmore} A.~P.,  {Tanvir} N.~R.,
  {O'Brien} P.~T.,  2013, \mn@doi [\mnras] {10.1093/mnras/stt175}, \href
  {http://adsabs.harvard.edu/abs/2013MNRAS.431..394W} {431, 394}

\bibitem[\protect\citeauthoryear{{Wise}, {Demchenko}, {Halicek}, {Norman},
  {Turk}, {Abel}  \& {Smith}}{{Wise} et~al.}{2014}]{Wise2014}
{Wise} J.~H.,  {Demchenko} V.~G.,  {Halicek} M.~T.,  {Norman} M.~L.,  {Turk}
  M.~J.,  {Abel} T.,   {Smith} B.~D.,  2014, \mn@doi [\mnras]
  {10.1093/mnras/stu979}, \href
  {http://adsabs.harvard.edu/abs/2014MNRAS.442.2560W} {442, 2560}

\bibitem[\protect\citeauthoryear{{Wiseman}, {Perley}, {Schady}, {Prochaska},
  {de Ugarte Postigo}, {Kr{\"u}hler}, {Yates}  \& {Greiner}}{{Wiseman}
  et~al.}{2017}]{Wiseman2017}
{Wiseman} P.,  {Perley} D.~A.,  {Schady} P.,  {Prochaska} J.~X.,  {de Ugarte
  Postigo} A.,  {Kr{\"u}hler} T.,  {Yates} R.~M.,   {Greiner} J.,  2017,
  \mn@doi [\aap] {10.1051/0004-6361/201731065}, \href
  {http://adsabs.harvard.edu/abs/2017A%26A...607A.107W} {607, A107}

\bibitem[\protect\citeauthoryear{{Xu} et~al.,}{{Xu} et~al.}{2013}]{Xu2013}
{Xu} D.,  et~al., 2013, \mn@doi [\apj] {10.1088/0004-637X/776/2/98}, \href
  {http://adsabs.harvard.edu/abs/2013ApJ...776...98X} {776, 98}

\bibitem[\protect\citeauthoryear{{Xu}, {Tanvir}, {Malesani}, {Fynbo},
  {Jakobsson}  \& {Saario}}{{Xu} et~al.}{2015}]{Xu2015}
{Xu} D.,  {Tanvir} N.~R.,  {Malesani} D.,  {Fynbo} J.~P.~U.,  {Jakobsson} P.,
  {Saario} J.,  2015, GRB Coordinates Network, \href
  {http://adsabs.harvard.edu/abs/2015GCN..18696...1X} {18696}

\bibitem[\protect\citeauthoryear{{Xu}, {Wise}, {Norman}, {Ahn}  \&
  {O'Shea}}{{Xu} et~al.}{2016a}]{HXu2016}
{Xu} H.,  {Wise} J.~H.,  {Norman} M.~L.,  {Ahn} K.,   {O'Shea} B.~W.,  2016a,
  \mn@doi [\apj] {10.3847/1538-4357/833/1/84}, \href
  {http://adsabs.harvard.edu/abs/2016ApJ...833...84X} {833, 84}

\bibitem[\protect\citeauthoryear{{Xu}, {Fynbo}, {Malesani}, {de Ugarte
  Postigo}, {Petrushevska}, {Saario}  \& {Telting}}{{Xu}
  et~al.}{2016b}]{Xu2016}
{Xu} D.,  {Fynbo} J.~P.~U.,  {Malesani} D.,  {de Ugarte Postigo} A.,
  {Petrushevska} T.,  {Saario} J.,   {Telting} J.,  2016b, GRB Coordinates
  Network, \href {http://ukads.nottingham.ac.uk/abs/2016GCN..19109...1X}
  {19109}

\bibitem[\protect\citeauthoryear{{Yoon} \& {Langer}}{{Yoon} \&
  {Langer}}{2005}]{Yoon2005}
{Yoon} S.-C.,  {Langer} N.,  2005, \mn@doi [\aap] {10.1051/0004-6361:20054030},
  \href {http://adsabs.harvard.edu/abs/2005A%26A...443..643Y} {443, 643}

\bibitem[\protect\citeauthoryear{{Yoon}, {Langer}  \& {Norman}}{{Yoon}
  et~al.}{2006}]{Yoon2006}
{Yoon} S.-C.,  {Langer} N.,   {Norman} C.,  2006, \mn@doi [\aap]
  {10.1051/0004-6361:20065912}, \href
  {http://adsabs.harvard.edu/abs/2006A%26A...460..199Y} {460, 199}

\bibitem[\protect\citeauthoryear{{Yuan} et~al.,}{{Yuan}
  et~al.}{2016}]{Yuan2016}
{Yuan} W.,  et~al., 2016, \mn@doi [\ssr] {10.1007/s11214-016-0274-z}, \href
  {http://adsabs.harvard.edu/abs/2016SSRv..202..235Y} {202, 235}

\bibitem[\protect\citeauthoryear{{Zackrisson}, {Inoue}  \&
  {Jensen}}{{Zackrisson} et~al.}{2013}]{Zackrisson2013}
{Zackrisson} E.,  {Inoue} A.~K.,   {Jensen} H.,  2013, \mn@doi [\apj]
  {10.1088/0004-637X/777/1/39}, \href
  {http://adsabs.harvard.edu/abs/2013ApJ...777...39Z} {777, 39}

\bibitem[\protect\citeauthoryear{{Zafar}, {Watson}, {Tanvir}, {Fynbo},
  {Starling}  \& {Levan}}{{Zafar} et~al.}{2011}]{Zafar2011}
{Zafar} T.,  {Watson} D.~J.,  {Tanvir} N.~R.,  {Fynbo} J.~P.~U.,  {Starling}
  R.~L.~C.,   {Levan} A.~J.,  2011, \mn@doi [\apj] {10.1088/0004-637X/735/1/2},
  \href {http://adsabs.harvard.edu/abs/2011ApJ...735....2Z} {735, 2}

\bibitem[\protect\citeauthoryear{{de Barros} et~al.,}{{de Barros}
  et~al.}{2016}]{DeBarros2016}
{de Barros} S.,  et~al., 2016, \mn@doi [\aap] {10.1051/0004-6361/201527046},
  \href {http://adsabs.harvard.edu/abs/2016A%26A...585A..51D} {585, A51}

\bibitem[\protect\citeauthoryear{{de Ugarte Postigo} \& {Tomasella}}{{de Ugarte
  Postigo} \& {Tomasella}}{2015}]{deUP2015}
{de Ugarte Postigo} A.,  {Tomasella} L.,  2015, GRB Coordinates Network, \href
  {http://ukads.nottingham.ac.uk/abs/2015GCN..17710...1D} {17710}

\bibitem[\protect\citeauthoryear{{de Ugarte Postigo} et~al.,}{{de Ugarte
  Postigo} et~al.}{2012}]{deUP2012}
{de Ugarte Postigo} A.,  et~al., 2012, \mn@doi [\aap]
  {10.1051/0004-6361/201219894}, \href
  {http://adsabs.harvard.edu/abs/2012A%26A...548A..11D} {548, A11}

\bibitem[\protect\citeauthoryear{{de Ugarte Postigo}, {Thoene}, {Gorosabel},
  {Sanchez-Ramirez}, {Fynbo}, {Tanvir}, {Cabrera-Lavers}  \& {Garcia}}{{de
  Ugarte Postigo} et~al.}{2013}]{deUP2013}
{de Ugarte Postigo} A.,  {Thoene} C.~C.,  {Gorosabel} J.,  {Sanchez-Ramirez}
  R.,  {Fynbo} J.~P.~U.,  {Tanvir} N.~R.,  {Cabrera-Lavers} A.,   {Garcia} A.,
  2013, GRB Coordinates Network, \href
  {http://adsabs.harvard.edu/abs/2013GCN..15470...1D} {15470}

\bibitem[\protect\citeauthoryear{{de Ugarte Postigo}, {Blazek}, {Janout},
  {Sprimont}, {Th{\"o}ne}, {Gorosabel}  \& {S{\'a}nchez-Ram{\'{\i}}rez}}{{de
  Ugarte Postigo} et~al.}{2014a}]{adUP2014}
{de Ugarte Postigo} A.,  {Blazek} M.,  {Janout} P.,  {Sprimont} P.,
  {Th{\"o}ne} C.~C.,  {Gorosabel} J.,   {S{\'a}nchez-Ram{\'{\i}}rez} R.,
  2014a, in Software and Cyberinfrastructure for Astronomy III. p. 91520B,
  \mn@doi{10.1117/12.2055774}

\bibitem[\protect\citeauthoryear{{de Ugarte Postigo}, {Th\"one}, {Tanvir},
  {Gorosabel}, {Fynbo}, {Lombardi}, {Reverte-Paya}  \& {Perez}}{{de Ugarte
  Postigo} et~al.}{2014b}]{deUgarte2014}
{de Ugarte Postigo} A.,  {Th\"one} C.~C.,  {Tanvir} N.~R.,  {Gorosabel} J.,
  {Fynbo} J.,  {Lombardi} G.,  {Reverte-Paya} D.,   {Perez} D.,  2014b, GRB
  Coordinates Network, \href
  {http://adsabs.harvard.edu/abs/2014GCN..16968...1D} {16968}

\bibitem[\protect\citeauthoryear{{de Ugarte Postigo}, {Tanvir}, {Cano}, {Izzo},
  {Fynbo}, {Sanchez-Ramirez}, {Thoene}  \& {Pesev}}{{de Ugarte Postigo}
  et~al.}{2016}]{deUP2016}
{de Ugarte Postigo} A.,  {Tanvir} N.~R.,  {Cano} Z.,  {Izzo} L.,  {Fynbo}
  J.~P.~U.,  {Sanchez-Ramirez} R.,  {Thoene} C.~C.,   {Pesev} P.,  2016, GRB
  Coordinates Network, \href
  {http://adsabs.harvard.edu/abs/2016GCN..19245...1D} {19245}

\bibitem[\protect\citeauthoryear{{de Ugarte Postigo}, {Kann}, {Thoene}, {Izzo},
  {Tanvir}, {Lombardi}  \& {Marante}}{{de Ugarte Postigo}
  et~al.}{2017a}]{deUP2017b}
{de Ugarte Postigo} A.,  {Kann} D.~A.,  {Thoene} C.,  {Izzo} L.,  {Tanvir}
  N.~R.,  {Lombardi} G.,   {Marante} A.,  2017a, GRB Coordinates Network, \href
  {http://adsabs.harvard.edu/abs/2017GCN..20990...1D} {20990}

\bibitem[\protect\citeauthoryear{{de Ugarte Postigo}, {Izzo}, {Kann}, {Thoene},
  {Cano}, {Fynbo}  \& {Garcia Alvarez}}{{de Ugarte Postigo}
  et~al.}{2017b}]{deUP2017c}
{de Ugarte Postigo} A.,  {Izzo} L.,  {Kann} D.~A.,  {Thoene} C.,  {Cano} Z.,
  {Fynbo} J.~P.~U.,   {Garcia Alvarez} D.,  2017b, GRB Coordinates Network,
  \href {http://adsabs.harvard.edu/abs/2017GCN..21177...1D} {21177}

\bibitem[\protect\citeauthoryear{{de Ugarte Postigo}, {Cano}, {Izzo}, {Thoene},
  {Kann}, {Castro-Rodriguez}  \& {Valladares}}{{de Ugarte Postigo}
  et~al.}{2018}]{deUP2018}
{de Ugarte Postigo} A.,  {Cano} Z.,  {Izzo} L.,  {Thoene} C.~C.,  {Kann} D.~A.,
   {Castro-Rodriguez} N.,   {Valladares} D.~P.,  2018, GRB Coordinates Network,
  Circular Service, No.~22346, \#1, \href
  {http://adsabs.harvard.edu/abs/2018GCN.22346....1D} {22346}

\bibitem[\protect\citeauthoryear{{van Marle}, {Langer}, {Achterberg}  \&
  {Garc{\'{\i}}a-Segura}}{{van Marle} et~al.}{2006}]{vanMarle2006}
{van Marle} A.~J.,  {Langer} N.,  {Achterberg} A.,   {Garc{\'{\i}}a-Segura} G.,
   2006, \mn@doi [\aap] {10.1051/0004-6361:20065709}, \href
  {http://adsabs.harvard.edu/abs/2006A%26A...460..105V} {460, 105}

\bibitem[\protect\citeauthoryear{{van Marle}, {Langer}, {Yoon}  \&
  {Garc{\'{\i}}a-Segura}}{{van Marle} et~al.}{2008}]{vanMarle2008}
{van Marle} A.~J.,  {Langer} N.,  {Yoon} S.-C.,   {Garc{\'{\i}}a-Segura} G.,
  2008, \mn@doi [\aap] {10.1051/0004-6361:20078802}, \href
  {http://adsabs.harvard.edu/abs/2008A%26A...478..769V} {478, 769}

\bibitem[\protect\citeauthoryear{{van den Heuvel} \& {Portegies Zwart}}{{van
  den Heuvel} \& {Portegies Zwart}}{2013}]{vandenHeuvel2013}
{van den Heuvel} E.~P.~J.,  {Portegies Zwart} S.~F.,  2013, \mn@doi [\apj]
  {10.1088/0004-637X/779/2/114}, \href
  {http://adsabs.harvard.edu/abs/2013ApJ...779..114V} {779, 114}

\makeatother
\end{thebibliography}



\appendix

\section{Individual bursts}
\label{appendix}

In this appendix we provide information about selected GRBs.
In particular we present our \lya\ fits for
those cases where the column-densities have not previously been reported and are not included in \citet{Selsing2018}.
Unless stated otherwise, the precise redshift is taken from metal absorption lines
seen in the spectra, which reduces the free parameters, and improves accuracy of the fits.
Details of the fitting procedure are given in \citet{Selsing2018}.

We caution that the nature of target-of-opportunity observations of variable sources means
that the source magnitude is frequently poorly known prior to observation, in many cases
there was limited time available, and non-optimal conditions or sky location.
Thus, some of the spectra are 
unusually low signal-to-noise and/or suffer from imperfect flux calibration or other anomalies.
Furthermore, there are cases where host galaxy emission, including \lya\ emission lines,
contaminates the afterglow signal, and where intervening absorbers introduce metal lines.
However, fortunately for the purposes of this analysis, the red-wing of the  \lya\ absorption features can still be measured
to an adequate level of precision, with little overall systematic bias in determination of \NH.

Data are taken from various sources: some from observations we obtained ourselves and
in other cases from archives. Spectrographs used include the 
Gemini Multi-Object Spectrograph North and South (GMOS-N, GMOS-S),
the VLT FOcal Reducer and low dispersion Spectrograph (FORS1 and FORS2)
and UltraViolet Echelle Spectrograph (UVES), 
the William Herschel Telescope (WHT)
Intermediate dispersion Spectrograph and Imaging System (ISIS) and Auxiliary port CAMera (ACAM), 
the Gran Telescopio Canarias (GTC)
Optical System for Imaging and low-Intermediate-Resolution Integrated Spectroscopy (OSIRIS),
the Nordic Optical Telescope (NOT) Andalucia Faint Object Spectrograph and Camera (ALFOSC),
the Telescopio Nazionale Galileo (TNG)
Device Optimized for the Low Resolution (DOLoRes),
the Asiago Copernico Telescope (CT)
Asiago Faint Object Spectrograph and Camera (AFOSC),
the Keck Low Resolution Imaging Spectrograph (LRIS).

Spectra presented in this appendix will be made available in the GRBspec database \url{http://grbspec.iaa.es}
\citep{adUP2014}.

\subsection{GRB\,021004}
\label{app:021004}
The bright afterglow of GRB\,021004 was well studied, and spectroscopy revealed an unusually complex velocity structure
for the absorbing gas \citep{Fiore2005,Starling2005,CastroTirado2010}. 
Combined with saturation of the lines, establishing the \HI\ column-density was 
not straight-forward, and here we adopt a reasonable value and error based on the $\lognhc>18$ from the  
the absence of continuum below the Lyman limit \citep{Fynbo2005}
and $\lognhc<20$ from analysis of the multiple \lya\ lines themselves \citep{Moller2002}.

\subsection{GRB\,060522}
\label{app:060522}
GRB\,060522  was observed with Keck-I/LRIS, starting 14:24 UT on 22-May-2006.
A total exposure of $1800$\,s was obtained.
The fit to the \lya\ line is shown in Figure~\ref{fig:060522}; the inferred column $\lognhc=20.6\pm0.3$
is consistent with previous estimates  \citep{Cenko2006,Chary2007}. Note, the region redward of \lya\
is badly affected by fringing, impeding searches for metal absorption lines.
\begin{figure}
\includegraphics[angle=0,width=\columnwidth]{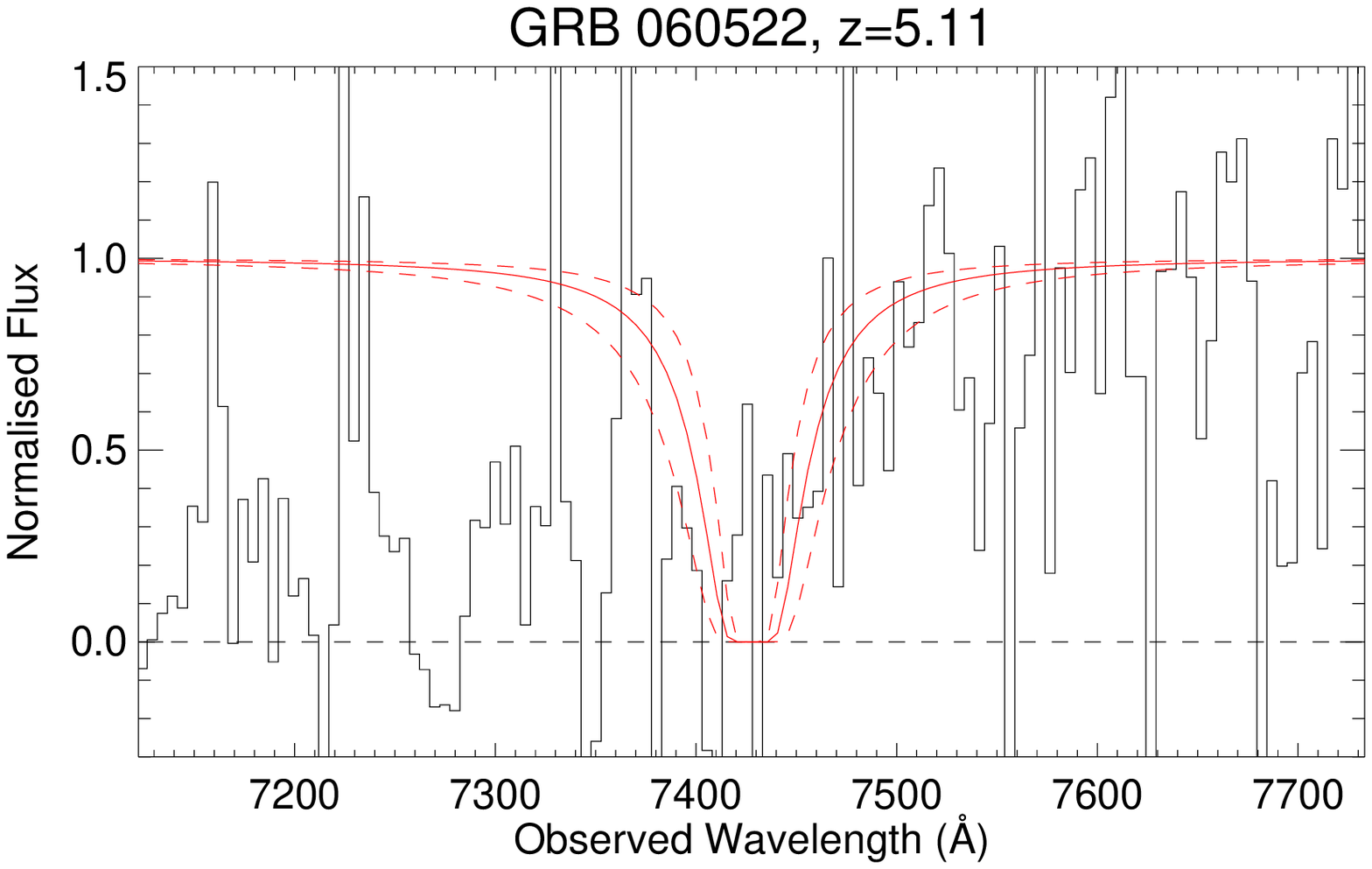}
    \caption{Fit of the red wing of \lya\ for the GRB\,060522 Keck-I/LRIS spectrum.}
    \label{fig:060522}
\end{figure}

\subsection{GRB\,070223}
\label{app:070223}
Near-IR imaging of GRB\,070223 was obtained with the WHT Long-slit Intermediate Resolution Infrared Spectrograph
(LIRIS) instrument in the $JHK$ filters between 2.7 and 3.7\,hr post-burst. 
A faint source was detected at the location of the X-ray afterglow \citep{Rol2007}, for which we find a
magnitude $K_{\rm AB}=21.9\pm0.3$ calibrated against 2MASS stars in the field
\citep[and corrected for small foreground extinction via][]{Schlafly2011}.
Subsequent imaging obtained 8\,day post-burst showed this source to have declined to $K_{\rm AB}=22.8\pm0.2$,
confirming the identification of the afterglow, but also indicating the presence of an underlying host galaxy
\citep[also detected in 3.6\,$\mu$m {\em Spitzer} imaging by ][]{Perley2016b}.

GRB\,070223 was observed rapidly with the 2\,m Liverpool Telescope (LT) in various optical filters, beginning only 18\,min post-burst \citep{Melandri2008}.
We  created a 30\,min stacked integration from $r$-band imaging taken between 3 and 3.7\,hr post-burst.
At the location of the $K$-band transient, there is a faint detection of a source with foreground corrected magnitude
$r=23.8\pm0.3$, calibrated against SDSS stars in the field.
This object was also seen in $R$-band early imaging by the MDM 1.3\,m telescope \citep{Rol2007}, but,
as detailed in the text, appears to be a constant source, presumably the host galaxy.

\subsection{GRB\,070810A}
GRB\,070810A  was observed with Keck-I/LRIS, starting 05:47 UT on 10-Aug-2007.
A total exposure of $2\times600$\,s was obtained \citep{Thoene2007}.
The fit to the \lya\ line is shown in Figure~\ref{fig:070810}. 
\begin{figure}
\includegraphics[angle=0,width=\columnwidth]{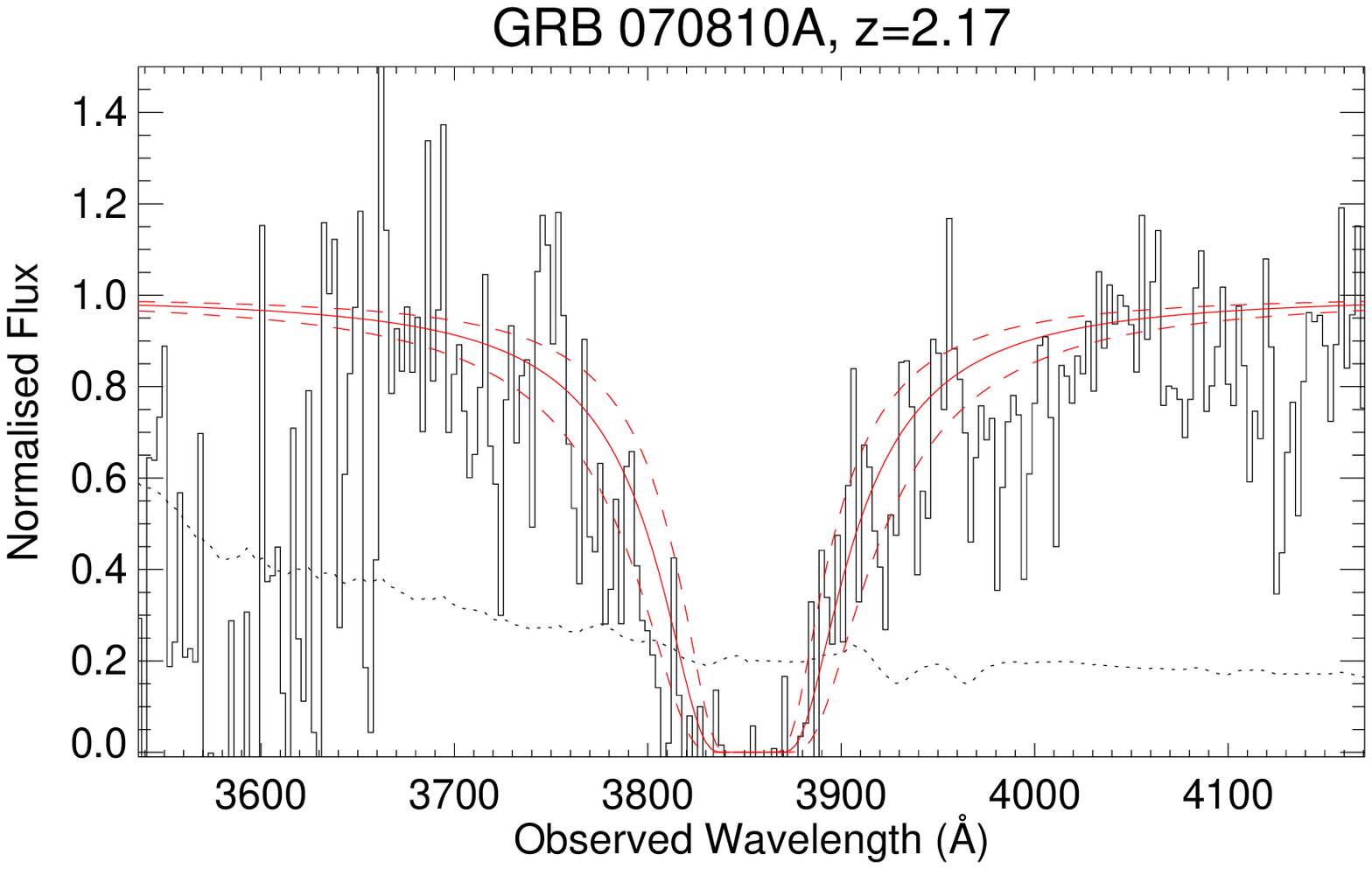}
    \caption{Fit of the red wing of \lya\ for the GRB\,070810A Keck-I/LRIS spectrum.}
    \label{fig:070810}
\end{figure}

\subsection{GRB\,080129}
GRB\,080129 was observed with VLT/FORS1, starting 05:24 UT on 30-Jan-2008.
A total exposure of $4\times1800$\,s was obtained with  the OG590 blocking filter and
300I grism, and
reduced with the standard ESO pipeline \citep{Greiner2009}.
The redshift is fixed to that of the metal absorption lines,
and the fit to the red wing of \lya\ (Figure~\ref{fig:080129}) is rather poor in this case, plausibly due to
velocity structure in low metallicity and low column-density gas close to the host.
\begin{figure}
\includegraphics[angle=0,width=\columnwidth]{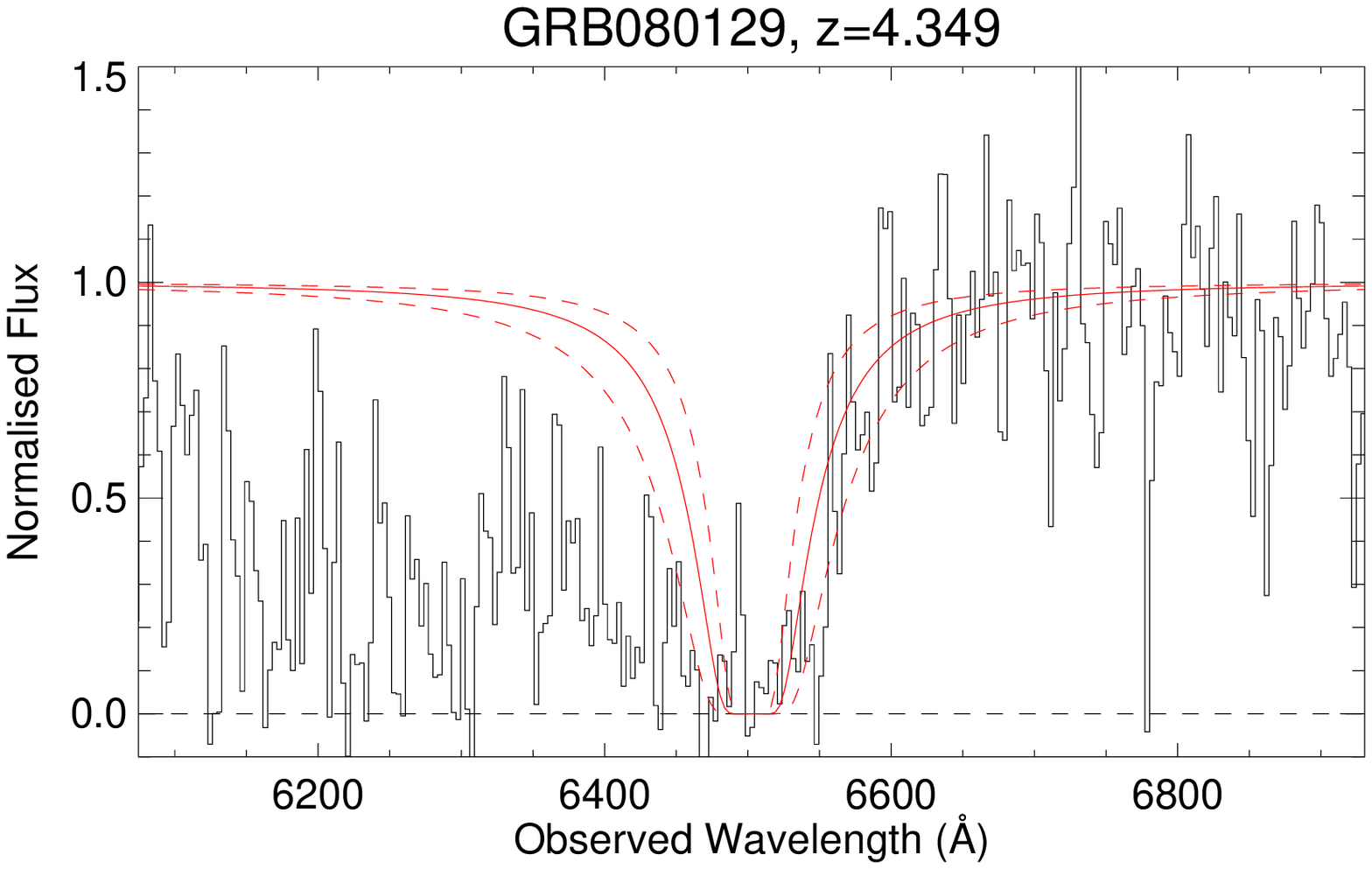}
    \caption{Fit of the red wing of \lya\ for the GRB\,080129 VLT/FORS1 spectrum.}
    \label{fig:080129}
\end{figure}

\subsection{GRB\,080810}
\label{app:080810}
GRB\,080810 was observed at high resolution by Keck/HIRES \citep{Prochaska2008}, 
and the spectrum showed a somewhat complex  \HI\ 
absorption system, with two main components separated by $\approx700$\,km\,s$^{-1}$.
The lower redshift system showed \ion{Si}{II*} fine-structure lines,  suggesting
this gas was closer to the GRB location, and likely that the other system is infalling on the near-side.
\citet{Page2009} found an upper limit for the combined absorber of $\lognhc<19.5$
and a lower limit for the higher redshift component of $\lognhc>17.7$  from the absence of emission below the Lyman limit.
Recently, \citet{Wiseman2017} reanalysed the HIRES spectrum, finding $\lognhc=18.10\pm0.25$ for this dominant component, which
we use here.

\subsection{GRB\,080905B}
GRB\,080905B was observed by VLT/FORS2, starting 01:16 UT on 6-Sep-2008.
A total exposure of $2\times600$\,s was obtained with  the GRIS\_300V grism, and
reduced with the standard ESO pipeline \citep{Vreeswijk2008}.
The fit to  the \lya\ line is shown in Figure~\ref{fig:080905B}.
\begin{figure}
\includegraphics[angle=0,width=\columnwidth]{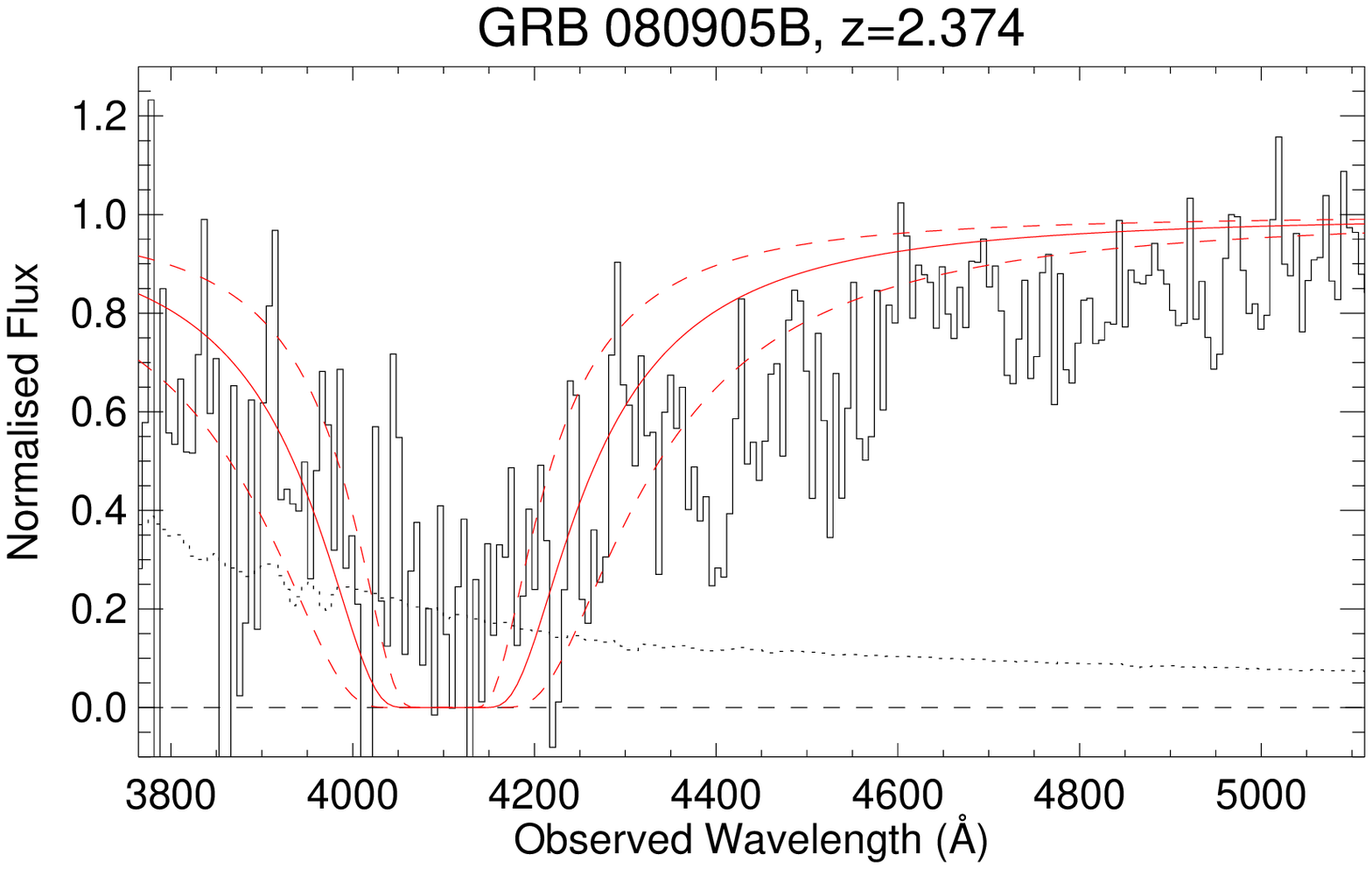}
    \caption{Fit of the red wing of \lya\ for the GRB\,080905B VLT/FORS2 spectrum.}
    \label{fig:080905B}
\end{figure}

\subsection{GRB\,080913}
\label{app:080913}
GRB\,080913 was observed with VLT/FORS2, being found to be at redshift $z\approx6.7$ from the location
of the \lya\ break by \citet{Greiner2009b}.
Lacking a precise metal line redshift, \citet{Greiner2009b} were only able to place weak constraints
on the \HI\ column-density concluding $20.3<\lognhc<21.4$.
A subsequent reanalysis of the spectrum by \citet{Patel2010} located a weak \ion{S}{II}+\ion{Si}{II} blend,
establishing a firmer redshift of $z=6.733$.  This allowed a more precise determination of  the \HI\ column-density,
$\lognhc=19.84$ assuming no neutral component of the IGM, and $\lognhc=19.6$ in a fit that allowed
the neutral fraction of the IGM to be a free parameter (specifically they found $x_{\HI}=0.06$).
Given this uncertainty, and the poor S/N of the spectrum, we adopt $\lognhc=19.6\pm0.3$ here.

\subsection{GRB\,081029}
GRB\,081029 was observed by Gemini/GMOS-S, starting 07:04 UT on 29-Oct-2008 \citep{Cucchiara2008b}.
A total exposure of $2\times900$\,s was obtained
with the R400 grating set at 6000\,\AA\ central
wavelength and reduced using the standard Gemini reduction tools within \iraf.
The fit to the \lya\ line is shown in Figure~\ref{fig:081029}.
\begin{figure}
\includegraphics[angle=0,width=\columnwidth]{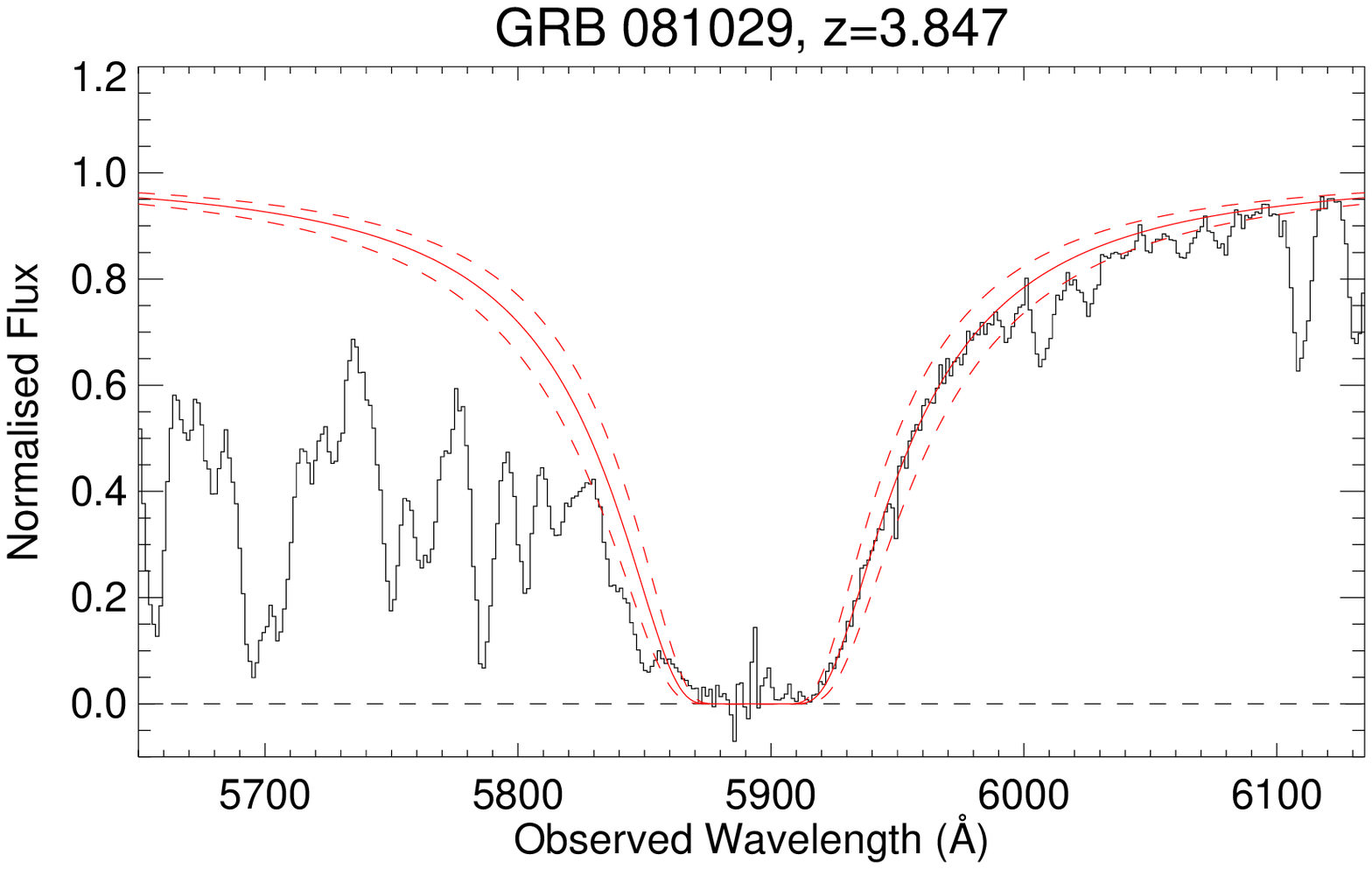}
    \caption{Fit of the red wing of \lya\ for the GRB\,081029 Gemini/GMOS-S spectrum.}
    \label{fig:081029}
\end{figure}

\subsection{GRB\,081118}
GRB\,081118 was observed by VLT/FORS2, starting 02:48 UT on 19-Nov-2008.
A total exposure of $2\times1800$\,s was obtained with  the GRIS\_300V grism, and
reduced with the standard ESO pipeline \citep{DElia2008}.
The fit to  the \lya\ line is shown in Figure~\ref{fig:081118}.
\begin{figure}
\includegraphics[angle=0,width=\columnwidth]{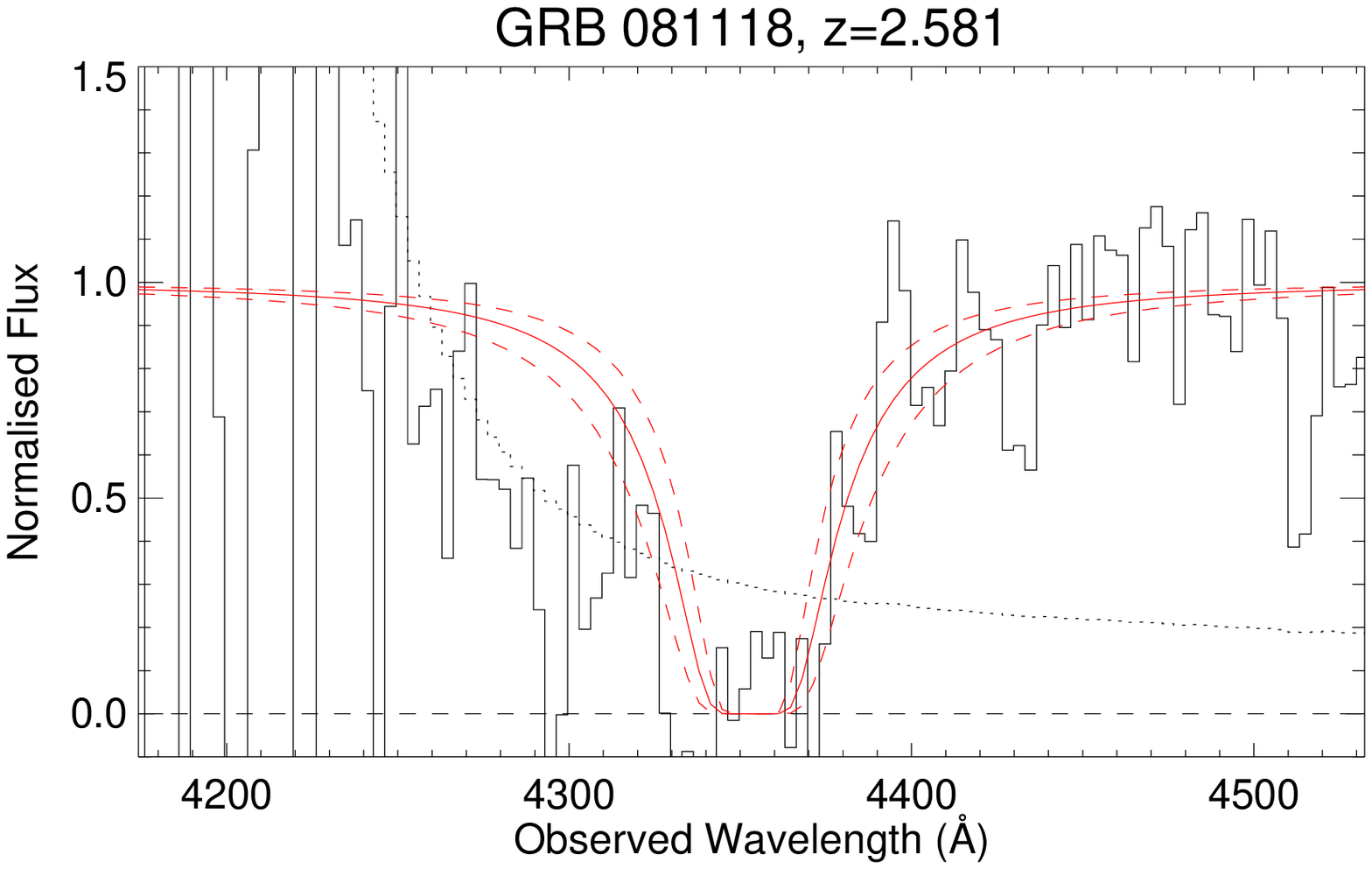}
    \caption{Fit of the red wing of \lya\ for the GRB\,081118 VLT/FORS2 spectrum.}
    \label{fig:081118}
\end{figure}

\subsection{GRB\,081222}
GRB\,081222 was observed by Gemini/GMOS-S, starting 01:02 UT on 23-Dec-2008 \citep{Cucchiara2008}.
A total exposure of $2\times900$\,s was obtained
with the R400 grating set at 6000\,\AA\ central
wavelength, and reduced using the standard Gemini reduction tools within \iraf. 
The fit to  the \lya\ line is shown in Figure~\ref{fig:081222}.
\begin{figure}
\includegraphics[angle=0,width=\columnwidth]{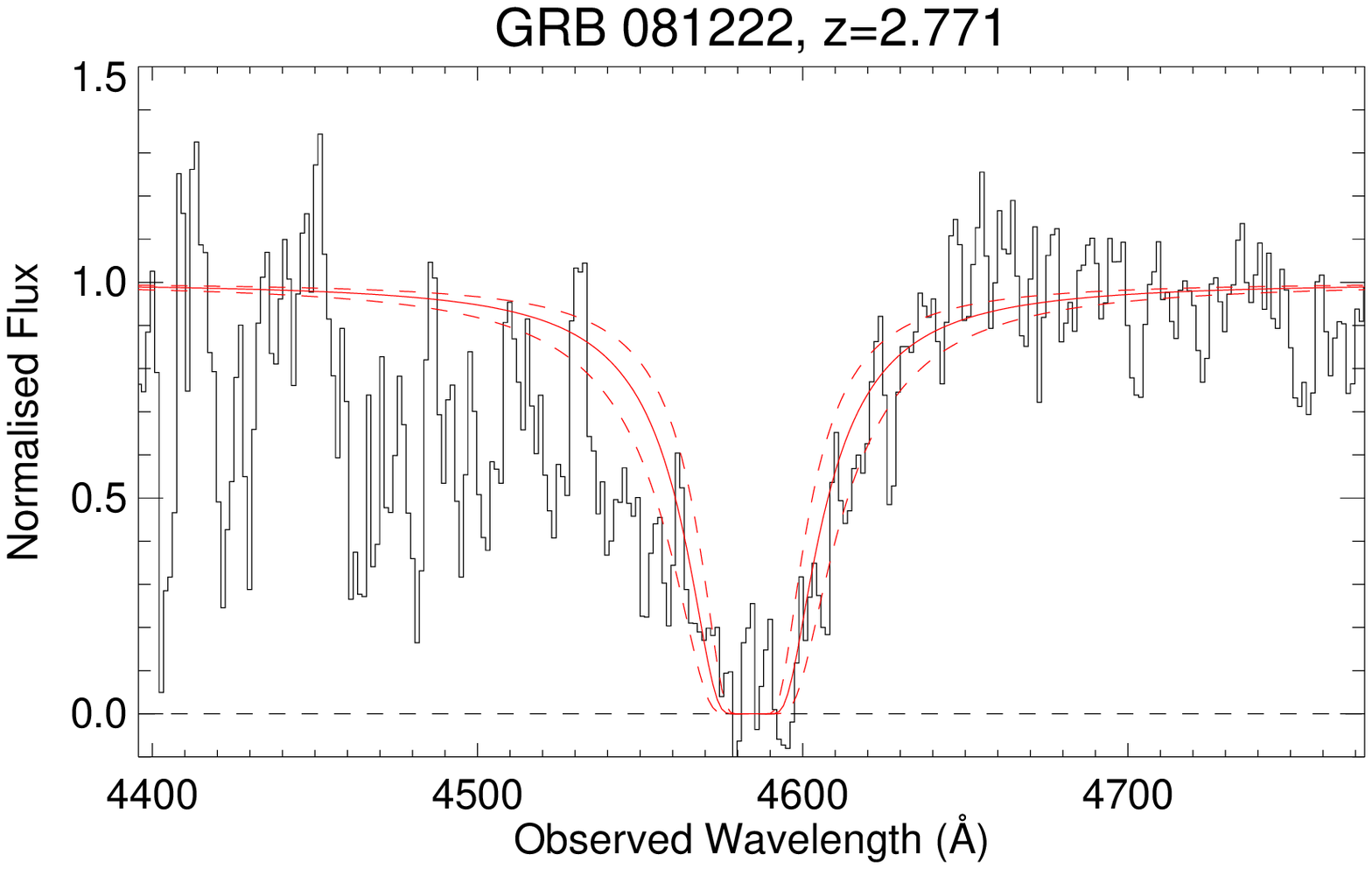}
    \caption{Fit of the red wing of \lya\ for the GRB\,081222 Gemini/GMOS-S spectrum.}
    \label{fig:081222}
\end{figure}

\subsection{GRB\,090313}
GRB\,090313 was observed by Gemini/GMOS-S, starting 04:20 UT on 14-Mar-2009 \citep{Chornock2009}.
A total exposure of $2\times600$\,s was obtained  with the R400 grating set at 6000\,\AA\ central
wavelength, and reduced using the standard Gemini reduction tools within \iraf. 
The fit to  the \lya\ line is shown in Figure~\ref{fig:090313}.
\begin{figure}
\includegraphics[angle=0,width=\columnwidth]{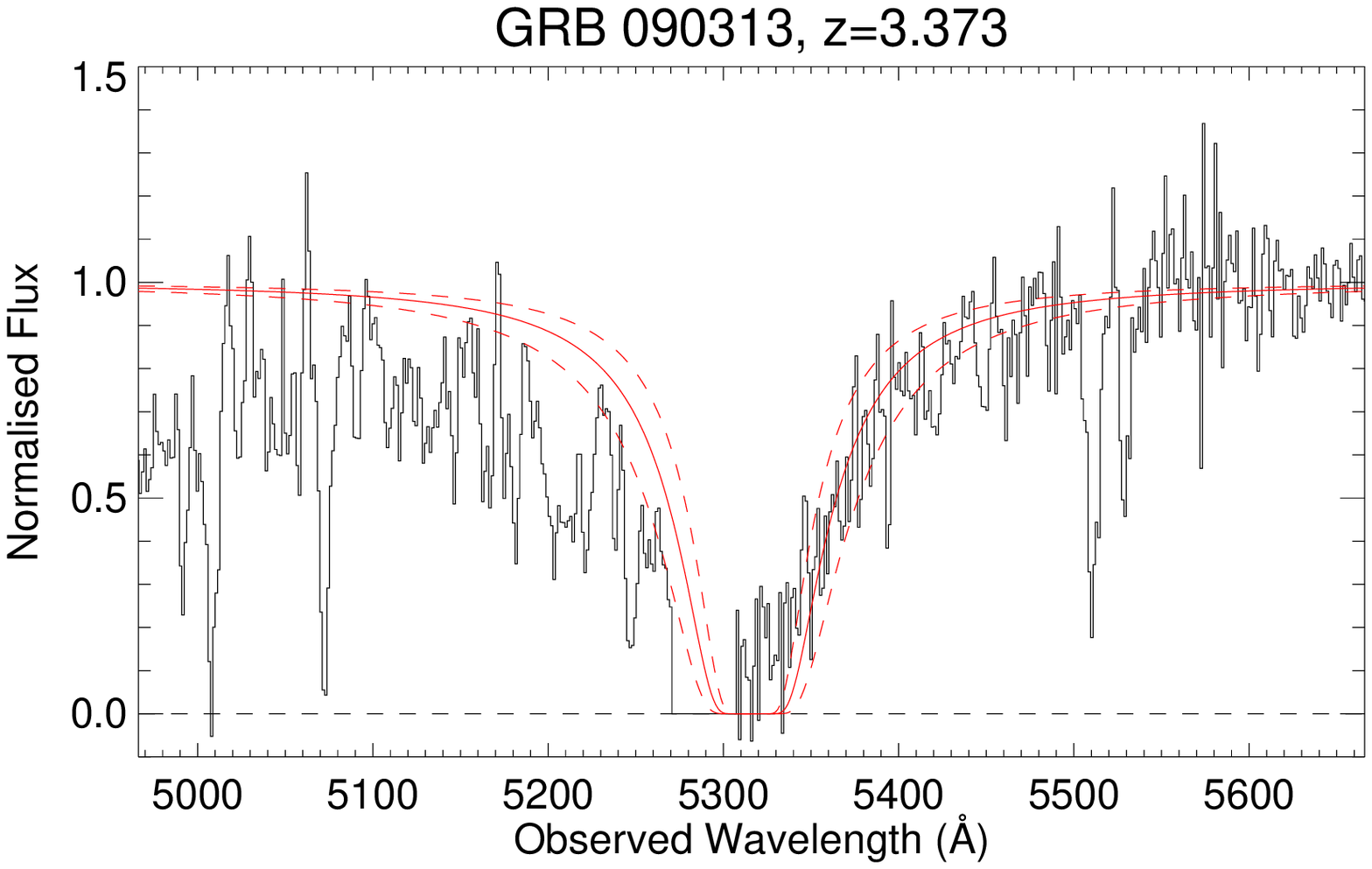}
    \caption{Fit of the red wing of \lya\ for the GRB\,090313 Gemini/GMOS-S spectrum. Note that there is a gap 
    between chips of the camera around 5280\,\AA.  This produces an anomaly in the trace
    which was ignored in the fit.}
    \label{fig:090313}
\end{figure}

\subsection{GRB\,090323}
\label{app:090323}
GRB\,090323 was unusual exhibiting two absorption systems separated by $\approx700$\,km\,s$^{-1}$,
and a relatively high metal abundance \citep{Savaglio2012}. In this case both systems showed
 \ion{Si}{II*} fine structure lines, likely indicating fairly close proximity to the GRB. The value for $\NH$ used
here was obtained by summing the column-densities of the two systems.

\subsection{GRB\,090426}
\label{app:090426}
The prompt duration of GRB\,090426 was $T_{90}\approx1.3$\,s, suggesting it could be a short-duration burst,
particularly given that cosmological time-dilation makes this less than 0.4\,s in the source-frame.
However since it was intrinsically bright and took place in an interacting star-forming system \citep{Thoene2011}, we include
it in our sample as a possible long-duration GRB \citep[see also][]{Nicuesa2011}.  
The \HI\ column-density in this case was seen to vary, and we take here 
 that measured at 1.1\,hr post-burst by \citet{Levesque2010} using Keck/LRIS.
Removing GRB\,090426 from the sample would not have a significant affect on any of the conclusions.

\subsection{GRB\,090519}
GRB\,090519 was observed by VLT/FORS2       starting at
01:03  on 20-May-2009 UT 
\citep{Thoene2009}.
A total exposure of $3\times1800$\,s was obtained, covering the wavelength range 3500--9200\AA.
The fit to  the \lya\ line is shown in Figure~\ref{fig:090519}.
 The afterglow was faint, and the redshift is estimated from the \lya\ and \lyb\
 breaks since no clear metal lines were seen.
\begin{figure}
\includegraphics[angle=0,width=\columnwidth]{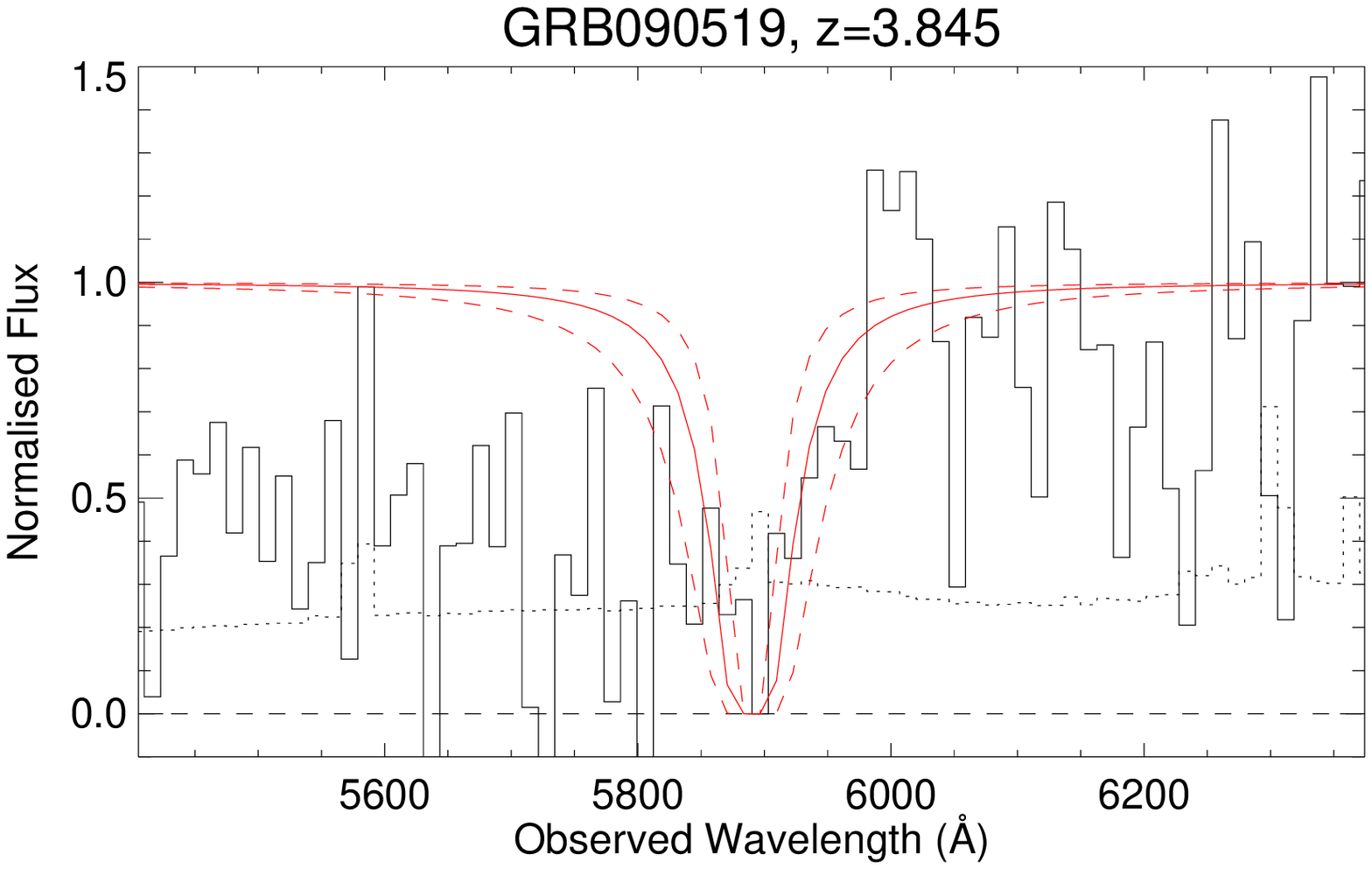}
    \caption{Fit of the red wing of \lya\ for the GRB\,090519 VLT/FORS2  spectrum.}
    \label{fig:090519}
\end{figure}

\subsection{GRB\,090529}
GRB\,090529 was observed by VLT/FORS2       starting at
01:52  on 31-May-2009 UT, roughly 1.5\,days post-burst 
\citep{Malesani2009}.
A total exposure of $2\times1800$\,s was obtained, covering the wavelength range 3500--9200\AA.
The fit to  the \lya\ line is shown in Figure~\ref{fig:090529}.
Although the S/N is unusually poor (largely due to the lateness of the observation), the fit benefits
from the redshift being fixed by metal absorption lines.
\begin{figure}
\includegraphics[angle=0,width=\columnwidth]{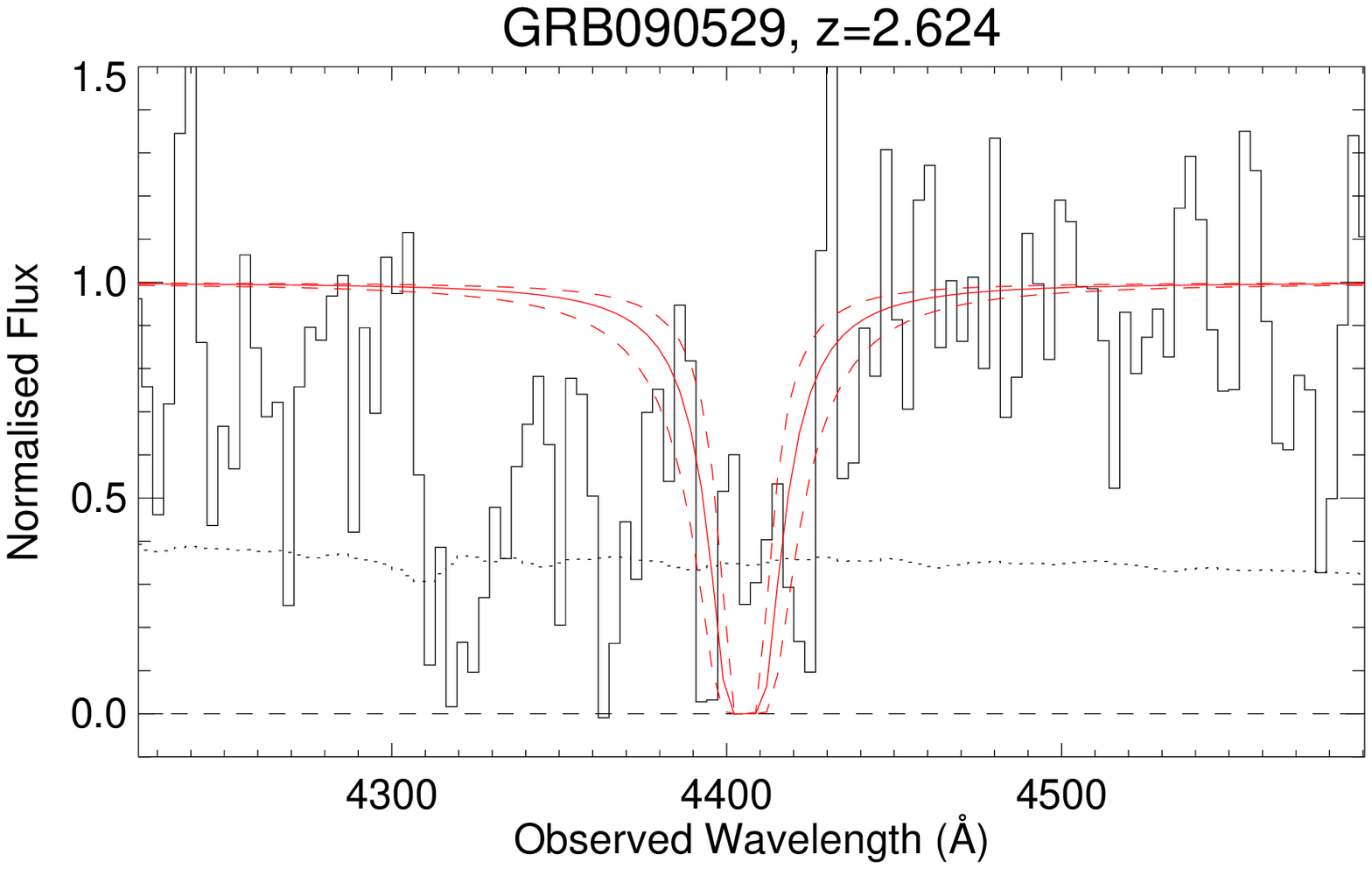}
    \caption{Fit of the red wing of \lya\ for the GRB\,090529 VLT/FORS2  spectrum.}
    \label{fig:090529}
\end{figure}

\subsection{GRB\,090715B}
GRB\,090715B was observed with the WHT/ISIS 
starting 23:46 UT on 15-Jul-2009
\citep{Wiersema2009}.
This spectrograph has a blue and a red arm, separated by a dichroic; we used the 300B and 316R gratings.
Spectroscopic observations consisted of $4\times900$\,s exposure time.
The data were reduced using standard techniques in
{\sc IRAF}. Several metal absorption lines give a redshift $z = 3.01$. 
The fit to  the \lya\ line is shown in Figure~\ref{fig:090715B}.

\begin{figure}
\includegraphics[angle=0,width=\columnwidth]{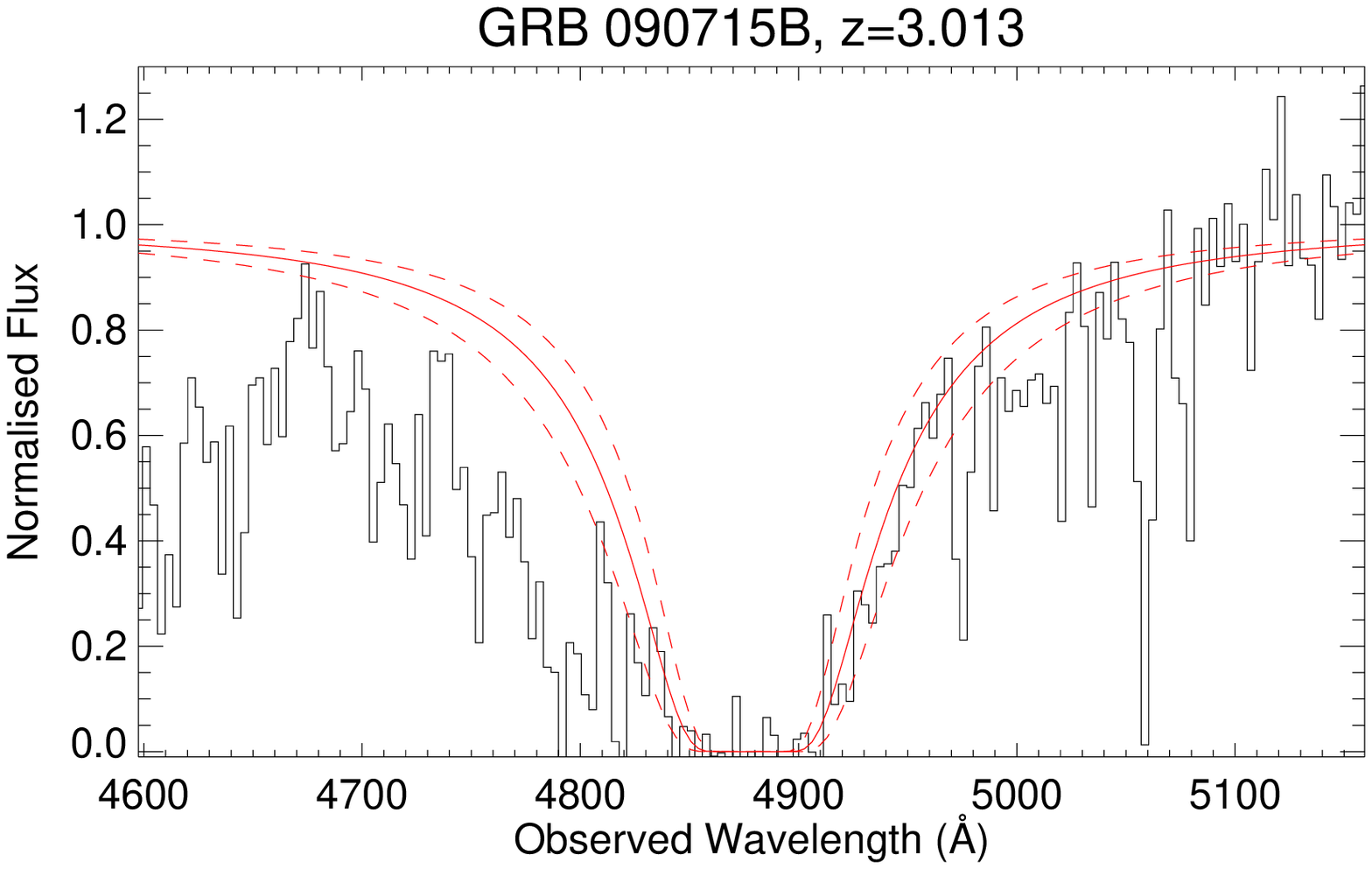}
    \caption{Fit of the red wing of \lya\ for the GRB\,090715B WHT/ISIS spectrum.}
    \label{fig:090715B}
\end{figure}

\subsection{GRB\,090726}
GRB\,090726 was observed by the SAO RAS 6-m telescope using the SCORPIO spectrograph starting at
00:15  on 27-Jul-2009 UT 
\citep{Fatkhullin2009}.
A total exposure of $600$\,s was obtained, covering the wavelength range 3700--7800\AA.
The fit to  the \lya\ line is shown in Figure~\ref{fig:090726}.
\begin{figure}
\includegraphics[angle=0,width=\columnwidth]{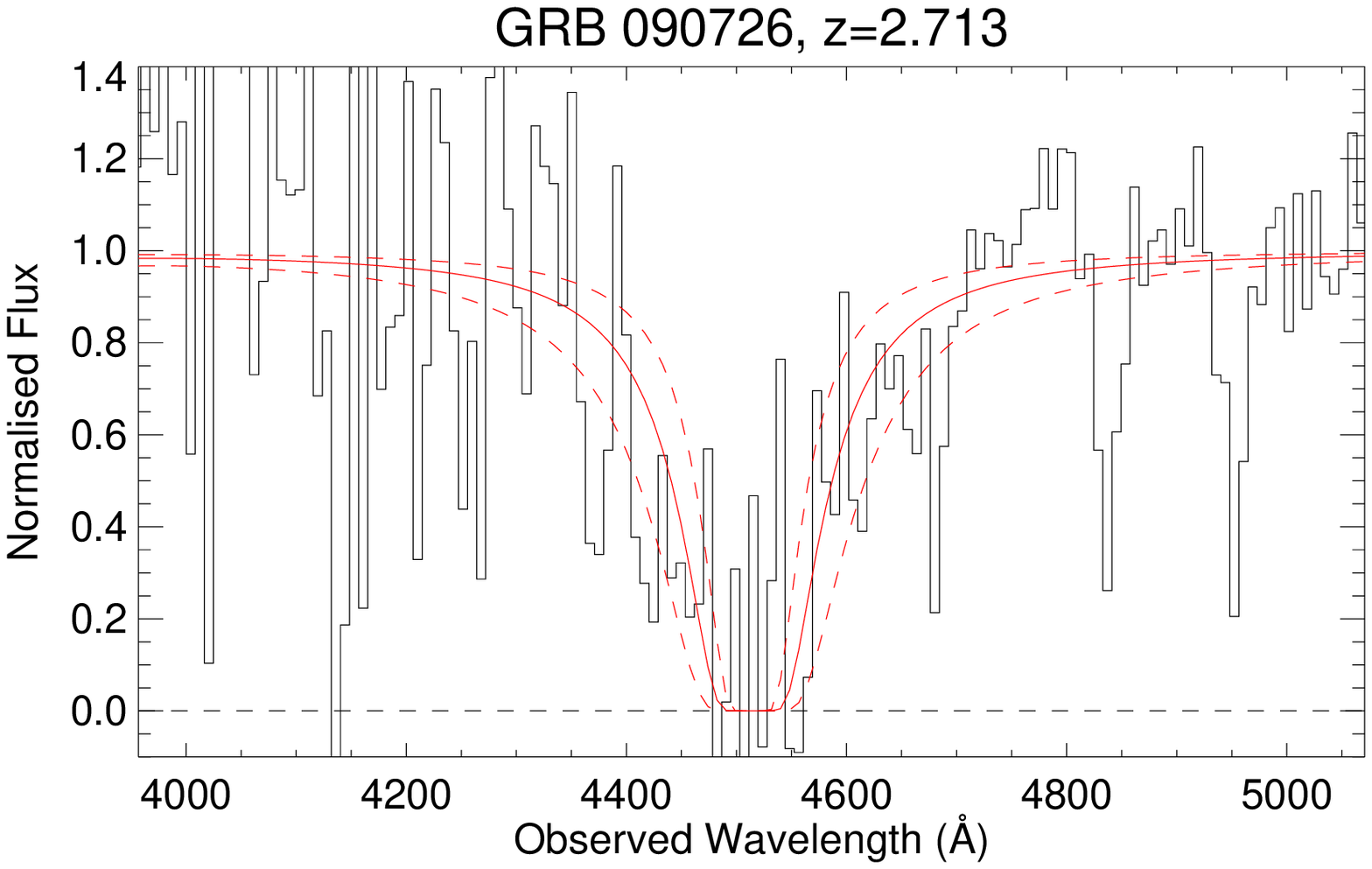}
    \caption{Fit of the red wing of \lya\ for the GRB\,090726  SAO RAS 6-m  / SCORPIO spectrum.}
    \label{fig:090726}
\end{figure}

\subsection{GRB\,091029}
GRB\,091029 was observed by Gemini/GMOS-S, starting 06:05 UT on 29-Oct-2009 \citep{Chornock2009b}.
A total exposure of $4\times600$\,s was obtained
 with the R400 grating set at 6000\,\AA\ central
wavelength and reduced using the standard Gemini reduction tools within \iraf.
The fit to  the \lya\ line is shown in Figure~\ref{fig:091029}.
\begin{figure}
\includegraphics[angle=0,width=\columnwidth]{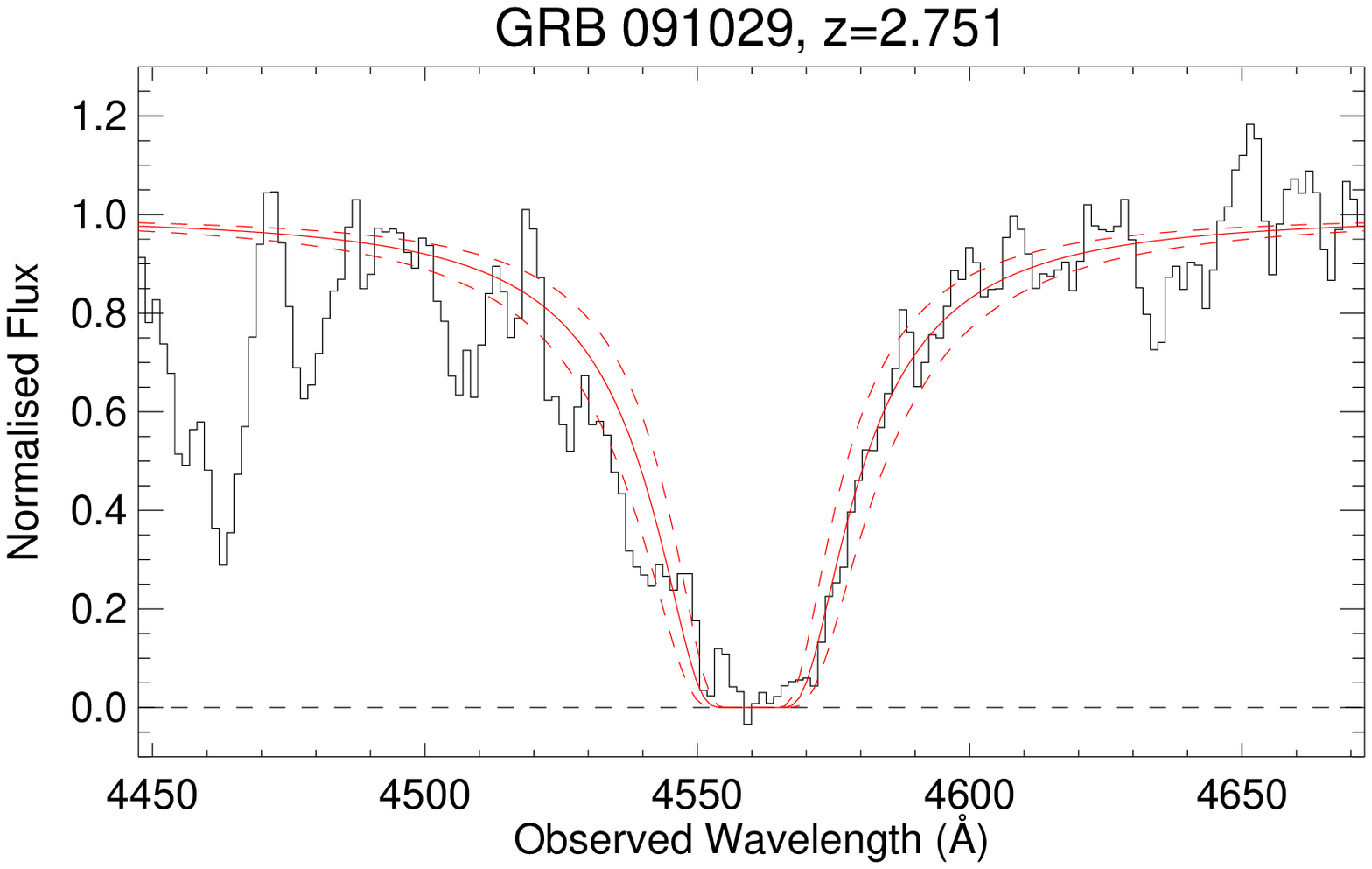}
    \caption{Fit of the red wing of \lya\ for the GRB\,091029 Gemini/GMOS-S spectrum.}
    \label{fig:091029}
\end{figure}

\subsection{GRB\,100302A}
GRB\,100302A was observed by Gemini/GMOS-N, starting 09:42  UT on 03-Mar-2010 \citep{Chornock2010}.
A total exposure of $1200$\,s was obtained with the R400 grating.
The continuum S/N is rather poor, but weak metal lines indicate a redshift of $z=4.813$. The 
fit to  the \lya\ red wing is shown in Figure~\ref{fig:100302A}.
\begin{figure}
\includegraphics[angle=0,width=\columnwidth]{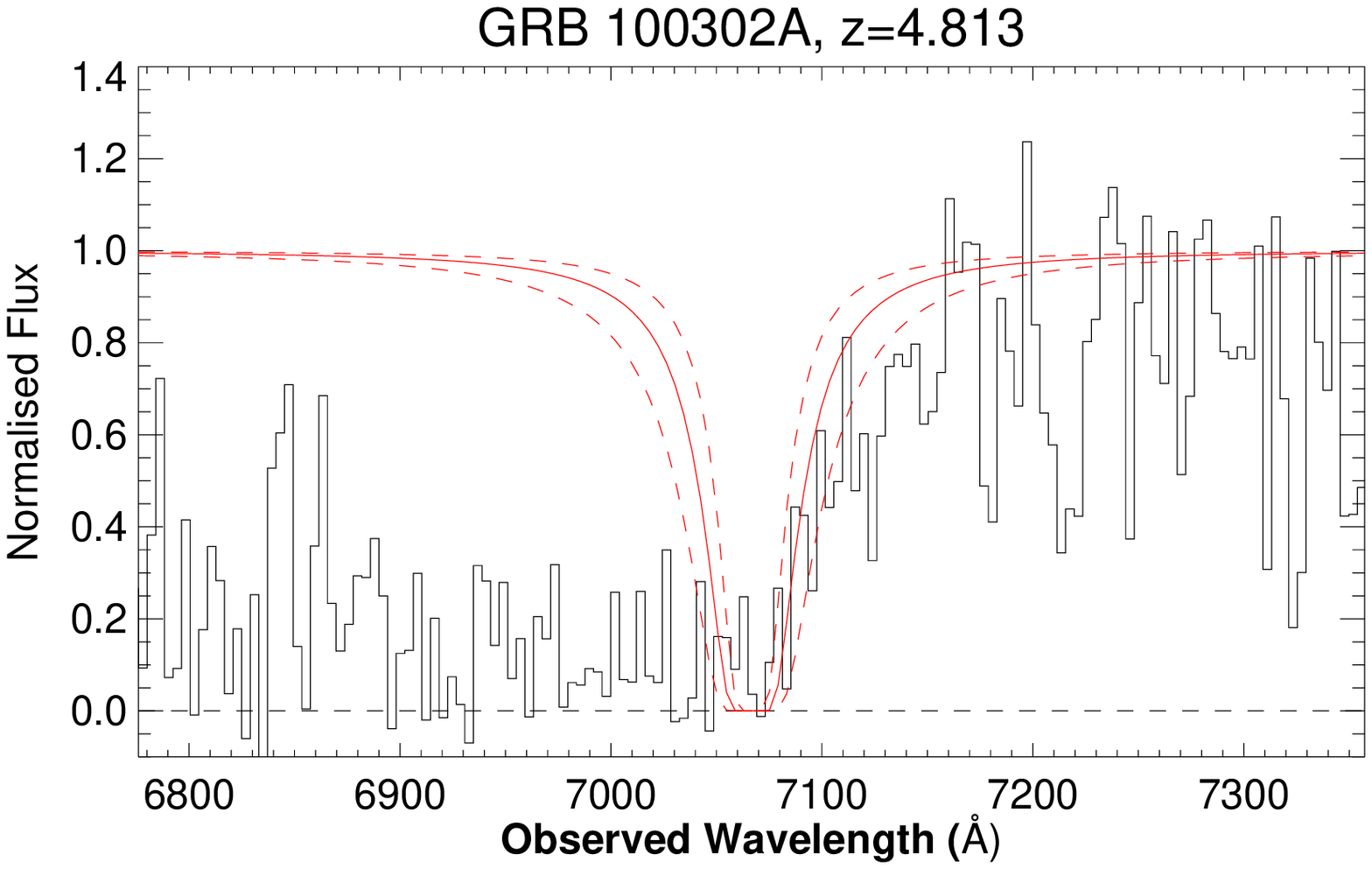}
    \caption{Fit of the red wing of \lya\ for the GRB\,100302A Gemini/GMOS spectrum. }
    \label{fig:100302A}
\end{figure}

\subsection{GRB\,100316A}
GRB\,100316A was observed by GTC/OSIRIS, starting 06:13 UT on 16-Mar-2010 \citep{SanchezRamirez2013b}.
A total exposure of $2\times900$\,s was obtained with the R300B grating.
The fit to  the \lya\ line is shown in Figure~\ref{fig:100316A}.
The precise redshift is known from the \lya\  emission line in a late-time spectrum of the host.
\begin{figure}
\includegraphics[angle=0,width=\columnwidth]{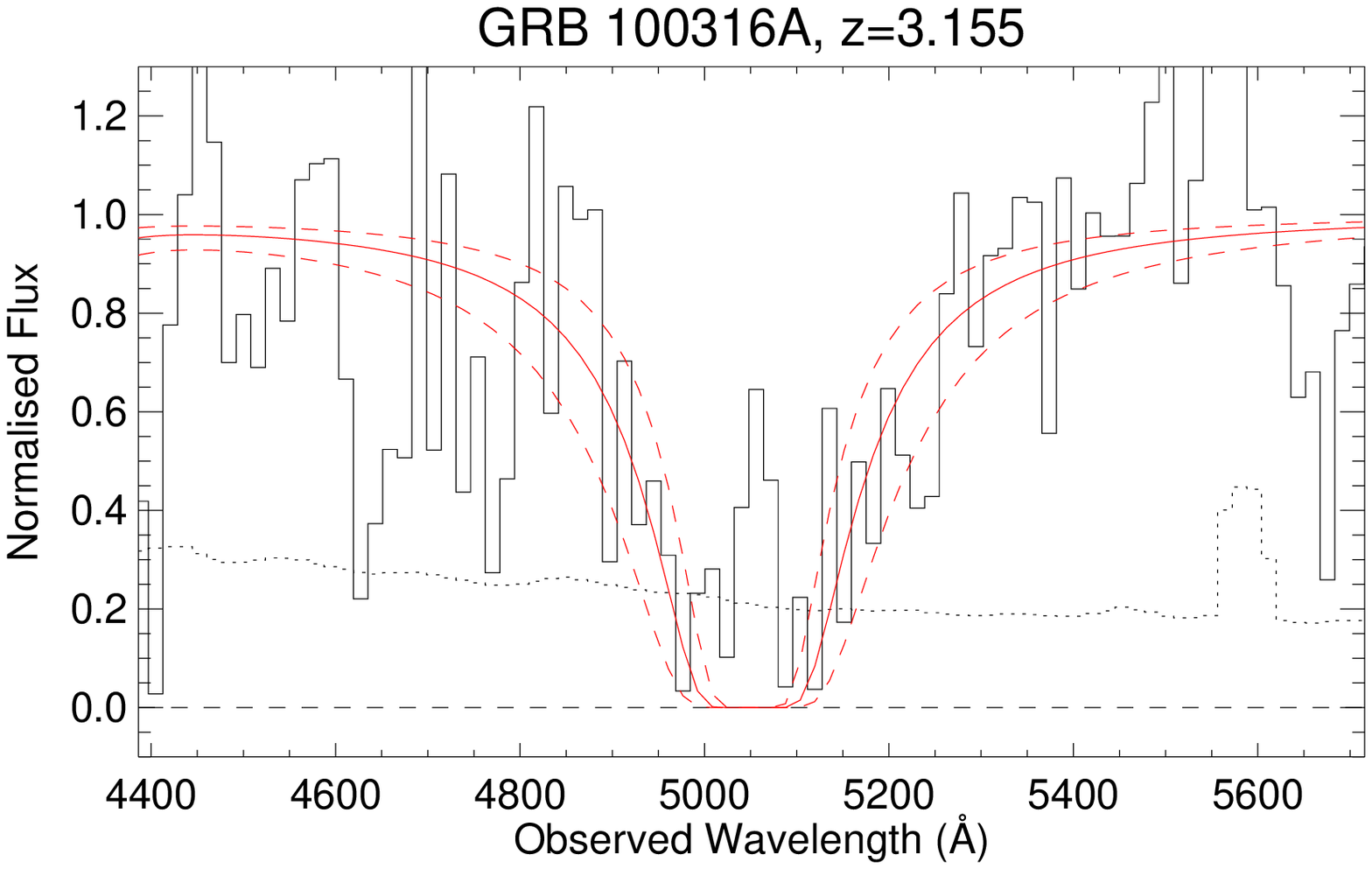}
    \caption{Fit of the red wing of \lya\ for the GRB\,100316A GTC/OSIRIS spectrum.
   \lya\ line emission from the host galaxy is evident in the absorption trough, but does not
    affect the fit.}
    \label{fig:100316A}
\end{figure}

\subsection{GRB\,100513A}
GRB\,100513A was observed by Gemini/GMOS-N, starting 06:13 UT on 13-May-2010 \citep{Cenko2010}.
A total exposure of $2\times1200$\,s was obtained
 with the R400 grating set at 8000\,\AA\ central
wavelength and reduced using the standard Gemini reduction tools within \iraf.
The fit to  the \lya\ line is shown in Figure~\ref{fig:100513A}.
\begin{figure}
\includegraphics[angle=0,width=\columnwidth]{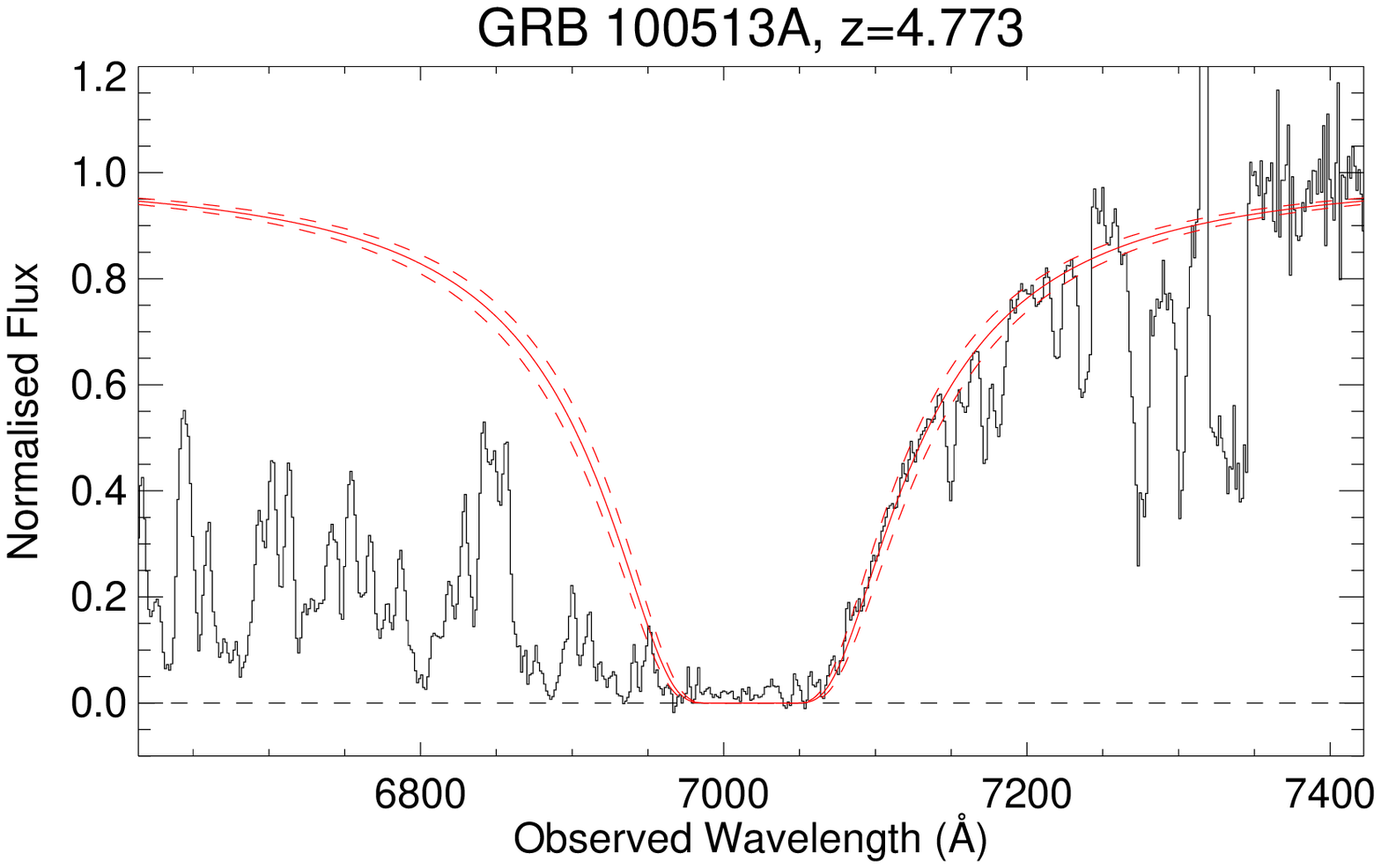}
    \caption{Fit of the red wing of \lya\ for the GRB\,100513A Gemini/GMOS-N spectrum.
 The large dips redward of the \lya\ line are residuals due to gaps between the detectors in the spectrograph.}
    \label{fig:100513A}
\end{figure}

\subsection{GRB\,110731A}
GRB\,110731A was observed by Gemini/GMOS-N, starting 09:08 UT on 01-Aug-2011 \citep{Tanvir2011}.
A total exposure of $4\times900$\,s was obtained
 with the B600 grating set at 5250\,\AA\ central
wavelength and reduced using the standard Gemini reduction tools within \iraf.
The fit to  the \lya\ line is shown in Figure~\ref{fig:110731A}.
\begin{figure}
\includegraphics[angle=0,width=\columnwidth]{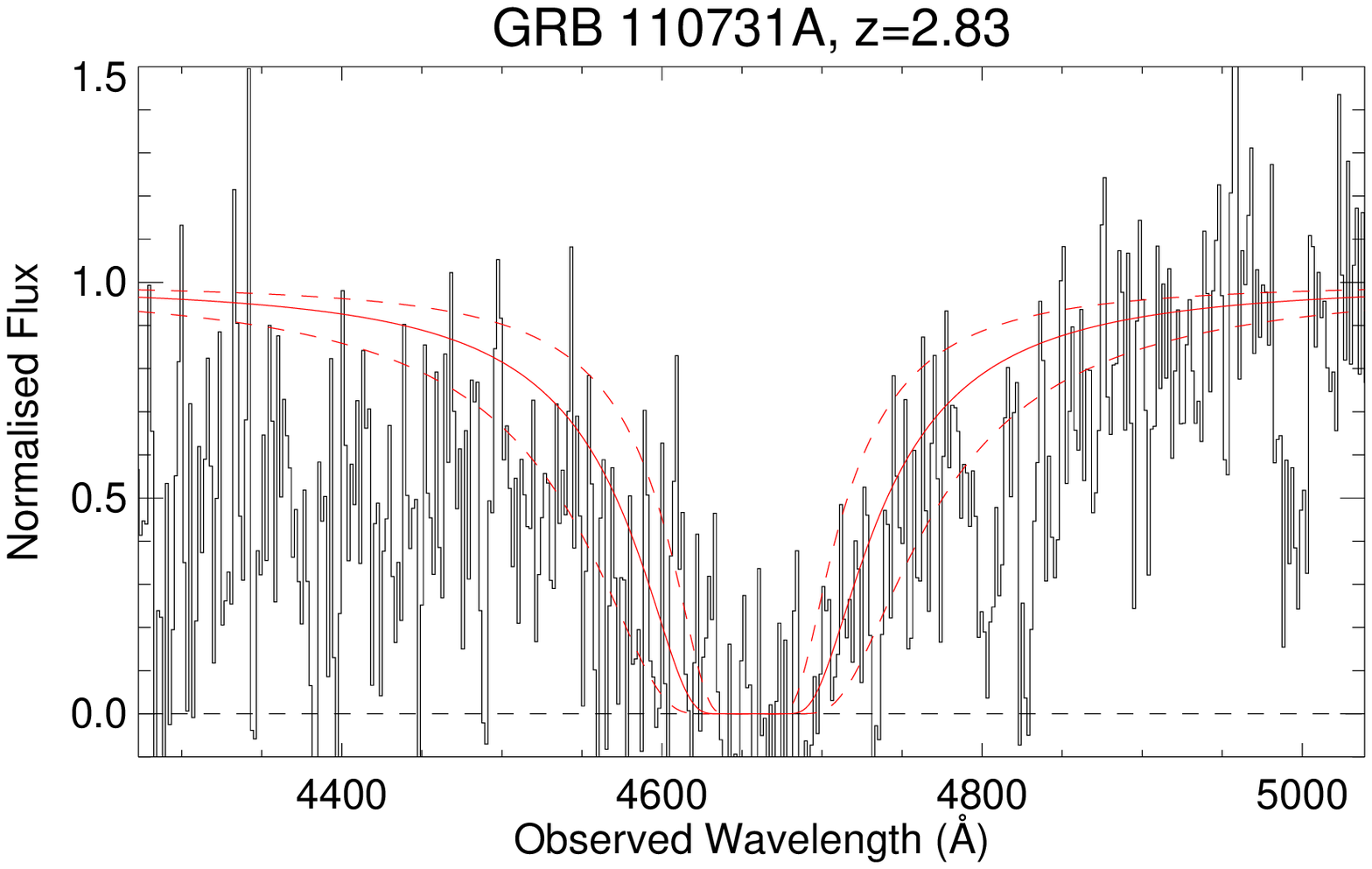}
    \caption{Fit of the red wing of \lya\ for the GRB\,110731A Gemini/GMOS-N spectrum.
The apparent feature at 4800\,\AA\ is due to the gap between the detectors in the spectrograph.}

    \label{fig:110731A}
\end{figure}

\subsection{GRB\,120811C}
GRB\,120811C was observed by the GTC/OSIRIS, starting 15:35 UT on 11-Aug-2012 \citep{Thoene2012}.
A total exposure time of 2400\,s was obtained, spanning a wavelength range 3640--7875\,\AA.
The fit to  the \lya\ line is shown in Figure~\ref{fig:120811C}.
\begin{figure}
\includegraphics[angle=0,width=\columnwidth]{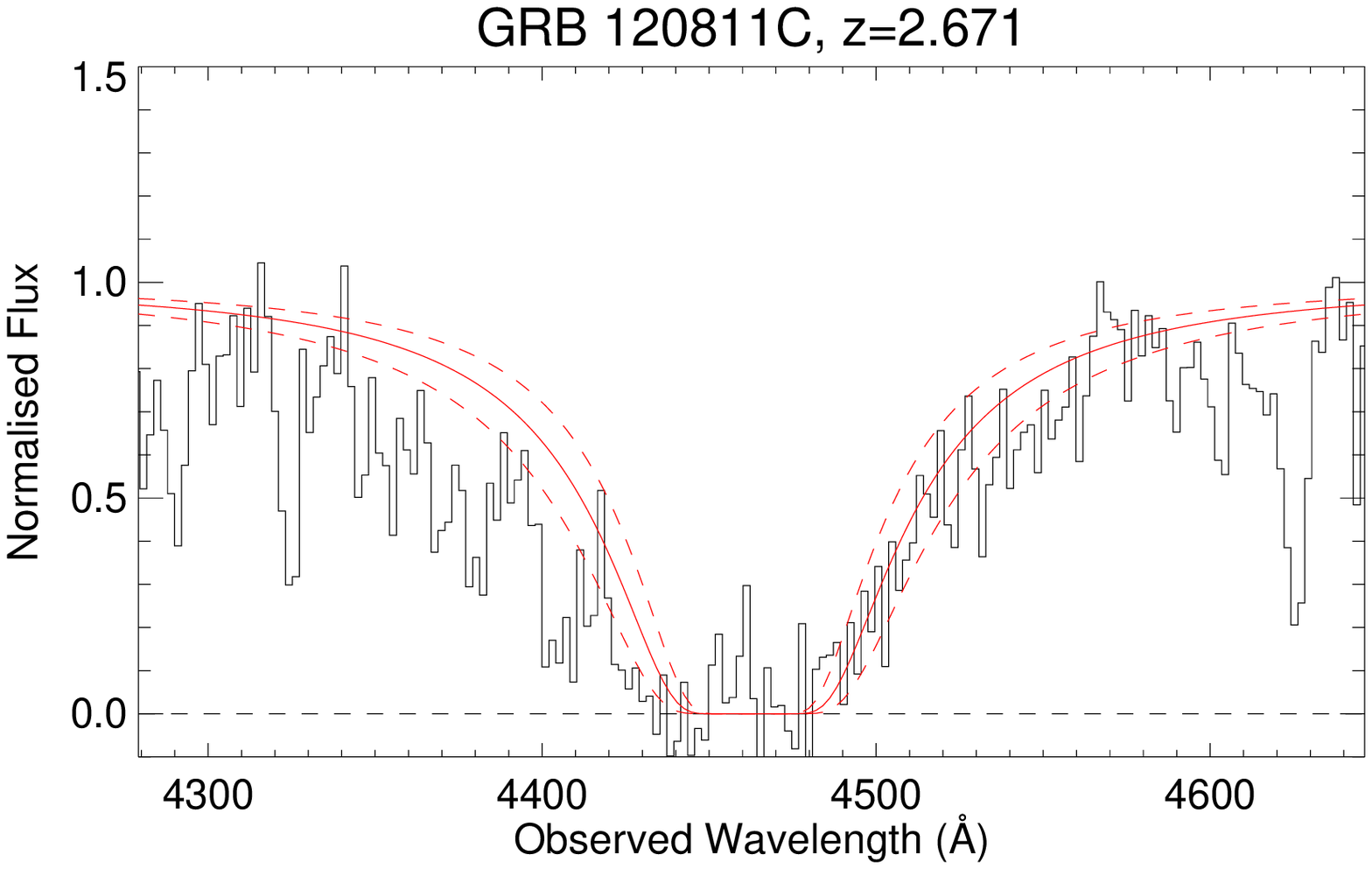}
    \caption{Fit of the red wing of \lya\ for the GRB\,120811C GTC/OSIRIS spectrum.}
    \label{fig:120811C}
\end{figure}

\subsection{GRB\,121027A}
\label{app:121027A}
GRB\,121027A has been suggested as a member of the `ultra-long' class of GRBs,
whose exact nature remains uncertain, but since they also appear to be associated
with massive stars in low metallicity galaxies \citep{Levan2014,Greiner2015b,Kann2017}, we include it in our sample.

\subsection{GRB\,121128A}
\label{app:121128A}
GRB\,121128A was observed by Gemini/GMOS-N, starting 06:28 UT on 28-Nov-2012 \citep{Tanvir2012b}.
A total exposure of $4\times400$\,s was obtained
 with the B600 grating set at 5250\,\AA\ central
wavelength and reduced using the standard Gemini reduction tools within \iraf.
The fit to  the \lya\ line is shown in Figure~\ref{fig:121128A}.
\begin{figure}
\includegraphics[angle=0,width=\columnwidth]{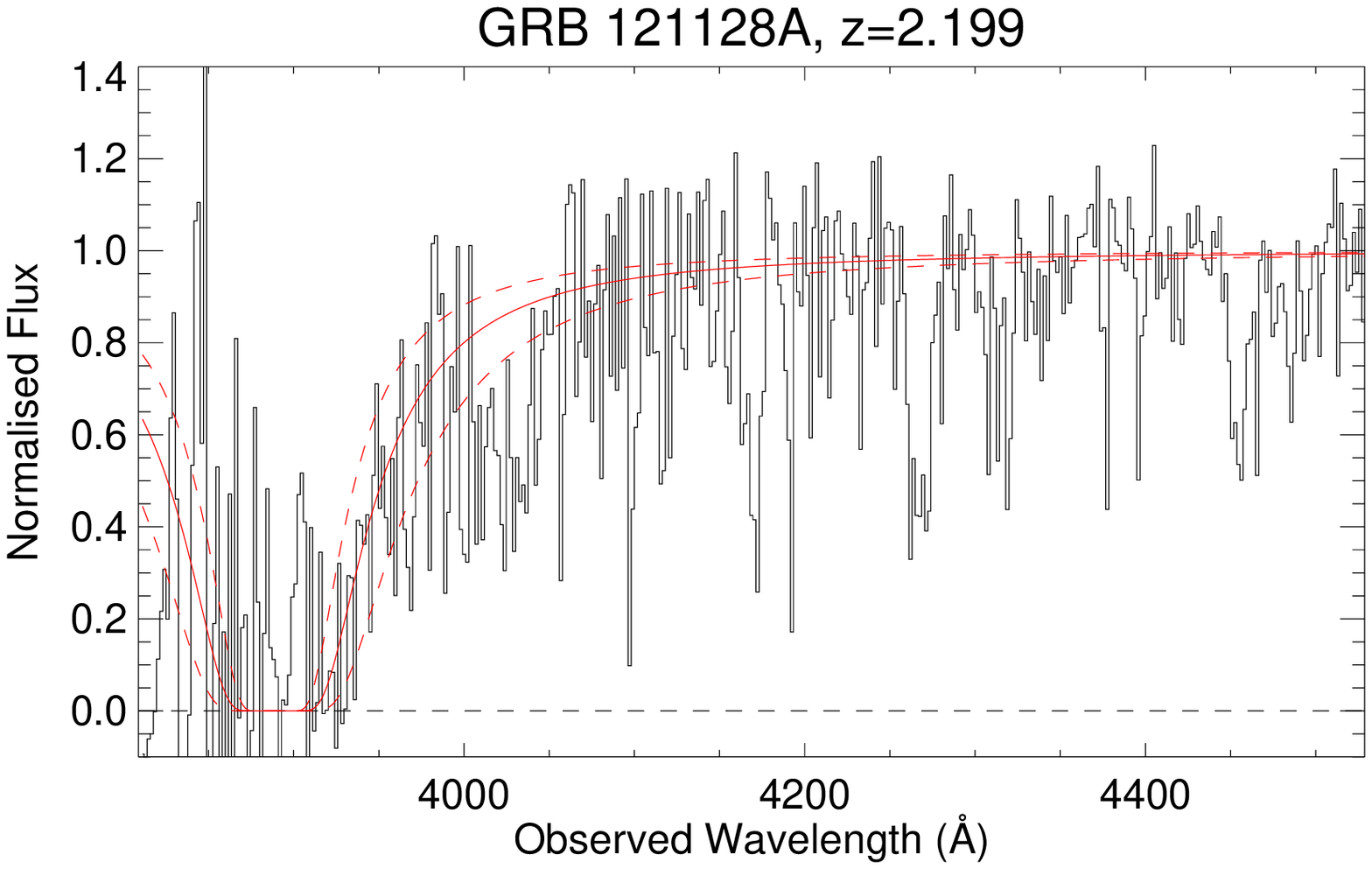}
    \caption{Fit of the red wing of \lya\ for the GRB\,121128A Gemini/GMOS-N spectrum.}
    \label{fig:121128A}
\end{figure}

\subsection{GRB\,130518A}
GRB\,130518A was observed at high resolution with GTC/OSIRIS, starting 04:47 UT on 20-May-2013. 
A total exposure time of 840\,s was obtained, spanning a wavelength range 3700--7800\,\AA\ and
was originally reported in \citet{SanchezRamirez2013}.
The fit to  the \lya\ line is shown in Figure~\ref{fig:130518A}.
\begin{figure}
\includegraphics[angle=0,width=\columnwidth]{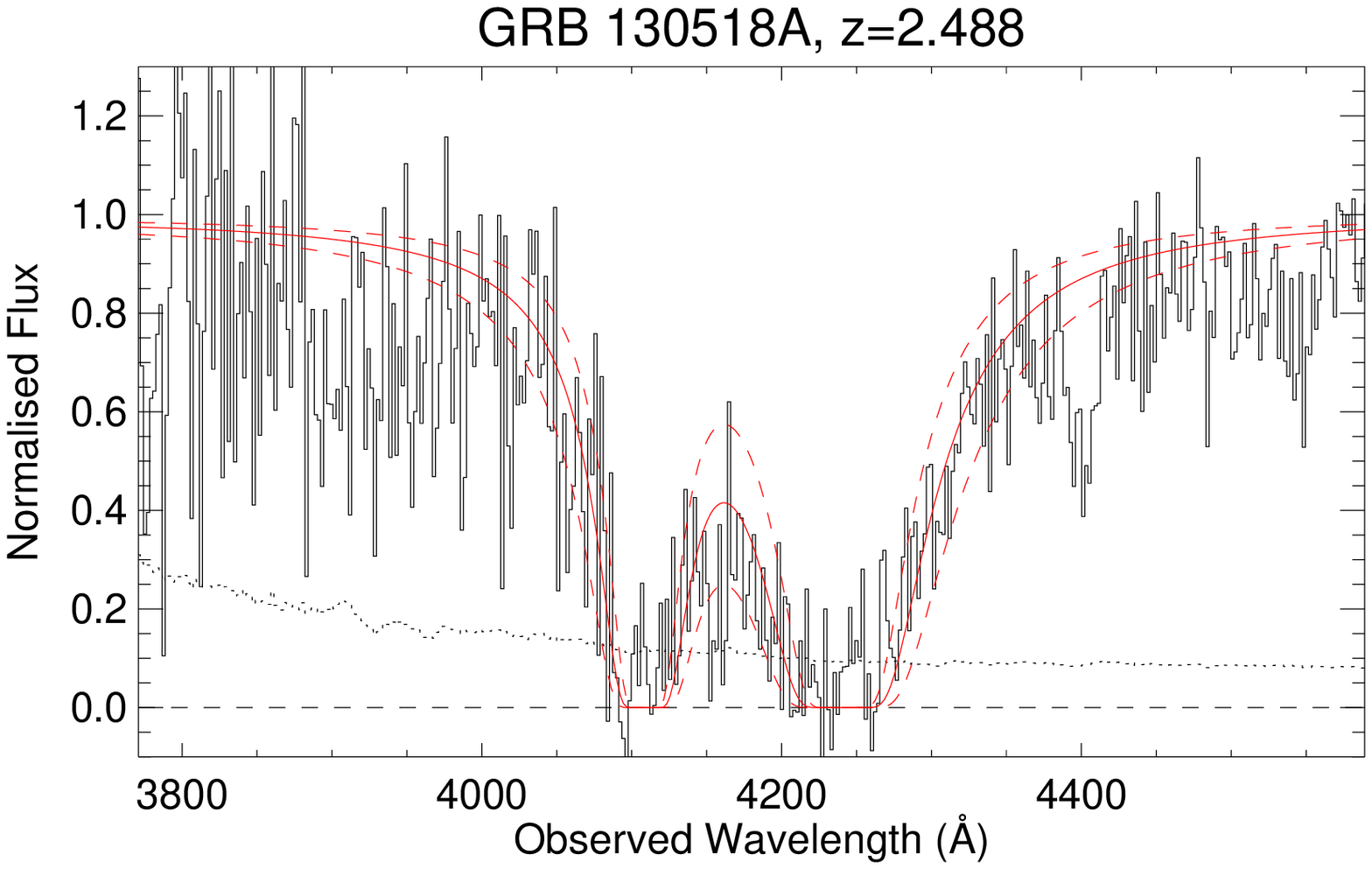}
    \caption{Fit of the red wing of \lya\ for the GRB\,130518A GTC/OSIRIS spectrum. A
    second, intervening DLA, at $z=2.38$ is apparent.}
    \label{fig:130518A}
\end{figure}

\subsection{GRB\,130610A}
GRB\,130610A was observed at high resolution with VLT/UVES, starting 03:25 UT on 10-Jun-2013 \citep{Smette2013}.
A series of exposures were obtained totalling 6500\,s.
The fit to  the \lya\ line is shown in Figure~\ref{fig:130610A}. 
 \begin{figure}
\includegraphics[angle=0,width=\columnwidth]{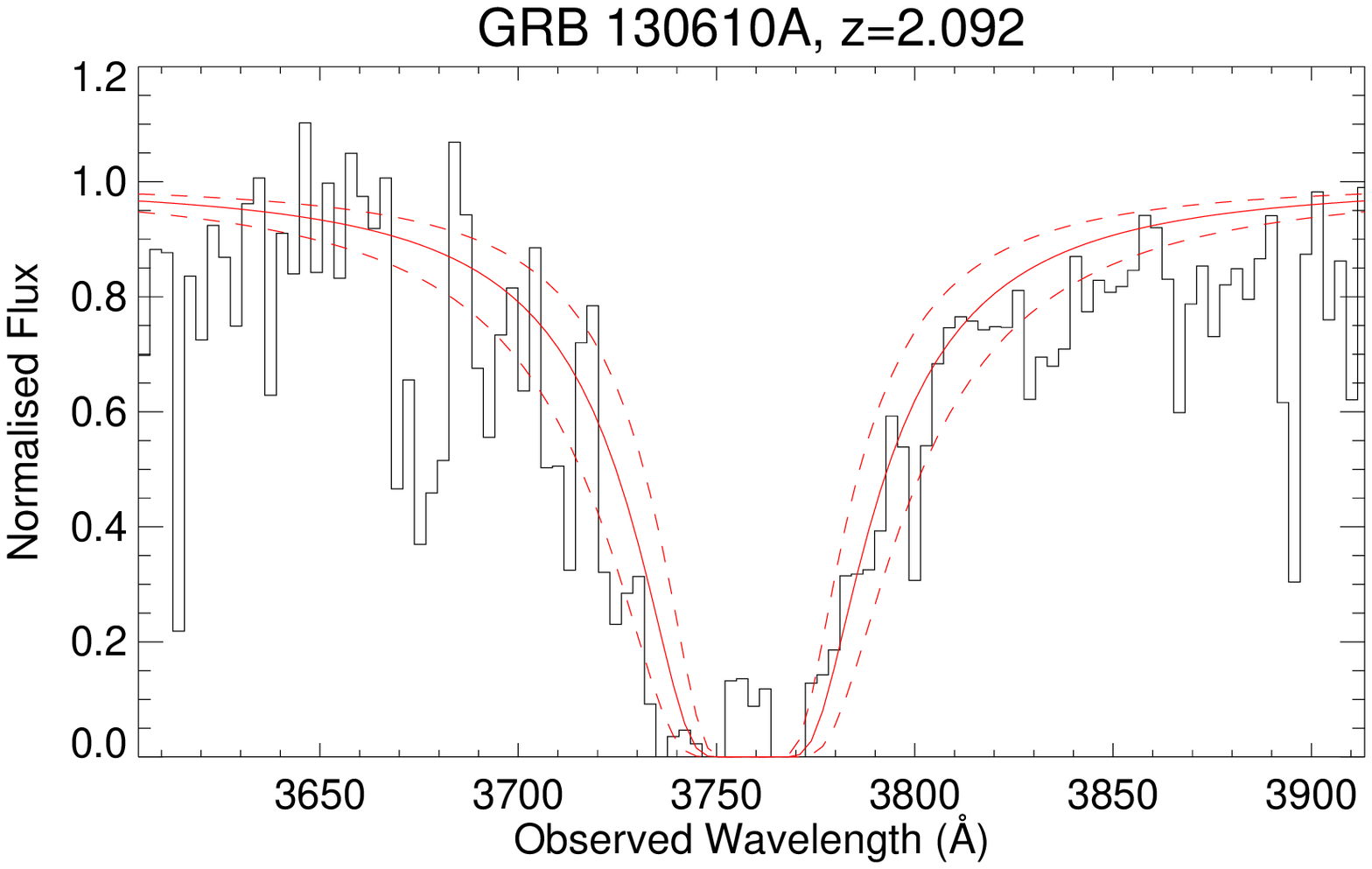}
    \caption{Fit of the red wing of \lya\ for the GRB\,130610A VLT/UVES spectrum.}
    \label{fig:130610A}
\end{figure}

\subsection{GRB\,131108A}
GRB\,131108A was observed with the GTC/OSIRIS, starting 20:42 UT on 08-Nov-2013 \citep{deUP2013}.
A total exposure of 1800\,s was obtained covering a spectral range 3700--7870\,\AA.
The fit to  the \lya\ line is shown in Figure~\ref{fig:131108A}.
\begin{figure}
\includegraphics[angle=0,width=\columnwidth]{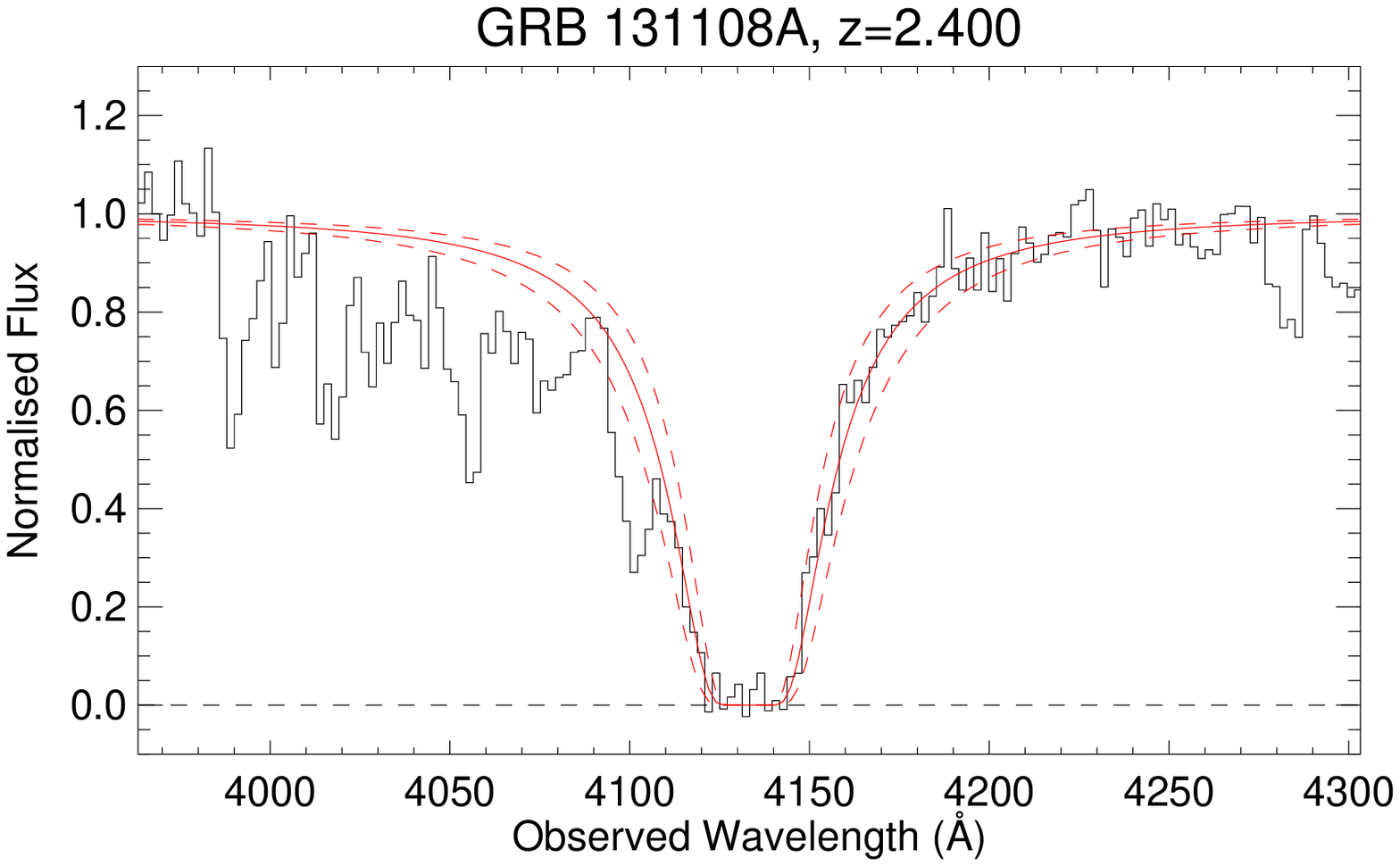}
    \caption{Fit of the red wing of \lya\ for the GRB\,131108A GTC/OSIRIS spectrum.}
    \label{fig:131108A}
\end{figure}

\subsection{GRB\,140206A}
GRB\,140206A was observed with the NOT/ALFOSC starting 19:56 UT
on 6-Feb-2014 \citep[][see also \citet{DElia2014} for TNG/DOLoRes spectroscopy]{Malesani2014}. 
A total exposure of 3600\,s was obtained covering a spectral range 3750--9000\,\AA.
The fit to  the \lya\ line is shown in Figure~\ref{fig:140206A}. 
\begin{figure}
\includegraphics[angle=0,width=\columnwidth]{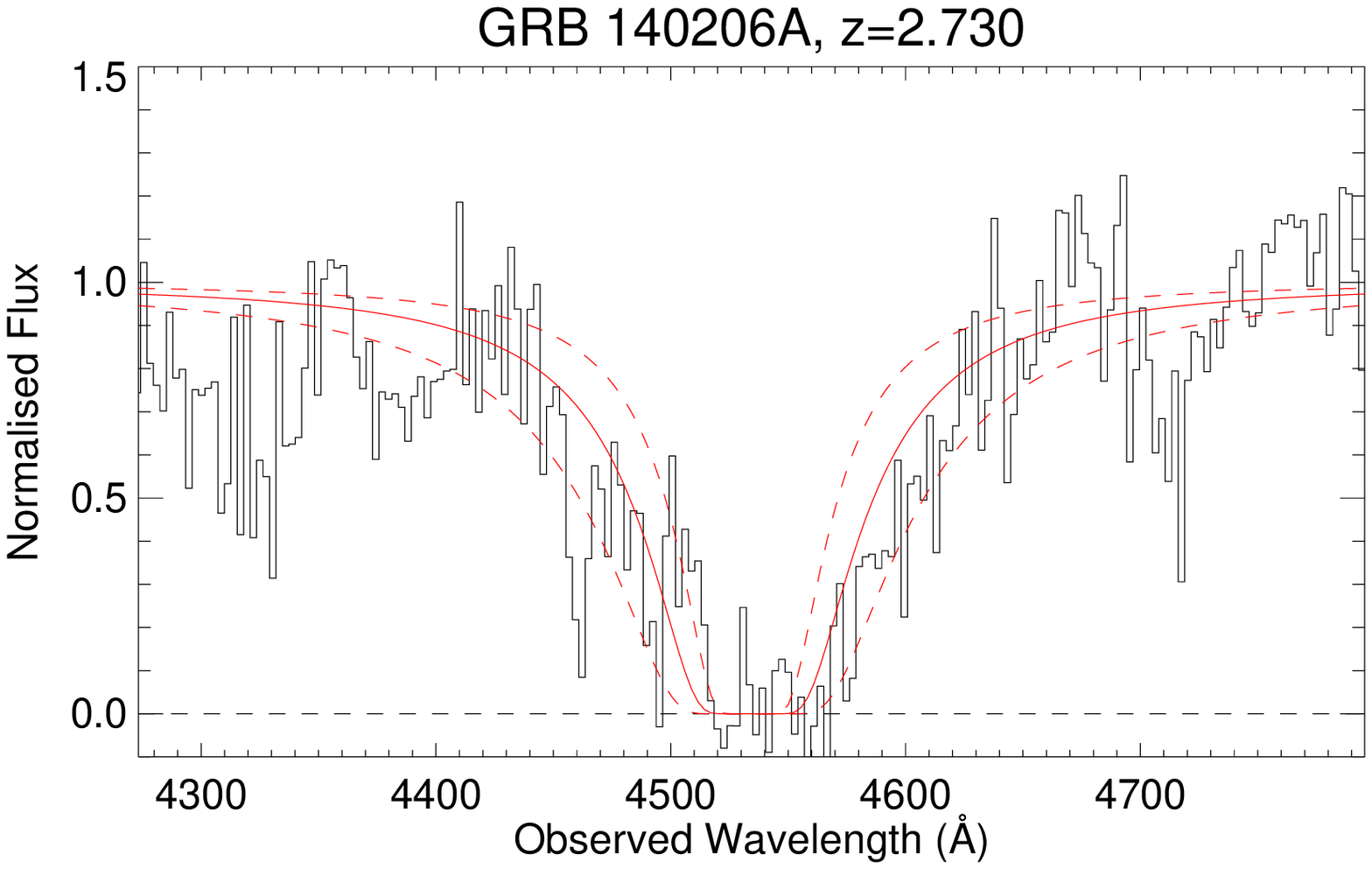}
    \caption{Fit of the red wing of \lya\ for the GRB\,140206A NOT/ALFOSC spectrum.}
    \label{fig:140206A}
\end{figure}

\subsection{GRB\,140515A}
\label{app:140515A}
GRB\,140515A was a high redshift burst observed at several facilities. No metal lines were
confidently detected so the redshift could only be estimated from the \lya\ break itself.
This limits the conclusions that can be drawn, since 
the damping wing must be decomposed into ISM and IGM contributions, which is less
certain in the absence of a precise redshift.
Nonetheless, the sharpness of the break clearly indicates a relatively low \HI\ column-density.
\citet{Chornock2014} obtained a value of $\lognhc=18.62\pm0.08$ assuming an ionized IGM
and $\lognhc=18.43$ in a joint fit including a neutral IGM component
from an early Gemini-N/GMOS spectrum. \citet{Melandri2015} analysed later GTC and VLT
spectroscopy and concluded $\lognhc<18.5$ whereas \citet{Selsing2018} estimate $\lognhc=19.0\pm0.5$.
In this paper we therefore adopt a compromise value of $\lognhc=18.5\pm0.3$.

\subsection{GRB\,140629A}
GRB\,140629A was observed with  the TNG/DOLoRes, starting 02:07 UT on 30-Jun-2014 \citep{DAvanzo2014}.
A total exposure of 1200\,s was obtained, covering a spectral range 3000--8000\,\AA.
The fit to  the \lya\ line is shown in Figure~\ref{fig:140629A}. 
\begin{figure}
\includegraphics[angle=0,width=\columnwidth]{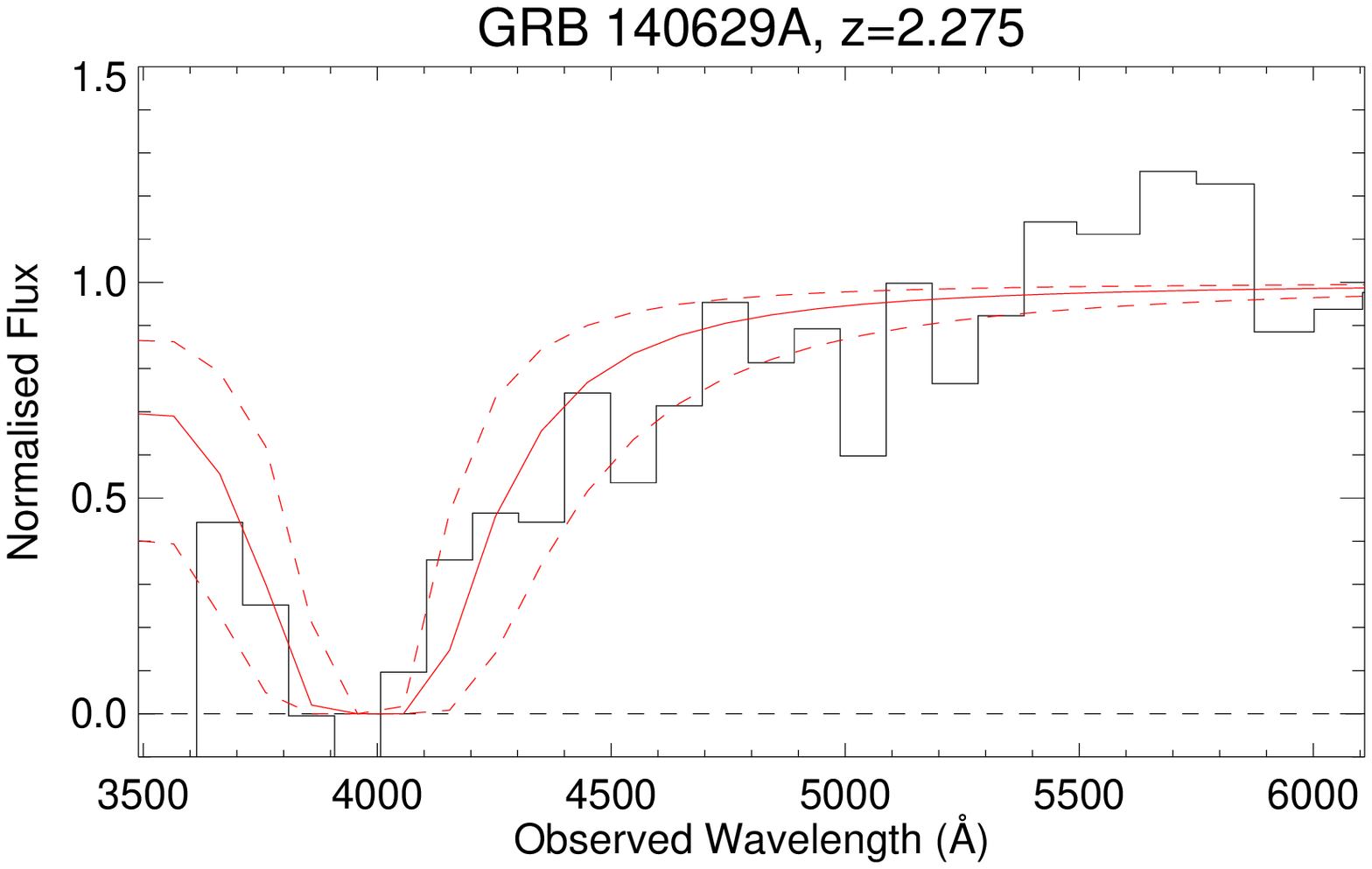}
    \caption{Fit of the red wing of \lya\ for the GRB\,140629A TNG/DOLoRes spectrum.
        }
    \label{fig:140629A}
\end{figure}

\subsection{GRB\,140703A}
GRB\,140703A was observed with  the GTC/OSIRIS, starting 03:16 UT on 3-Jul-2014.
A total exposure of 450\,s was obtained covering a spectral range 3700--10000\,\AA\ and
was originally reported in \citet{CastroTirado2014}.
The fit to  the \lya\ line is shown in Figure~\ref{fig:140703A}. 
\begin{figure}
\includegraphics[angle=0,width=\columnwidth]{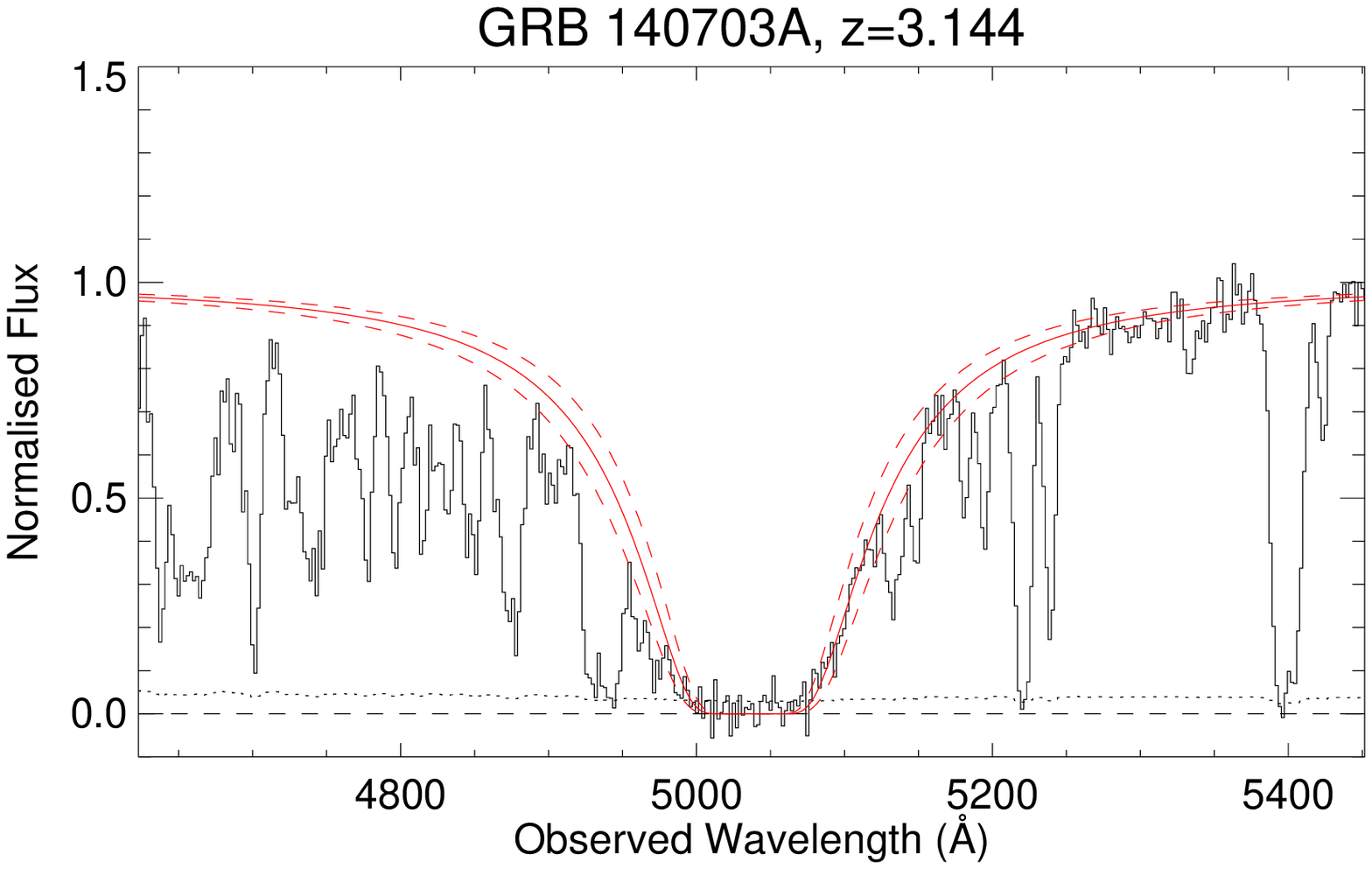}
    \caption{Fit of the red wing of \lya\ for the GRB\,140703A GTC/OSIRIS  spectrum.
        }
    \label{fig:140703A}
\end{figure}

\subsection{GRB\,140808A}
GRB\,140808A was observed with the GTC/OSIRIS, starting 00:54 UT on 08-Aug-2014 \citep{Gorosabel2014}.
A total exposure of 3600\,s was obtained covering a spectral range 3700--7800\,\AA.
The fit to  the \lya\ line is shown in Figure~\ref{fig:140808A}.
\begin{figure}
\includegraphics[angle=0,width=\columnwidth]{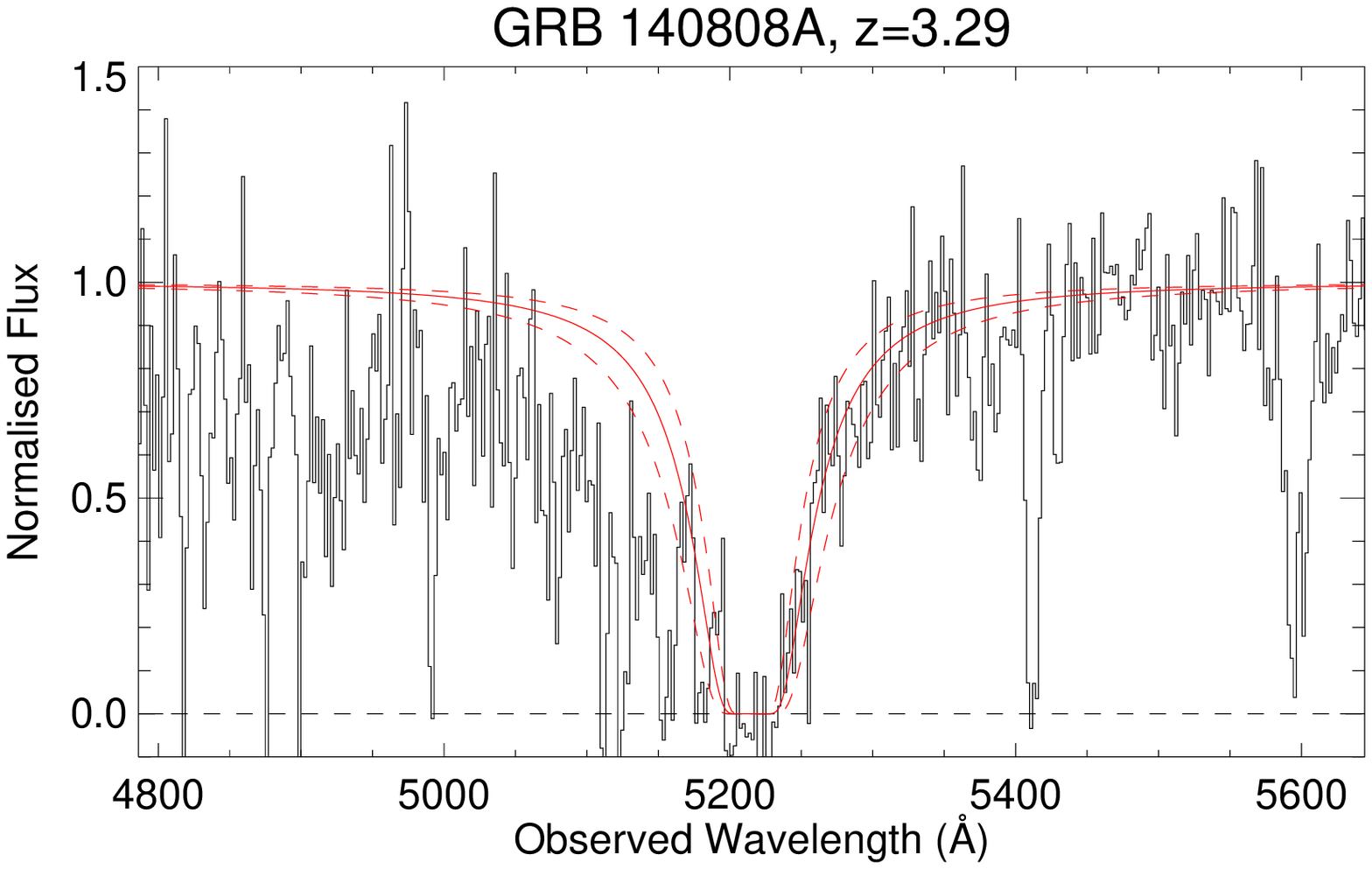}
    \caption{Fit of the red wing of \lya\ for the GRB\,140808A GTC/OSIRIS spectrum.}
    \label{fig:140808A}
\end{figure}

\subsection{GRB\,150413A}
GRB\,150413A was observed with the Asiago(CT)/AFOSC, starting 20:53 UT on 13-Apr-2015 \citep{deUP2015}.
A total exposure of 1800\,s was obtained covering a wavelength range 3400--8200\,\AA.
The fit to  the \lya\ line is shown in Figure~\ref{fig:150413A}.
\begin{figure}
\includegraphics[angle=0,width=\columnwidth]{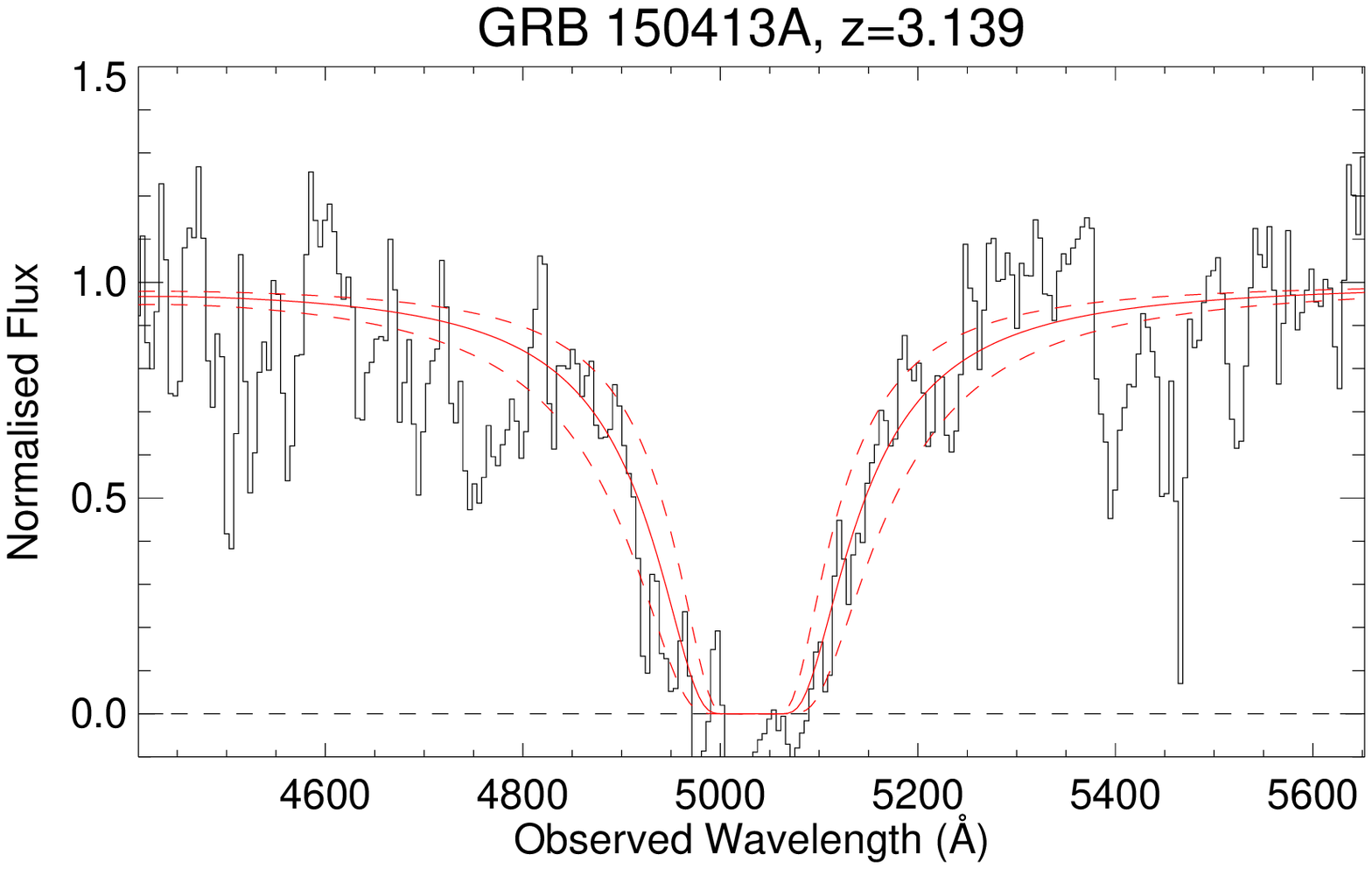}
    \caption{Fit of the red wing of \lya\ for the GRB\,150413A CT/AFOSC spectrum.}
    \label{fig:150413A}
\end{figure}

\subsection{GRB\,151215A}
GRB\,151215A was observed with the NOT/ALFOSC, starting 04:13 UT on 15-Dec-2015 \citep{Xu2015}. 
A total exposure of $3\times1200$\,s was obtained covering a spectral range 3200--9000\,\AA.
The fit to  the \lya\ line is shown in Figure~\ref{fig:151215A}. 
\begin{figure}
\includegraphics[angle=0,width=\columnwidth]{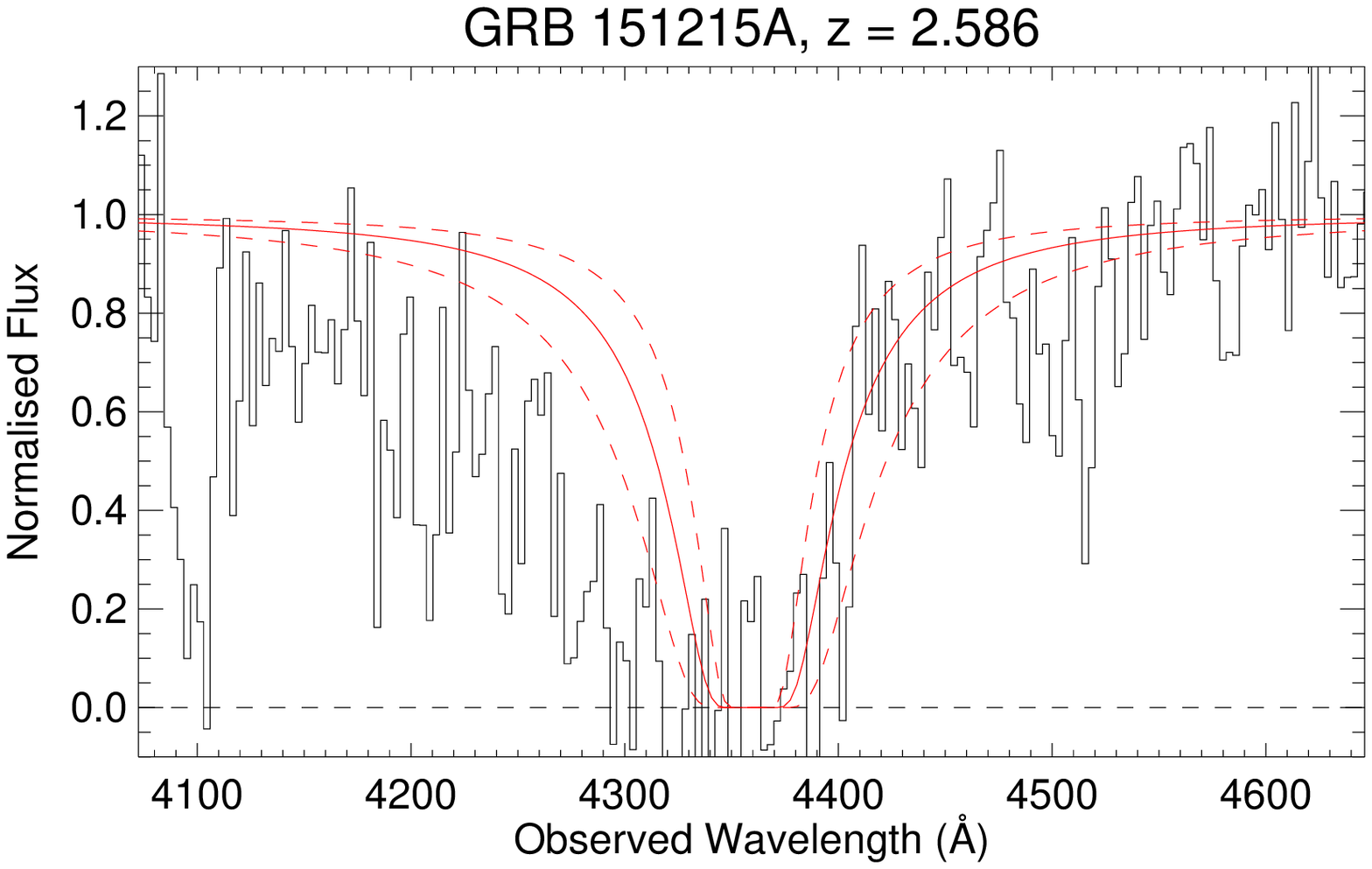}
    \caption{Fit of the red wing of \lya\ for the GRB\,151215A NOT/ALFOSC spectrum.}
    \label{fig:151215A}
\end{figure}

\subsection{GRB\,160227A}
GRB\,160227A was observed with the NOT/ALFOSC, starting at 20:19 UT on 27-Feb-2016 \citep{Xu2016}.
A total exposure of 4800\,s was obtained covering a wavelength range 3200--9000\,\AA.
The fit to  the \lya\ line is shown in Figure~\ref{fig:160227A}. 
\begin{figure}
\includegraphics[angle=0,width=\columnwidth]{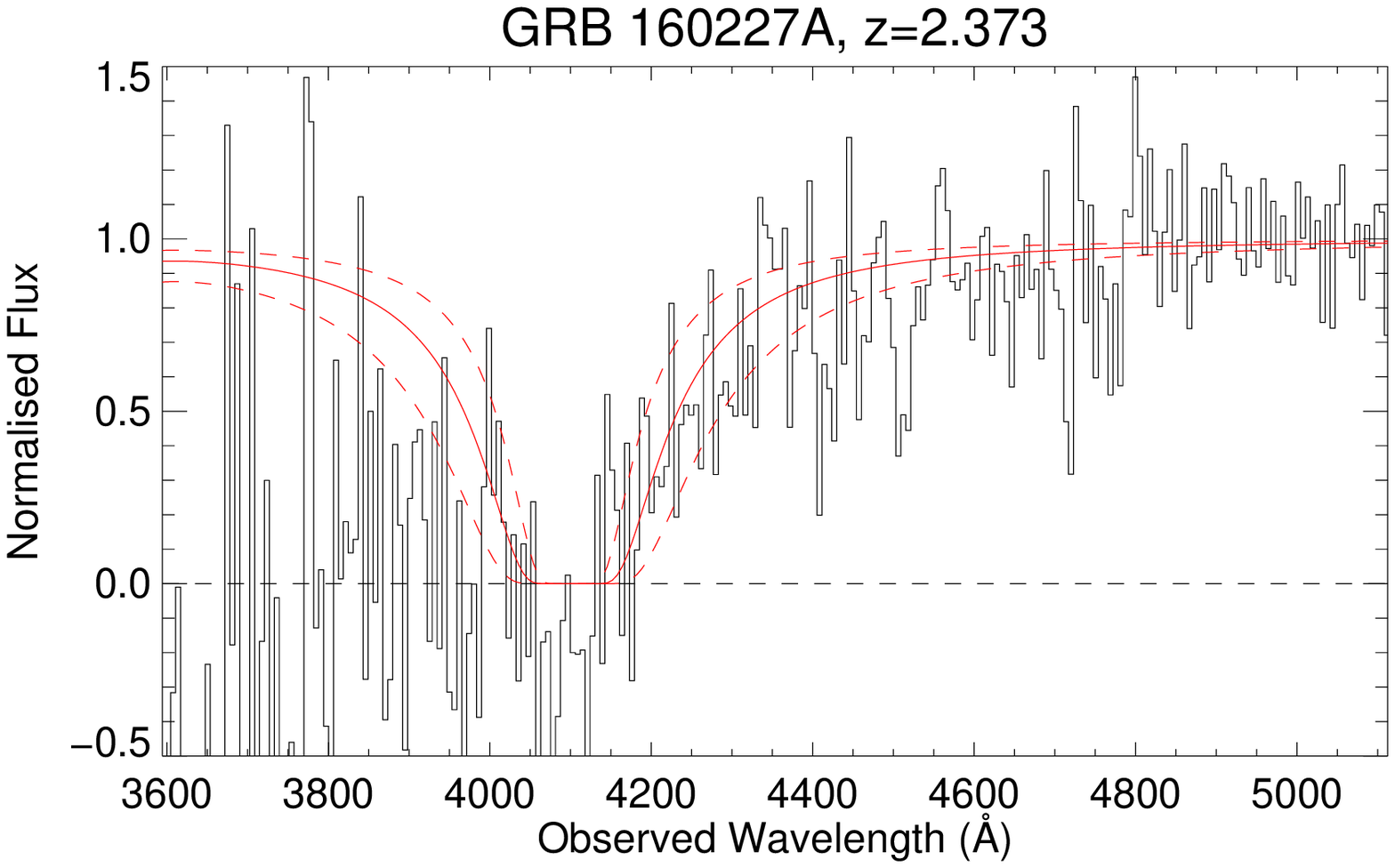}
    \caption{Fit of the red wing of \lya\ for the GRB\,160227A NOT/ALFOSC spectrum.}
    \label{fig:160227A}
\end{figure}

\subsection{GRB\,160629A}
GRB\,160629A was observed with the GTC/OSIRIS, starting at 04:40 UT on 30-Jun-2016 \citep{CastroTirado2016}.
A total exposure of 600\,s was obtained covering a wavelength range 3700--7880\,\AA.
The fit to  the \lya\ line is shown in Figure~\ref{fig:160629A}. 
\begin{figure}
\includegraphics[angle=0,width=\columnwidth]{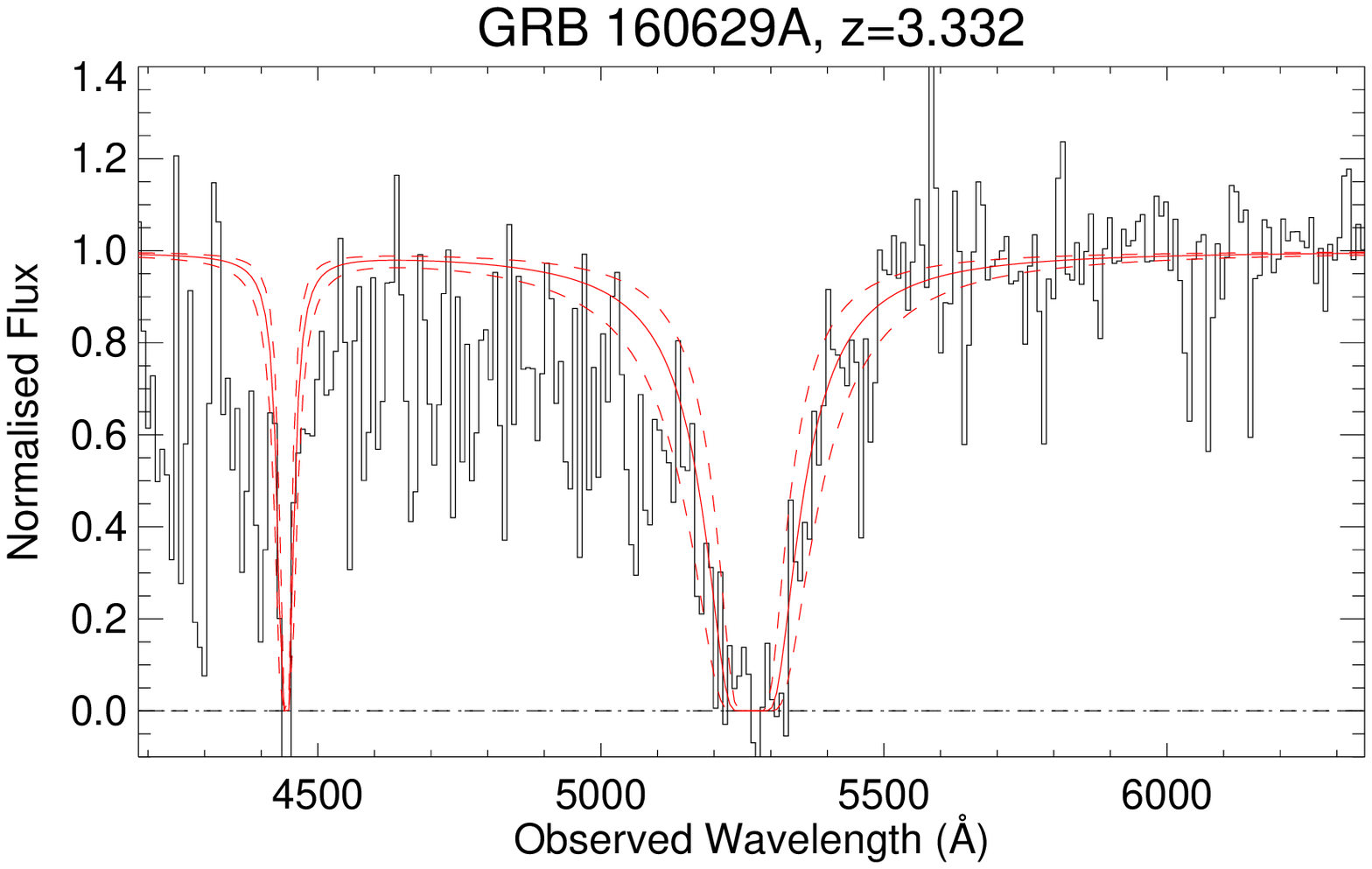}
    \caption{Fit of the red wing of \lya\ for the GRB\,160629A GTC/OSIRIS spectrum.}
    \label{fig:160629A}
\end{figure}

\subsection{GRB\,161017A}
GRB\,161017A was observed with the TNG/DOLoRes, starting at  04:27 UT on 18-Oct-2016 \citep{DAvanzo2016}.
A total exposure of 1200\,s was obtained covering a wavelength range 3500--8000\,\AA.
The fit to  the \lya\ line is shown in Figure~\ref{fig:161017A}. 
\begin{figure}
\includegraphics[angle=0,width=\columnwidth]{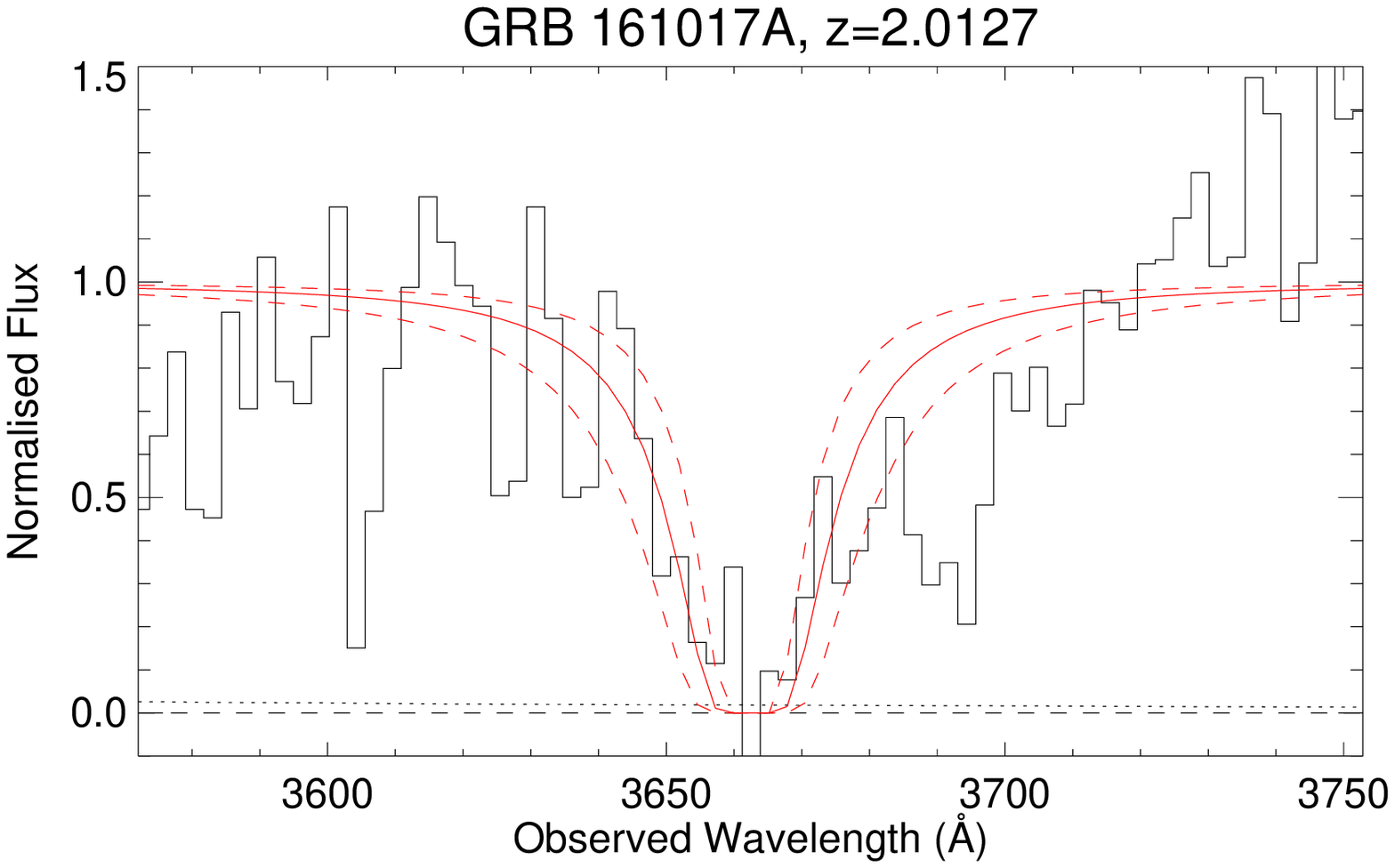}
    \caption{Fit of the red wing of \lya\ for the GRB\,161017A  TNG/DOLoRes spectrum.}
    \label{fig:161017A}
\end{figure}

For this GRB we also conducted a host search using the WHT/ACAM on 7 April 2017.
In seeing of 1.1\,arcsec we obtained a 45\,min integration in the $g$-band.
No source was detected at the GRB position down to a 2$\sigma$ limiting magnitude
of $g=24.70$, which is corrected for foreground Milky Way extinction \citep{Schlafly2011}.

\subsection{GRB\,170405A}
GRB\,170405A was observed with the GTC/OSIRIS, starting at 02:14 UT on 6-Apr-2017 \citep{deUP2017b}. 
A total exposure of $3\times900$\,s was obtained covering a wavelength range 3700--7800\,\AA.
The fit to  the \lya\ line is shown in Figure~\ref{fig:170405A}. 
\begin{figure}
\includegraphics[angle=0,width=\columnwidth]{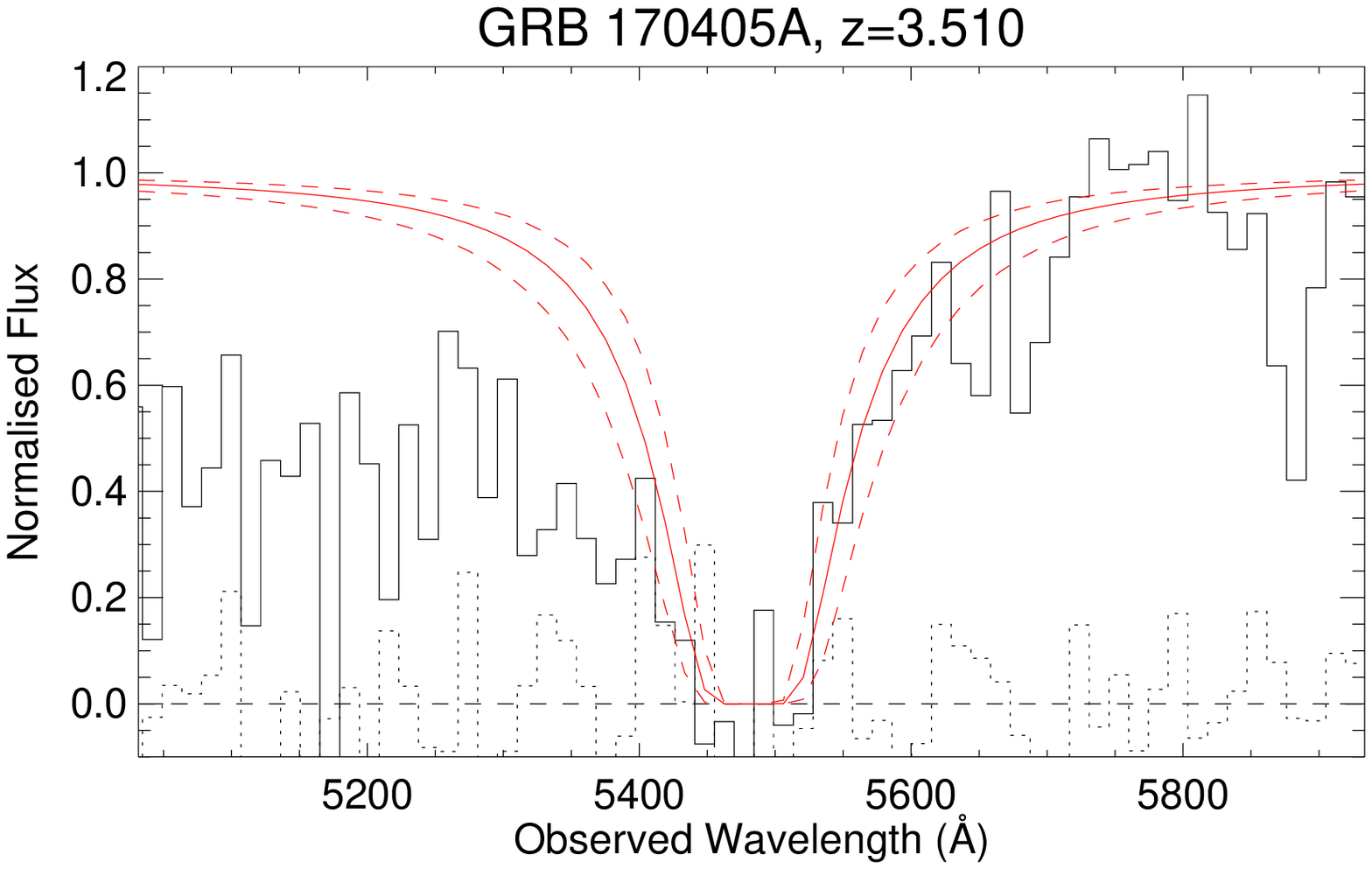}
    \caption{Fit of the red wing of \lya\ for the GRB\,170405A GTC/OSIRIS spectrum.}
    \label{fig:170405A}
\end{figure}

\subsection{GRB\,170531B}
GRB\,170531B was observed with the GTC/OSIRIS, starting at 02:47 UT on 1-Jun-2017 \citep{deUP2017c}. 
A total exposure of $3\times900$\,s was obtained covering a wavelength range 3700--7880\,\AA.
The fit to  the \lya\ line is shown in Figure~\ref{fig:170531B}. 
\begin{figure}
\includegraphics[angle=0,width=\columnwidth]{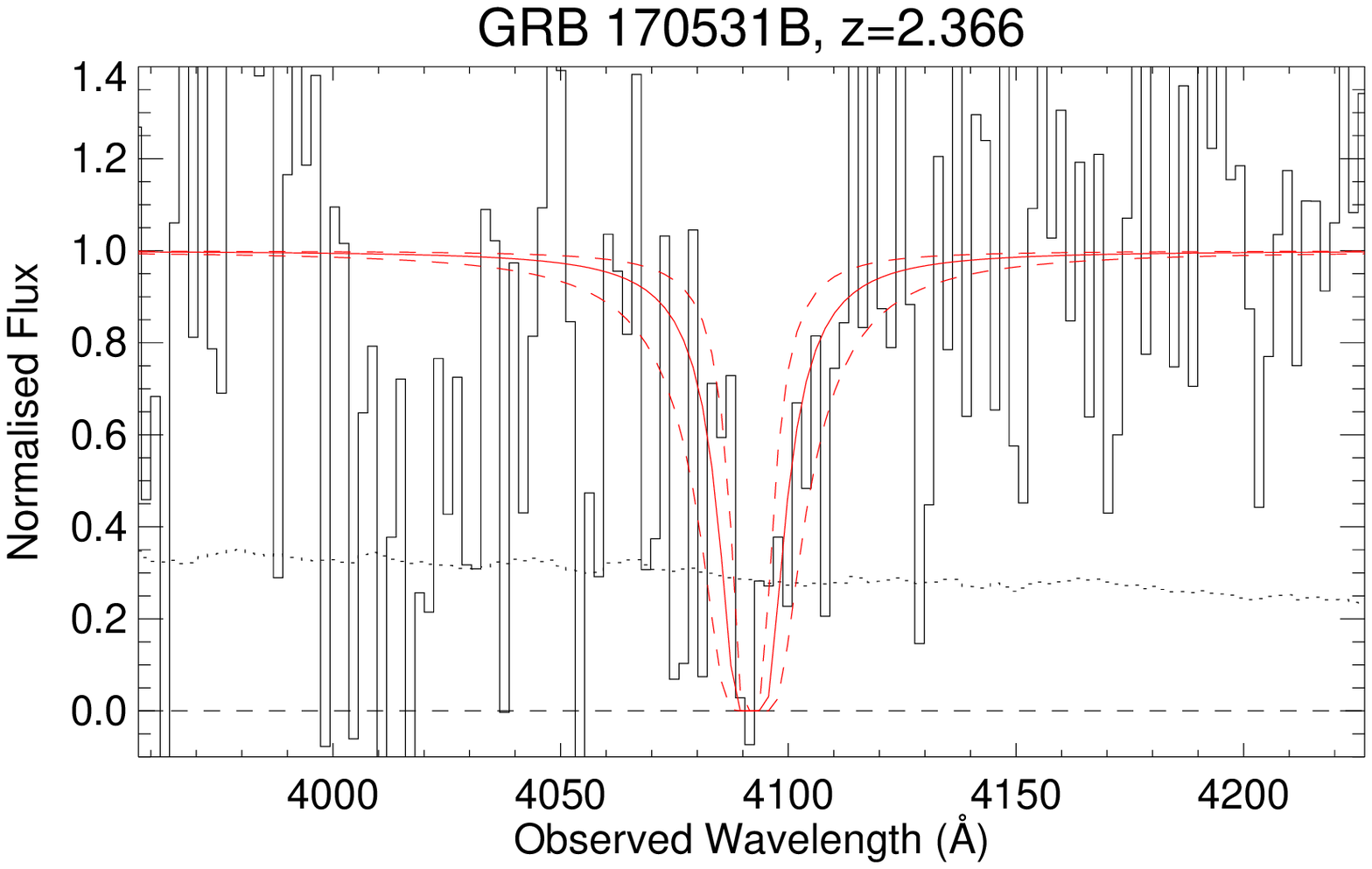}
    \caption{Fit of the red wing of \lya\ for the GRB\,170531B GTC/OSIRIS spectrum.}
    \label{fig:170531B}
\end{figure}

\subsection{GRB\,180115A}
GRB\,180115A was observed with the GTC/OSIRIS, starting at 20:32 UT on 15-Jan-2018 \citep{deUP2018}. 
A total exposure of $3\times900$\,s was obtained covering a wavelength range 3700--7880\,\AA.
The fit to  the \lya\ line is shown in Figure~\ref{fig:180115A}.  In this case, no metal lines were detected,
so the redshift is based solely on \lya.
\begin{figure}
\includegraphics[angle=0,width=\columnwidth]{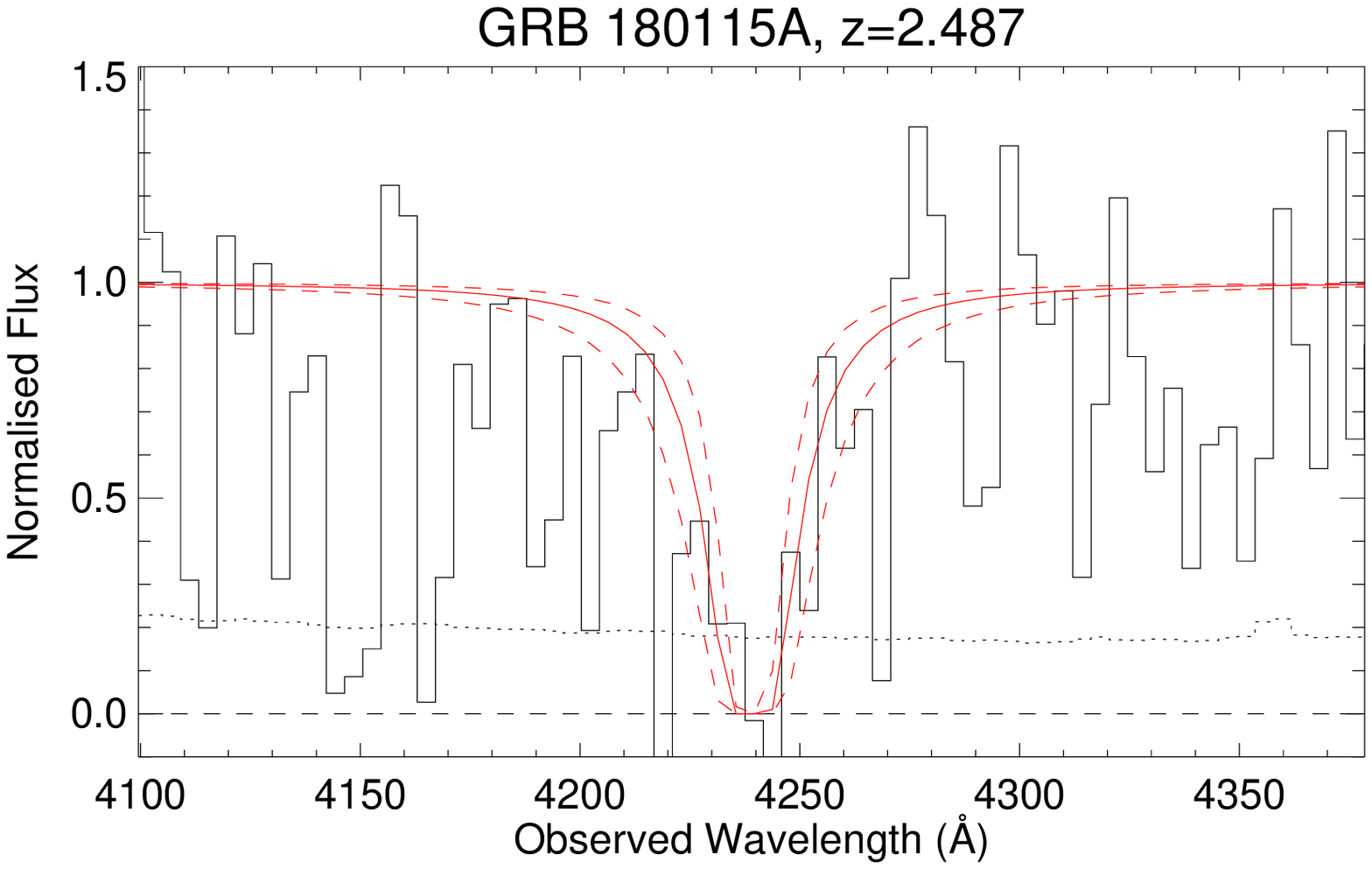}
    \caption{Fit of the red wing of \lya\ for the GRB\,180115A GTC/OSIRIS spectrum.}
    \label{fig:180115A}
\end{figure}

\subsection{GRB\,180329B}
GRB\,180329B was observed with the VLT/X-shooter, starting at 00:10 UT on 30-Mar-2018 \citep{Izzo2018}. 
A total exposure of $2\times600$\,s was obtained covering a wavelength range 3000--21000\,\AA.
The fit to  the \lya\ line is shown in Figure~\ref{fig:180329B}. 
\begin{figure}
\includegraphics[angle=0,width=\columnwidth]{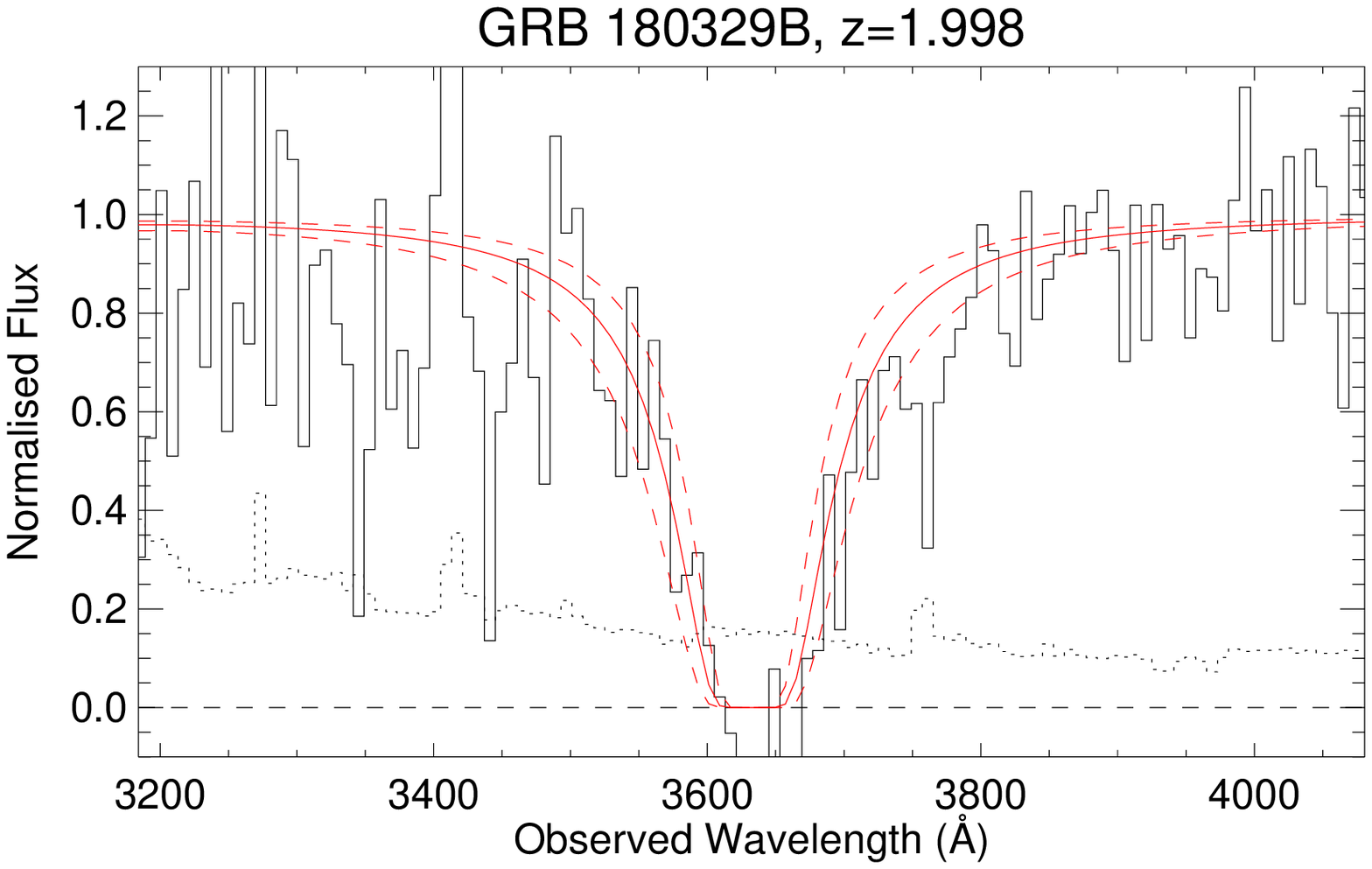}
    \caption{Fit of the red wing of \lya\ for the GRB\,180329B VLT/X-shooter spectrum.}
    \label{fig:180329B}
\end{figure}

\bsp	
\label{lastpage}
\end{document}